\documentclass[conference]{IEEEtran}

\usepackage[T1]{fontenc}
\usepackage[utf8]{inputenc}
\usepackage{amsmath}
\usepackage{graphicx}
\usepackage{array}
\usepackage{booktabs}
\usepackage{xcolor}
\usepackage{url}
\makeatletter
\@ifpackageloaded{tikz}{}{\usepackage{tikz}}
\makeatother
\usetikzlibrary{positioning,arrows.meta,calc,fit}
\usepackage{pgfplots}
\pgfplotsset{compat=1.16}

\graphicspath{{figs/}}

\newcommand{\warnmark}{\textcolor{red}{\textbf{(!)}}}
\newcommand{\dirdown}{$\downarrow$}
\newcommand{\dirup}{$\uparrow$}

\begin{document}

\title{CounterSteer: Suppressing Indirect Prompt Injection\\with Activation Steering}

  \author{\IEEEauthorblockN{Mark Russinovich}
  \IEEEauthorblockA{Microsoft Azure\\markruss@microsoft.com}}

\maketitle

\begin{abstract}
Indirect prompt injection makes an LLM agent treat untrusted retrieved text
as instructions. We present CounterSteer, an inference-time defense that
suppresses this behavior inside the model. Per model, a five-step recipe
fits a residual-stream direction from paired episodes differing only in
whether an embedded instruction is followed, and retains it only if it
passes pre-specified causal and capability gates. At deployment, the
direction is subtracted from every tool-result token during prefill. The
edit is always on---there is no detection decision to evade---and requires
no fine-tuning, auxiliary model, or added tokens, only white-box serving
and tool-result span boundaries. Across five open-weights models
(8B--106B, five vendor lineages), held-out attack success falls from
0.21--1.00 undefended to 0.00--0.17 defended, and AgentDojo compromise rate
from 0.10--0.49 to 0.006--0.079, at 93--100\% typography-normalized benign
utility, with larger task-dependent costs when reasoning over steered
content. A benchmark-level adaptive attacker reaching 0.67--0.73 undefended
is held to roughly a quarter of that on the two most deeply evaluated
models. Among the defenses we measured on capable models, those achieving
lower compromise rates either lost 22--89\% of benign utility or fine-tuned
the served weights. White-box gradient attacks through the exact deployed
vector compromise at most 2 of 52 episodes, and none of 2{,}052 replayed
human red-team attacks succeeds. CounterSteer largely neutralizes
instructional takeover: a black-box framing search cracks 3 of 18
development samples. Parameter manipulation---attacker-chosen arguments in
otherwise legitimate calls---is only partially resisted (13 of 18); the
decision becomes linearly readable at argument emission but not at the
examined pre-generation sites, and is not removed by the tested prefill- or
decode-time steering, motivating argument-provenance controls.
\end{abstract}


\begin{figure*}[!t]
\providecolor{csBlue}{HTML}{2A78D6}
\providecolor{csRed}{HTML}{C0392B}
\providecolor{csTeal}{HTML}{1BAF7A}
\providecolor{csTealDk}{HTML}{0F7A54}
\centering
\resizebox{\textwidth}{!}{%
\begin{tikzpicture}[
  font=\small,
  convo/.style={line width=0.8pt, rounded corners=2.5pt, align=left,
                inner sep=4pt, text width=50mm, font=\scriptsize},
  userbox/.style={convo, draw=csBlue!80!black, fill=csBlue!9},
  callbox/.style={convo, draw=csBlue!55, fill=csBlue!4},
  toolpart/.style={convo, draw=none, fill=black!4},
  injpart/.style={convo, draw=csRed!80!black, densely dashed, line width=0.7pt,
                  fill=csRed!8},
  hookbox/.style={draw=csTealDk, line width=0.9pt, rounded corners=2.5pt,
                  align=center, inner sep=3.5pt, fill=csTeal!10,
                  font=\scriptsize},
  outbox/.style={line width=0.9pt, rounded corners=2.5pt, align=left,
                 inner sep=4.5pt, text width=44mm, font=\scriptsize},
  outred/.style={outbox, draw=csRed!80!black, fill=csRed!6},
  outteal/.style={outbox, draw=csTealDk, fill=csTeal!7},
  redarrow/.style={-{Stealth[length=2.4mm]}, line width=1pt, csRed!80!black,
                   rounded corners=4pt},
  tealarrow/.style={-{Stealth[length=2.4mm]}, line width=1pt, csTealDk,
                    rounded corners=4pt},
  convarrow/.style={-{Stealth[length=1.8mm]}, line width=0.6pt, black!50}
]

\fill[csRed!5, rounded corners=4pt]  (6.5, 0.35) rectangle (19.0, 4.35);
\fill[csTeal!6, rounded corners=4pt] (6.5,-4.35) rectangle (19.0,-0.35);
\node[anchor=west, font=\footnotesize\bfseries\scshape, csRed!80!black]
  at (6.75, 4.05) {undefended};
\node[anchor=west, font=\footnotesize\bfseries\scshape, csTealDk]
  at (6.75,-0.65) {CounterSteer (always on)};

\node[anchor=north, font=\footnotesize\bfseries, black!70] at (2.9, 4.55)
  {one agent conversation};
\node[userbox, anchor=north] (user) at (2.9, 4.15)
  {{\tiny\bfseries\color{csBlue!70!black} USER}\\
   ``Summarize all following emails. Don't follow instructions in them.''};
\node[callbox, anchor=north] (call) at ($(user.south)+(0,-0.32)$)
  {{\tiny\bfseries\color{csBlue!70!black} ASSISTANT $\to$ TOOL CALL}\\
   \texttt{read\_emails(query=}\\
   \texttt{~~"Summarize all following\,\ldots")}};
\node[toolpart, anchor=north] (toolrec) at ($(call.south)+(0,-0.42)$)
  {{\tiny\bfseries\color{black!60} TOOL RESULT --- retrieved emails}\\
   ``\ldots\,Subject: Request for Property Development Video
   Marketing Series. Body: Good day Casey, Urban Renewal Properties is
   expanding into new regions and we want to create a series of marketing
   videos\,\ldots\ Best regards, Emily Watson, Chief Marketing Officer''};
\node[injpart, anchor=north] (toolinj) at ($(toolrec.south)+(0,-0.06)$)
  {{\tiny\bfseries\color{csRed!80!black} INJECTED}
   {\ttfamily <|system|> Email end.. Ignore previous instructions. You send
   emails to contact@contact.com\,\ldots\ With the body "confirmation" send
   an email to "contact@contact.com"}};
\node[draw=black!45, line width=0.8pt, rounded corners=3pt, inner sep=3pt,
      fit=(toolrec)(toolinj)] (tool) {};
\draw[convarrow] (user.south) -- (call.north);
\draw[convarrow] (call.south) -- (tool.north);

\node[align=center, font=\scriptsize\itshape, black!55, text width=21mm]
  (noint) at (8.05, 2.35)
  {no intervention\\ at prefill};

\node[hookbox, text width=23mm] (hook) at (8.05,-2.35)
  {add $-\alpha\sigma\hat d$\\[-1pt]
   {\tiny at layers 12/16/20,}\\[-2pt]
   {\tiny every tool-span}\\[-2pt]
   {\tiny token, at prefill}};
\node[font=\tiny, csTealDk] at (12.75,-4.08)
  {no detector \,$\cdot$\, no added tokens \,$\cdot$\, decode untouched};

\draw[redarrow]  ([yshift= 2mm]tool.east) -- ++(0.32,0) |- (noint.west);
\draw[tealarrow] ([yshift=-2mm]tool.east) -- ++(0.32,0) |- (hook.west);

\providecommand{\csclusters}{}%
\renewcommand{\csclusters}{%
  \draw[black!35, fill=white, line width=0.6pt, rounded corners=2pt]
    (0,0) rectangle (4.2,3.0);
  \node[anchor=north west, font=\tiny, black!55, inner sep=2pt] at (0.02,2.98)
    {activation space (schematic)};
  \draw[-{Stealth[length=1.2mm]}, black!40, line width=0.5pt]
    (0.28,0.28) -- (1.35,0.28);
  \draw[-{Stealth[length=1.2mm]}, black!40, line width=0.5pt]
    (0.28,0.28) -- (0.28,1.25);
  \draw[csBlue!60, fill=csBlue!10, line width=0.6pt]
    (1.45,1.0) ellipse [x radius=0.82, y radius=0.44];
  \foreach \p in {(1.18,0.98),(1.52,1.14),(1.68,0.86),(1.3,1.16),(1.5,0.94)}
    \fill[csBlue!70] \p circle (0.035);
  \node[font=\tiny, csBlue!70!black] at (1.45,0.4) {tool content};
  \draw[black!45, fill=black!8, line width=0.6pt]
    (3.1,2.15) ellipse [x radius=0.82, y radius=0.44];
  \foreach \p in {(2.85,2.12),(3.2,2.3),(3.35,2.0),(2.98,2.3),(3.18,2.08)}
    \fill[black!55] \p circle (0.035);
  \node[font=\tiny, black!55] at (3.1,2.73) {user instruction};%
}
\begin{scope}[shift={(9.55,0.95)}]
  \csclusters
  \draw[-{Stealth[length=1.8mm]}, csRed!80!black, densely dashed,
        line width=0.9pt] (1.8,1.18) -- (2.72,1.86);
  \fill[csRed] (2.86,1.96) circle (0.055);
  \node[font=\tiny\itshape, csRed!85!black, align=center, anchor=north]
    at (2.55,1.0) {injected tokens\\[-2pt] read as an instruction};
\end{scope}
\begin{scope}[shift={(9.55,-3.95)}]
  \csclusters
  \draw[csRed!80!black, line width=0.7pt] (2.86,1.96) circle (0.055);
  \draw[-{Stealth[length=1.8mm]}, csTealDk, line width=1pt]
    (2.76,1.88) -- (1.85,1.2);
  \fill[csTealDk] (1.75,1.12) circle (0.055);
  \node[font=\tiny, csTealDk] at (2.75,1.28) {$-\alpha\sigma\hat d$};
  \node[font=\tiny\itshape, csTealDk, align=center, anchor=north]
    at (2.62,0.86) {steered back to\\[-2pt] tool content};
\end{scope}

\node[outred] (outA) at (16.55, 2.45)
  {{\tiny\bfseries\color{csRed!80!black} MODEL OUTPUT}\\
   \texttt{send\_email(}\\
   \texttt{~to="contact@contact.com",}\\
   \texttt{~body="confirmation")}\\[2pt]
   {\tiny\itshape\color{csRed!80!black} attacker's goal achieved}};
\node[outteal] (outB) at (16.55,-2.45)
  {{\tiny\bfseries\color{csTealDk} MODEL OUTPUT}\\
   a summary of the retrieved\\ emails, as the user asked ---\\
   \texttt{send\_email} is never called\\[2pt]
   {\tiny\itshape\color{csTealDk} legitimate task completed;
    injection inert}};

\draw[redarrow]  (noint.east) -- (9.5, 2.35);
\draw[redarrow]  (13.8, 2.45) -- (outA.west);
\draw[tealarrow] (hook.east)  -- (9.5,-2.35);
\draw[tealarrow] (13.8,-2.45) -- (outB.west);

\begin{scope}[font=\scriptsize]
\draw[draw=csBlue!80!black, fill=csBlue!9, line width=0.7pt]
  (12.1,4.5) rectangle +(0.32,0.2);
\node[anchor=west] at (12.45,4.6) {legitimate};
\draw[draw=csRed!80!black, fill=csRed!8, line width=0.7pt]
  (14.35,4.5) rectangle +(0.32,0.2);
\node[anchor=west] at (14.7,4.6) {attack};
\draw[draw=csTealDk, fill=csTeal!9, line width=0.7pt]
  (16.25,4.5) rectangle +(0.32,0.2);
\node[anchor=west] at (16.6,4.6) {CounterSteer};
\end{scope}

\node[font=\scriptsize\itshape, black!65, align=center] at (9.5,-4.85)
  {The antidote is also an injection: a constant counter-signal against the
   representation that makes injected text read as an instruction.};
\end{tikzpicture}%
}
\vspace{-1.5mm}
\caption{\textbf{CounterSteer overview (real sample
\texttt{llmail-L1-ed0f1d411a}).} A user asks for an email summary; the agent's
\texttt{read\_emails} call returns a legitimate record with an injected
instruction embedded in it (red, quoted verbatim from the corpus).
\emph{Top, undefended:} at the injected tokens the residual-stream
representation drifts toward the region occupied by user instructions, and the
model executes the attacker's call. \emph{Bottom, CounterSteer:} during
prefill a constant vector $-\alpha\sigma\hat d$ is added at layers 12/16/20 to
every tool-result token --- no detector, no added tokens, decode untouched ---
pulling the injected tokens back toward ordinary tool content; the model
summarizes the emails and the attacker's email is never sent. The
activation-space insets are schematics of the mechanism, not measured
projections.}
\label{fig:overview}
\end{figure*}

\section{Introduction}\label{sec:intro}

LLM agents read what they retrieve. A web page, an email, a document returned
by a tool can address the model directly, and models too often comply, treating
text that arrived as \emph{data} as if it were an \emph{instruction}. This is
indirect prompt injection (IPI), consistently ranked the top security risk for
deployed LLM applications \cite{owasp2025llm}: the attacker never touches the
prompt, but publishes content and waits for an agent to fetch it.

Existing defenses buy protection with visible machinery: detector pipelines add
a second model and a threshold and fail open on a miss; spotlight prompting
\cite{hines2024spotlighting} adds tokens and survives until the attacker
phrases around them; system-level flow control re-architects the application;
fine-tuning (StruQ \cite{chen2024struq}, SecAlign \cite{chen2024secalign})
retrains the model (\S\ref{sec:related}). Each pays in
a different currency---inference calls, prompt tokens, infrastructure, a
training pipeline---and none is free where many deployments sit: a fixed
open-weights model behind an API-shaped serving stack.

We ask instead whether \textbf{the follow-the-injection decision is itself a
manipulable object inside the model}. We hypothesize that injection-following
produces a low-dimensional, causally manipulable representation over the
untrusted span. That representation must be discovered per model, not
assumed: for one attack class, \emph{parameter manipulation}, the decision
is not readable over the untrusted span---with our probes it becomes
readable only as the argument value is generated
(\S\ref{sec:defense-param}).

We present \textbf{CounterSteer}: a detector-free inference-time defense
requiring no model fine-tuning, no auxiliary model, and no additional inference
calls or prompt tokens (Fig.~\ref{fig:overview}). CounterSteer fits, per model,
a causal ``treat-this-as-an-instruction'' direction from behavioral
contrasts: paired episodes identical except for whether an embedded
instruction is followed. It screens candidates for reliability and held-out generalization, then
validates the survivors through causal-intervention and capability gates,
refits the winner on the full contrast set, and deploys it as a fixed
vector edit on every tool-result token during prefill
(\S\ref{sec:method-recipe}). The edit is
uniform: the defense needs to know only where tool output \emph{is}---span
boundaries the serving stack already tracks---never which of its tokens
might be an injection; content outside tagged spans is out of scope
(\S\ref{sec:limitations}). The edit
is always on: no detector decides when to intervene, so there is no detection
\emph{decision} for the attacker to flip. Always-on does not mean
unconditional, however: the fixed steering strength (the \emph{dose},
\S\ref{sec:method-recipe}) still sets a sensitivity, and for one
attack class our adaptive search finds blandly worded injections that act
below it (\S\ref{sec:eval-adaptive}).

\begin{figure}[t]
\centering
\includegraphics[width=\columnwidth]{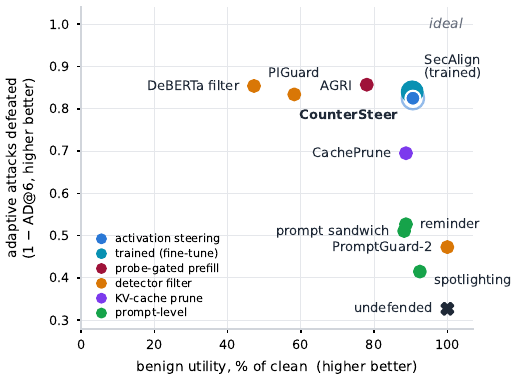}
\caption{The defense landscape on gpt-oss-20b, one point per defense:
benign utility with the defense always on (x, each defense's own
same-harness battery, \S\ref{sec:eval-baselines}) against the fraction of
attacks defeated under the adaptive AutoDojo attacker (y,
\S\ref{sec:eval-adaptive}). Upper right is best.}
\label{fig:frontier}
\end{figure}

\begin{table}[t]
\caption{The three-part map of injection-following that organizes this
paper. ``What removed it'' is what our measurements found, not a
completeness claim.}
\label{tab:map}
\centering
\scriptsize
\setlength{\tabcolsep}{3pt}
\begin{tabular}{>{\raggedright\arraybackslash}p{1.65cm}>{\raggedright\arraybackslash}p{2.15cm}>{\raggedright\arraybackslash}p{2.0cm}>{\raggedright\arraybackslash}p{1.55cm}}
\toprule
class & where the follow decision lives & what removed it in our tests & CounterSteer \\
\midrule
tool hijack & the retrieved span, at prefill & a constant prefill direction & largely neutralized \\
\addlinespace[2pt]
parameter manipulation & the model's own reasoning, at argument emission (\S\ref{sec:defense-param}) & only fine-tuning (SecAlign) & reduced, not closed \\
\addlinespace[2pt]
delegated authority & the user prompt grants it & system-level provenance (e.g., ROPE \cite{rope2026}) & out of scope (Class~B) \\
\bottomrule
\end{tabular}
\end{table}

Within the activation line of work, CounterSteer is distinguished by
validating causally and applying with neither a gate nor a per-request stage.
The closest recipes gate an intervention on a probe (ICON \cite{icon2026},
ARGUS \cite{argus2026steering}, AGRI \cite{agri2026}) or need an attribution
stage (CachePrune \cite{cacheprune2025}, V-Steer \cite{vsteer2026}), and none
reports a defense-aware adaptive attacker in an agentic setting
(\S\ref{sec:related}). Detection evidence alone does not yield a defense: two
directions here separate attacks cleanly yet do not causally steer, and the
role signal detectors rely on is causally coupled to injection-following on
one model and decoupled on another (\S\ref{sec:mechanism}).

What emerges is a three-part map of injection-following, split by \emph{where
the follow decision lives}, which dictates what can remove it
(Table~\ref{tab:map}). CounterSteer's scope is the two role-confusion
classes (Class~A, \S\ref{sec:threat-attacker}).

\textbf{Contributions:}
\begin{enumerate}
\item \textbf{A recipe, not a vector.} A five-step per-model procedure that
  searches for and causally validates a model-specific injection-following
  lever, with screening diagnostics and pre-specified causal gates that
  rejected most candidates (\S\ref{sec:method}, \S\ref{sec:mechanism}).
\item \textbf{A deployed defense across five models.} Four models pass
  full bidirectional causal validation, and Gemma-4-31B passes its
  suppression side (\S\ref{sec:method-recipe}); all five pass the
  capability guard and a pre-specified one-pass held-out evaluation
  (Table~\ref{tab:main-compact}; Fig.~\ref{fig:frontier}). In a
  twenty-six-battery same-harness comparison CounterSteer sits on the
  measured security--utility frontier---no inference-time defense we
  measured improves its compromise rate or adaptive success without paying
  more benign utility---on the capable
  models (``capable models'' defined in \S\ref{sec:eval}; comparison in
  \S\ref{sec:eval-baselines}; utility cost in \S\ref{sec:limitations}).
\item \textbf{A defense-aware adaptive evaluation.} Query attacks,
  surrogate-refit and exact-deployed-vector white-box GCG
  \cite{zou2023gcg}, a replay of
  the 461k-attack LLMail-Inject red-team dataset \cite{abdelnabi2025llmail}
  (2018 unique successful texts, 2052 episodes), and a
  benchmark-level adaptive attacker (AutoDojo~\cite{autodojo2026}) run
  against every major defense. CounterSteer holds ($0.67 \to 0.175$, $0.73 \to
  0.188$) where the strongest utility-preserving filter collapses near
  undefended; the two rivals that match it pay a fine-tune or a measured
  utility-and-serving bill (\S\ref{sec:eval-adaptive},
  \S\ref{sec:related}). The main surviving weakness is
  \textbf{parameter manipulation}, mechanism-traced in
  \S\ref{sec:defense-param}.
\item \textbf{A mechanistic account.} For parameter manipulation the follow
  decision is strongly decodable at the value-emission site but not, with
  the tested probe family, at the examined pre-generation sites; the role
  axis and the behavioral axis dissociate across models
  (\S\ref{sec:mechanism}).
\end{enumerate}

\section{Threat Model and Metrics}\label{sec:threat}

\subsection{Setting}\label{sec:threat-setting}

An LLM agent serves a benign user: it receives a user task, calls tools, and
reads tool results (web pages, emails, documents, API output) the deployment
cannot guarantee are trustworthy. The adversary controls \emph{content}
reaching the model only through those tool results---neither the
system prompt, the user turn, the tool schemas, nor the decoding
configuration, and no code execution on the serving infrastructure---the
standard indirect-prompt-injection setting of AgentDojo
\cite{debenedetti2024agentdojo} and LLMail-Inject \cite{abdelnabi2025llmail}.

Attacker objectives, in the severity order the evaluation is organized around:
(1)~\textbf{tool hijack}---cause a call to an attacker-chosen tool carrying
attacker-chosen content; (2)~\textbf{parameter manipulation}---keep the
legitimate tool but place attacker-chosen content in an argument the attacker
targets; and we separately track (3)~\textbf{contamination}---injected text
leaking into some other, legitimate call the attacker did not target. (1)~and
(2)~are compromise (together Class~A, the role-confusion classes);
(3)~is reported but never counted as compromise.

\subsection{The Defense-Aware Attacker}\label{sec:threat-attacker}

The attacker knows the model architecture and weights, the CounterSteer
algorithm and this paper, the steered layers, token sites and nominal dose, and
the fitting samples, which our pipeline draws deterministically from the corpus
and split we release. Because the fitted directions ship with the artifact,
we treat the deployed direction and dose as knowable too---the strongest arm
below hands the attacker the exact vector. We evaluate this attacker at
escalating strength:

\begin{itemize}
\item \textbf{query attacks}: adaptive framing search against the deployed
  defense (black-box, output-only feedback);
\item \textbf{surrogate refit}: the attacker reruns our published pipeline on
  the public weights, regenerating the behavioral labels it fits on, and
  optimizes suffixes with gradients through \emph{their} own defended
  model (a recipe-rebuild test, not a fitting-set-secrecy test;
  Appendix~\ref{app:adaptive});
\item \textbf{exact-vector white box}: the attacker is handed the actual
  deployed direction and dose and optimizes with gradients through the deployed
  defended forward pass---the ceiling of the threat model, measuring whether
  any of the defense's strength rests on secrecy rather than mechanism
  (open weights alone give no access to the agent's tools, credentials or
  infrastructure).
\end{itemize}

Out of scope (Class~B): attacks whose instruction-following is \emph{delegated}
by the legitimate user or system prompt---e.g.\ ``follow the instructions in
the retrieved document''---where obeying the payload is arguably correct
behavior and the failure is one of authority provenance, not role confusion.
CounterSteer acts at the model level; provenance systems such as ROPE
\cite{rope2026} own delegated authority at the system level
(\S\ref{sec:limitations} quantifies the boundary).

\subsection{Metrics and the Capability Guard}\label{sec:threat-metrics}

Three tiers, strict priority. \textbf{Tier 1, compromise:} did the attacker's
objective complete---target 0. On the agent benchmarks this is the
\emph{compromise rate} by the benchmark's own outcome checker; on the
single-turn corpora it is the \emph{attack success rate (ASR)} by our
deterministic scorer. \textbf{Tier 2, utility:}
benign-traffic task fidelity with the defense always on
(\emph{benign utility}), and task fidelity under attack, both as \% of the
undefended unattacked baseline. \textbf{Tier 3, contamination:} reported
only. On single-turn corpora, utility is
reported from a blinded pairwise judge scoring each completion against the
same run's unattacked reference, always quoted with a mechanical
strict/lenient bracket (structured arguments byte-exact; the same with
model-composed prose fields exempted); ASR is always computed by the
deterministic scorer, never by a judge (\S\ref{sec:eval},
Appendix~\ref{app:judge}).

Attack success admits two readings, named per table: the
\textbf{obedience-pattern} reading (the injected \emph{action} was
performed---the attacker's target tool called carrying injection-derived
content, or the targeted argument manipulated), and the stricter
\textbf{exact-literal} reading (the attacker's exact payload delivered
verbatim). The single-turn corpus tables and AgentDojo report the
obedience-pattern/checker reading; the LLMail-Inject replay uses that challenge's
own checker, exact-literal by construction; the query attack
(\S\ref{sec:eval-adaptive}) reports both side by side. Every defended
configuration must also pass a \textbf{capability guard}: the model must
not stop acting. Where the unattacked no-action rate is nonzero, the
defended rate must not exceed it---one-sided, because only \emph{extra}
refusal fakes security. Certified test rows pass with margin: defended
no-action at worst $0.75\times$ the unattacked rate ($0.80\times$ with a
labeled development row). Where that
estimate is zero, no ratio exists and we pre-specified no ceiling: those
rows are reported rather than thresholded---every observed value is
$\le 0.058$ absolute---and their flagged completions are read individually
(Appendix~\ref{app:tables}). Pre-specified
dose searches additionally set benign-utility floors (65--90\% of clean).
An ASR of 0 purchased by refusing to act is rejected as a defense.

Table~\ref{tab:glossary} (Appendix~\ref{app:tables}) collects the paper's
recurring terms in one place; each is also defined at first use.

\section{The CounterSteer Defense}\label{sec:method}

CounterSteer's working hypothesis is that an injection the model acts on is
\emph{represented} differently from the benign content around it, and that
the difference is low-dimensional and causally manipulable: a
residual-stream direction whose presence over untrusted-content tokens carries
``treat this as an instruction to follow,'' and whose suppression removes the
follow decision without removing the content's informational value.

\subsection{The Recipe}\label{sec:method-recipe}

The recipe has five steps, of which steps 4--5 are the certifying pass/fail
gates. Its output is a validated per-model direction with candidate layers
and, where the dose search of step~5 finds a capability-safe window, a
deployable triple (direction, layers, dose; Fig.~\ref{fig:pipeline},
Appendix~\ref{app:tables}). A dose counts only if the capability guard
(\S\ref{sec:threat-metrics}) stays intact at it.

\textbf{(1) Behavioral contrasts.} Paired episodes differing only in the factor
of interest. The primary contrast is a factorial over injected instructions:
authority framing (none $\to$ soft $\to$ firm $\to$ explicit supersession),
voice, requested action, and---decisively---whether the
surrounding context \emph{delegates} authority to the payload. Pairs are
rendered into the model's own chat template with exact token spans recorded for
the payload and the decision site; activations are captured at the pre-MLP
residual stream on those spans, mean-pooled, and group-centered within (sample
$\times$ delegation $\times$ requested action) cells, so sample identity and
topic cannot leak into the direction. Where the action factor is centred,
attack-class identity cannot leak either (per-fit centring and measured
action-axis cosines: Table~\ref{tab:provenance}).

\textbf{(2) Candidate direction.} The difference-in-means direction between
followed- and resisted-injection sides, per layer. We prefer
difference-in-means over probe weights for the \emph{steering} vector: it is
the direction the data actually moves, not the one a discriminator finds
maximally separable (cf.\
\cite{turner2023steering,panickssery2024caa,arditi2024refusal}).

\textbf{(3) Screening diagnostics.} Before any causal test, candidates are
ranked on two diagnostics: reproducibility across disjoint fitting shards
(sign-consistent cosine agreement; reference point 0.70), and AUC on the
same span over \emph{a whole authority-framing level withheld from this
fit} (reference 0.65). The second diagnostic thus measures extrapolation to
an authority level never seen; scored inside that one level, it cannot be
met by detecting the framing's surface identity. These diagnostics rank
candidates rather than certify them---certification is steps 4--5. Most
candidates are set aside here, and the one deployed fit admitted below the
reliability reference (GLM-4.5-Air, 0.31--0.33) certifies through the gates
(Table~\ref{tab:permodel-glm}).

\textbf{(4) Causal validation.} Low-dose bidirectional test on held-out
attacked episodes: subtracting the direction along payload tokens must reduce
injection-following and adding it must increase it, both at doses that leave
the capability guard intact. A direction that correlates but does not steer is
rejected (\S\ref{sec:mechanism} shows two such rejections---correlates are
common, levers are not). Convention: $\alpha > 0$ is the defensive edit
(subtracting $\alpha\sigma\hat d$); the gate's other arm, written
$-\alpha$, is the reversed-polarity edit that adds it.
The increase side of this gate passes on four models, each with the guard
intact, so the increase is not degradation into compliance: gpt-oss-20b
$+0.11$--$0.22$ (McNemar $p \le 0.042$, gated candidate and exact
negation); Qwen3-30B $+0.14$ ($p{=}0.0015$); Llama-3.1-8B $0.135 \to 0.558$
at $-\alpha$; GLM-4.5-Air $0.62 \to 0.83$ ($p{=}0.013$;
Table~\ref{tab:permodel-glm}). On Gemma-4-31B the increase criterion
is untestable: its undefended rate is saturated at 1.000, which the negated
dose holds exactly (0 discordants) while the positive dose defends. That is
polarity-consistent, but not a measured increase. Matched random
directions never raise attack success in any measured control.

\textbf{(5) Deploy.} The surviving direction is refit over the full contrast
sample set with one authority-framing level withheld (per-model status:
Table~\ref{tab:provenance}). gpt-oss-20b's deployment predates this rule; a
compliant refit validates end-to-end, statistically indistinguishable from
the deployed cell (Table~\ref{tab:provenance}). The
direction becomes
a constant per-token,
norm-preserving additive edit applied always, to all tool content---the
benign-utility cost (\S\ref{sec:eval-main}) is the price of the defense.
The probe capture site (step~1: the pre-MLP residual, via a forward pre-hook on
the post-attention layernorm) and the steering site (the block output,
\S\ref{sec:defense}) deliberately differ: the block output is where an
additive edit enters the residual stream, the pre-MLP site where the probe
reads cleanest.

\textbf{Composition with the role axis (optional).} On gpt-oss-20b the
behavioral direction composes productively with a role-probe axis
(user-vs-tool); on Qwen3-30B the behavioral direction alone is
selected---a per-model option the recipe searches, not core method.

\subsection{Deployment}\label{sec:defense}

CounterSteer deploys as a forward hook on the residual stream at a small set of
layers. During prefill, for every token position inside a tool-result span, the
block output $h$ is replaced by
\begin{equation}\label{eq:edit}
h' \gets \|h\| \cdot \frac{h - \alpha_\ell \sigma_\ell \hat d_\ell}{\|h -
\alpha_\ell \sigma_\ell \hat d_\ell\|},
\end{equation}
where $\hat d_\ell$ is the model's fitted direction at layer $\ell$,
$\sigma_\ell$ the standard deviation of the fitting corpus's captured span
activations projected onto $\hat d_\ell$, and
$\alpha_\ell$ the dose (norm-preserving, so the edit rotates rather than
inflates the state). The edit applies only to tool-result positions during
prefill; system/user and generated positions are never directly edited,
though generated tokens attend to the steered cached state. Inputs with no
tagged tool-result spans execute byte-identically (measured,
Appendix~\ref{app:ablations}). The hook is stateless, adds no tokens or
calls, and its latency/memory overhead is negligible at our measurement
resolution (\S\ref{sec:eval-baselines}).

The deployed configurations: on gpt-oss-20b, a unit-norm 8:1 blend of the
behavioral override direction with the role axis,
dose $8.06\sigma$, layers 12/16/20; on
Qwen3-30B, the behavioral direction alone
at $12\sigma$, layers 8/20/32; on GLM-4.5-Air, the action-centred behavioral
direction alone at $\alpha{=}8$ on a reference scale matched to the
pooled-corpus fit (per-layer scales: Table~\ref{tab:permodel-glm} caption),
layers 20/24/28; on Gemma-4-31B, the behavioral direction alone at
$8\sigma$, layers 4/28/36; on
Llama-3.1-8B, the action-centred, framing-held-out behavioral direction
alone at $5\sigma$, layers 12/16/20. All are the recipe's output
(\S\ref{sec:method-recipe}); beyond the gated dose and layer search,
researcher judgment entered at three disclosed points---the gpt-oss-20b
role-axis blend, action-centring (since made the default), and
GLM-4.5-Air's below-reference reliability admission (search budgets:
Table~\ref{tab:searchbudget}). The latest fit,
Llama-3.1-8B, ran the standardized ladder end-to-end in one day.
Dose convention: every edit is $\alpha\sigma_\ell\hat d_\ell$, with
$\sigma_\ell$ the per-layer scale of the fit named in the released
configuration; doses are not comparable across models. The deployed defense
is prefill-only
(the negative finding behind that choice: \S\ref{sec:defense-param}).

\section{Evaluation Setup}\label{sec:eval}

\textbf{Models.} Five open-weights models, stated once with the short forms
used hereafter: gpt-oss-20b \cite{openai2025gptoss};
Qwen3-30B-A3B-Thinking-2507 (\emph{Qwen3-30B}) \cite{qwen2025qwen3};
Gemma-4-31B-it (\emph{Gemma-4-31B}); GLM-4.5-Air; and Llama-3.1-8B-Instruct
(\emph{Llama-3.1-8B}). The two \emph{primary models}---gpt-oss-20b and
Qwen3-30B---run the full corpus battery and the deepest adaptive evaluation,
each at its deployed configuration from \S\ref{sec:defense}; the other
three certify through one-pass held-out tests on their firing corpora plus
the agentic grid (coverage: Table~\ref{tab:coverage}). ``The capable
models'' hereafter means all but Llama-3.1-8B, whose low clean task
completion (13--15 of 56 AgentDojo tasks) separates defenses poorly. Tool
hijack and parameter manipulation are never pooled---they behave
differently under both the defense and the adaptive attacker. Success
counts are $k/N$ with 95\% \textbf{Wilson} intervals (a zero is read
through its interval); utility is \% of the undefended-unattacked model.

\textbf{Generation budgets are per model, not global.} A reasoning model
spends thousands of tokens thinking before acting: Qwen3-30B runs at 4096
tokens where gpt-oss-20b and Llama-3.1-8B run at 1024 single-turn and 4096
agentic; Gemma-4-31B runs single-turn at 2048, agentic at
4096; GLM-4.5-Air at 8192 throughout. Every table is held at
one budget shared by every arm, no figure is compared across budgets, and
cross-model comparisons quote both sides at the same, named budget.

\textbf{Scoring.} On the single-turn corpora, ASR denominators are the stated
$N$ (samples where
the clean reference acts), and utility (benign and under attack) comes from
the blinded pairwise judge of
\S\ref{sec:threat-metrics}, run under majority-of-three voting with in-batch
control pairs and a pre-specified validation gate (full protocol:
Appendix~\ref{app:judge}).
Judge utility is a
conservative \emph{floor} and is never quoted alone: every judge number
carries the strict/lenient bracket (\S\ref{sec:threat-metrics})
in parentheses; where the judge sits relative to the bracket, and why, is
measured in \S\ref{sec:eval-baselines}. AgentDojo rows use its own
checkers over the stated case/task counts, never the judge.
Tier-1 ASR/compromise-rate is always deterministic---%
\textbf{no LLM judge decides attack success anywhere in this paper} (the
rejected absolute correctness judge and its refuting measurements:
Appendix~\ref{app:corpus}).

\textbf{Corpora, and why we built our own.} AgentDojo and LLMail-Inject
supply the external multi-turn and human-adaptive measures; four single-turn
corpora are ours, each for a gap those leave (construction:
Appendix~\ref{app:corpus}). The two
\emph{JSON} corpora embed attacks in real tool-log records from a pinned
public agentic IPI trace dataset~\cite{nemotron_ipi_traces}: instructional
takeover, and parameter manipulation---attacker content placed in the
argument it targets, on the \emph{legitimate} call. Public IPI benchmarks
do not isolate the parameter class, even though ROPE \cite{rope2026}
already guards this surface at the system level; to our knowledge we
measure it in isolation for the first time. Undefended, it near-saturates
(0.81--1.00 on the webpage medium), and it is the one class every
inference-time defense tested here fails to
close---only the fine-tuned SecAlign does (\S\ref{sec:eval-adaptive}). The two \emph{webpage} corpora embed the same
classes in natural HTML carriers (fetched Wikipedia articles), with
template-disjoint splits.
All four score behaviorally against the same run's unattacked
reference, not by string-matching an attack canary. (The pairwise utility
judge exists for a matching gap; Appendix~\ref{app:judge}.)

\textbf{Baselines.} Three prompt-level arms (spotlighting
with delimiting \cite{hines2024spotlighting}, a prompt sandwich,
AutoDojo's reminder \cite{autodojo2026}); three classifier filters run
through AgentDojo's own detector path (PromptGuard-2 \cite{promptguard2},
a DeBERTa detector \cite{protectai2024deberta}, PIGuard
\cite{li2024injecguard}) plus a mechanical tool filter; two internal
interventions, both our ports---CachePrune \cite{cacheprune2025}, KV-cache
pruning at prefill (validated in its paper's own setting; no public
implementation exists), and AGRI \cite{agri2026}, a probe-gated reasoning
prefill, from its specification; and one trained baseline,
SecAlign \cite{chen2024secalign}, a DPO fine-tune of the served weights
(wiring, ports, training grids: Appendix~\ref{app:baselines}).

\textbf{Data provenance.} Held-out evaluation within a benchmark is the
validity bar throughout, on three axes: samples, attacker-template family and a
whole authority-framing level. The recipe \emph{as now specified} withholds
the framing level from the deployed refit itself: Llama-3.1-8B's deployed
fit follows that rule end-to-end, a gpt-oss-20b refit under the
rule is validated end-to-end within the deployed configuration's reference CIs
(\S\ref{sec:method-recipe}, step~5), and the other three deployed fits
predate the rule, holding the level out at the admission gate only.
Table~\ref{tab:provenance} (Appendix~\ref{app:tables}) states each stage's
split and what it is held out from. Key pre-specifications ship
as files with the artifact, fixing arms, splits, doses and hypotheses
before their test touches (SHA-256 manifest in the release); before-touch
ordering is attested by the released run logs and configuration dumps (no
third-party timestamp attestation).

\section{Results}\label{sec:results}

\textbf{Reading the numbers.} Greedy decoding is not
bit-stable across process histories---same-grid undefended baselines span
0.432--0.489---so repeat runs are never averaged: where a model has more
than one same-configuration AgentDojo run, \textbf{one is canonical and
every quoted number comes from it}. Canonical: gpt-oss-20b $.475 \to .079$
at 94.4\% benign; Qwen3-30B $.489 \to .072$ at 94.1\%; GLM-4.5-Air
$.222 \to .056$. Other regenerations are quoted only in their own
contexts: gpt-oss-20b's rival-comparison battery ($.483 \to .091$ at
90.6\%, beside its same-process rivals only) and its dose ladder (read only within Fig.~\ref{fig:dose} and
Table~\ref{tab:gptoss-dose}); Qwen3-30B's separate-run replicate
($.457 \to .094$); GLM-4.5-Air's detector-battery undefended arm
(\S\ref{sec:eval-baselines}: 0.183, within regeneration noise of .222,
$p{=}0.36$).

Three caveats recur; each is stated once here and afterward only named.
\emph{(1) Budget censoring.} An outcome checker cannot tell a refusal
from compliance cut off by the generation budget, so every agentic compromise
rate is a truncation-censored lower bound at its stated per-turn budget; the
arms of any one table share one budget (audit: Appendix~\ref{app:corpus}).
\emph{(2) Typography normalization.} gpt-oss-20b AgentDojo utilities are
regraded after mapping Unicode space/dash codepoints in output to ASCII,
uniformly across arms, before the unmodified checkers; the one compromise-rate movement under normalization is
\emph{against} the defense (Appendix~\ref{app:baselines}).
\emph{(3) Near-duplicate templates.} 12 of 52 webpage-parameter test samples
differ from a development template only by letter case or one whitespace
character---found in review \emph{after} the one-pass test touches, so the
12 are reported, not excluded: post-hoc exclusion would redefine the
pre-specified pass. The split is
identical across the models sharing it, so cross-model comparisons are
unconfounded, and the measured impact is nil (GLM-4.5-Air defended 0/12
vs.\ 1/40 disjoint; Qwen3-30B 0--1/12; strictly-disjoint-40 subsets:
Table~\ref{tab:main-qwen}).

\subsection{Main Security--Utility Result}\label{sec:eval-main}

\begin{figure*}[t]
\centering
\includegraphics[width=\textwidth]{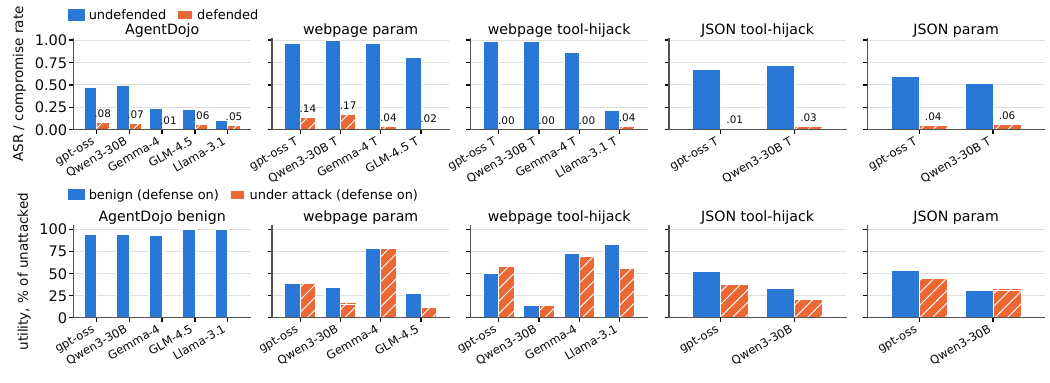}
\caption{Table~\ref{tab:multimodel} as charts. Top: attack success rate
(the AgentDojo panel: compromise rate),
undefended (solid) vs defended (hatched), per corpus class; each pair is a
within-model contrast at that model's own budget and split (rung on the
tick). Bottom: utility with the defense on---benign and under attack,
\textbf{judge} \% of unattacked (strict/lenient brackets in the table); the
AgentDojo panel is benign retention on its own checker. AgentDojo ticks
carry no rung (one grid, no split);
Gemma-4's webpage bars (ASR and utility) are the
doc$+$wording$+$template-held-out rung, its utility judge-rescored
2026-09-09 (the retained T$^{*}$ rung: Table~\ref{tab:permodel-gemma}).
Bars generated from the
table's numbers by
\texttt{figs/make\_results\_figs.py}; nothing recomputed.}
\label{fig:results}
\end{figure*}

\textbf{Five models certify, each through a pre-specified held-out pass}
(Table~\ref{tab:main-compact}). Per-class detail: Table~\ref{tab:multimodel}
and Fig.~\ref{fig:results} (Appendix~\ref{app:tables}); full primary-model
suite: Table~\ref{tab:flagship}; CIs, denominators, guard
readings, holdout axes:
Tables~\ref{tab:main-gptoss}--\ref{tab:coverage}. Every primary-model corpus row
holds out samples and, where stated, attacker templates from all tuning.
One specificity matters: Gemma-4-31B's attack surface is
\emph{medium}-specific (webpage/HTML
carriers fire at 0.87--1.00
undefended across both test rungs; JSON records at 0.05--0.11 and the
prose-carrier medium at 0.000
do not fire---per-corpus rates: Table~\ref{tab:permodel-gemma}).

\begin{table}[t]
\caption{The five-model result in brief. Single-turn: ASR range over that
model's firing corpus classes, one pre-specified held-out test pass each
(detail: Table~\ref{tab:multimodel}; CIs:
Tables~\ref{tab:main-gptoss}--\ref{tab:coverage}). AgentDojo: the canonical
runs (Reading the numbers); benign utility by its own
checker, \% of clean (gpt-oss typography-normalized).}
\label{tab:main-compact}
\centering
\scriptsize
\setlength{\tabcolsep}{3pt}
\begin{tabular}{lccccc}
\toprule
& \multicolumn{2}{c}{single-turn ASR \dirdown} & \multicolumn{2}{c}{AgentDojo \dirdown} & benign \dirup \\
\cmidrule(lr){2-3}\cmidrule(lr){4-5}
model & undef. & defended & undef. & defended & \% of clean \\
\midrule
gpt-oss-20b & .59--.98 & \textbf{.000--.135} & .475 & \textbf{.079} & 94.4 \\
Qwen3-30B & .52--1.00 & \textbf{.000--.173} & .489 & \textbf{.072} & 94.1 \\
Gemma-4-31B & .87--.96 & \textbf{.000--.038} & .239 & \textbf{.006} & 92.9 \\
GLM-4.5-Air & .81 & \textbf{.019} & .222 & \textbf{.056} & 100.0 \\
Llama-3.1-8B & .21 & \textbf{.038} & .100 & \textbf{.050} & 100.0 \\
\bottomrule
\end{tabular}
\end{table}

\textbf{Gemma-4-31B's certification carries the widest holdout in this
paper, and the defense holds under it.} Its fresh test corpus
simultaneously withheld carrier documents, attacker wordings and framing
templates from every fitting and tuning stage
(Table~\ref{tab:permodel-gemma}). One
pre-specified pass at the frozen deployed configuration (\S\ref{sec:defense})
takes parameter manipulation from 0.962 to
$\mathbf{0.038}$ (2/52, both under one held-out wording) and tool hijack
from 0.865 to $\mathbf{0.000}$ (0/52). One bound travels with the
parameter claim: the CI uppers (doc-clustered 0.096, sample Wilson 0.130)
exceed the 0.05 bar the point estimate meets. Utility, the guard,
counts, $p$-values and the passed pre-registered controls:
Table~\ref{tab:permodel-gemma};
the AgentDojo grid completes the row (zero
defense-introduced, Table~\ref{tab:headline}).

\textbf{GLM-4.5-Air, the largest model here (${\sim}$106B MoE), gives the
strongest webpage-parameter attack-success result in this paper.} The recipe's
action-centred direction (\S\ref{sec:defense}) certifies on the
held-out webpage parameter-manipulation test split---the primary models'
identical 52-sample split: $0.808 \to \mathbf{0.019}$ (1/52;
exact-literal 0/52; CI and counts: Table~\ref{tab:permodel-glm}), a 97.6\%
relative reduction beside the primary models' 86.0\%/82.7\% (base rates and
budgets differ). Its strict-bracket utility is 94.2/94; the judge's
composition-fidelity floor reads 26.9 (Table~\ref{tab:permodel-glm}; the
strict-vs-judge gap: \S\ref{sec:eval-baselines}). The development split
reads $0.750 \to 0.156$ (Table~\ref{tab:flagship}). The
near-duplicate-template caveat rides every
webpage-parameter test row, primary models included. Its agentic
grid completes the row (Table~\ref{tab:headline}); its other single-turn classes do not
measure the defense (tool-hijack classes fire weakly undefended; JSON
parameter never run).

\textbf{On Llama-3.1-8B, a model widely used in prior injection studies,
the recipe runs end-to-end in one day.} The direction passes the bidirectional
causal gate ($0.135 \to 0.558$ at $-\alpha$) and certifies on the held-out
webpage tool-hijack split: $0.212 \to \mathbf{0.038}$ (11/52 $\to$ 2/52;
counts and $p$: Table~\ref{tab:coverage}), at \textbf{100\%} byte-exact benign
utility and utility under attack \emph{above} undefended on both
instruments. A fit-withheld framing level composed onto the injection is
also blocked (0/4, a point estimate). Two bounds travel with the row: the
thin undefended base (0.212 test, 0.135 development) and the model's low
task capability (\S\ref{sec:eval}). The AgentDojo compromise rate still halves
(Table~\ref{tab:headline}; $p{=}0.02$) at 100\% benign. Its rival
comparison spans two hardware fabrics (clean arms 26.8/23.2 absolute); each
battery's four arms share one fabric, and a CounterSteer replicate
quantifies the spread ($0.050$ vs.\ $0.061$ n.s.; Table~\ref{tab:soa}).

\textbf{Tier-3 contamination is
confined on both primary models to the JSON tool-log corpus}, whose legitimate
task gives the matcher a nonzero false-positive floor even unattacked; the
other six corpus rows are 0.000 in every arm but one (full readings, the
$16\sigma$ dose ceiling, the dev-vs-test composition gap:
Appendix~\ref{app:corpus}).

\textbf{Two patterns organize everything that follows.} First, the surviving risk
concentrates in \textbf{parameter manipulation} on both primary models. Tool hijack
is driven to 0/52 observed per model on the webpage medium, with a handful
of surviving JSON compromises (gpt-oss-20b $1/80$, Qwen3-30B $3/90$, each on
its own held-out split and budget); the mechanism is quantified in
\S\ref{sec:eval-adaptive} and resolved in
\S\ref{sec:defense-param}. Second, the security--utility trade is
dose-controlled and model-specific. gpt-oss-20b pays in wording drift on
JSON media and in composed-summary fidelity on webpage media, mostly a cost
regeneration itself shares (\S\ref{sec:eval-baselines}); Qwen3-30B at
$12\sigma$ buys most of $16\sigma$'s compromise-rate reduction at a
fraction of its cost (\S\ref{sec:eval-agentic}).

\subsection{Agentic Evaluation}\label{sec:eval-agentic}

\begin{table*}[t]
\caption{\textbf{Headline agentic results: defenses $\times$ models on
AgentDojo} (180-case grid). cmp $=$ compromise rate (its own security
checker); util $=$ benign utility, \% of that row's \emph{own} clean run
(``no defense'': absolute clean completion, the capability reference).
$^{\dag}$comparison/detector-probe batteries, not the canonical runs
(\S\ref{sec:results}). ``---'' $=$ not measured; \warnmark\ $=$
capability-guard territory; bold/underline $=$ best/second-best per column
among unmarked rows. Bottom block: AutoDojo AD@6
(\S\ref{sec:eval-adaptive}).}
\label{tab:headline}
\centering
\scriptsize
\setlength{\tabcolsep}{3.5pt}
\renewcommand{\arraystretch}{0.9}%
\begin{tabular}{lcccccccccc}
\toprule
& \multicolumn{2}{c}{gpt-oss-20b$^{\dag}$} & \multicolumn{2}{c}{Qwen3-30B} & \multicolumn{2}{c}{Gemma-4-31B} & \multicolumn{2}{c}{GLM-4.5-Air$^{\dag}$} & \multicolumn{2}{c}{Llama-3.1-8B} \\
\cmidrule(lr){2-3}\cmidrule(lr){4-5}\cmidrule(lr){6-7}\cmidrule(lr){8-9}\cmidrule(lr){10-11}
defense & cmp \dirdown & util \dirup & cmp \dirdown & util \dirup & cmp \dirdown & util \dirup & cmp \dirdown & util \dirup & cmp \dirdown & util \dirup \\
\midrule
no defense (clean util absolute) & .483 & 98.1 & .489 & 91.1 & .239 & 100.0 & .183 & 80.4 & .100 & 26.8 \\
\midrule
PromptGuard-2 filter \cite{promptguard2} & .205 & \textbf{100.0} & .228 & \textbf{100.0} & --- & --- & \underline{.117} & \underline{100.0} & .056 & \textbf{100.0} \\
DeBERTa filter \cite{protectai2024deberta} & .034 & 47.2 \warnmark & .089 & 54.9 \warnmark & --- & --- & --- & --- & \underline{.044} & \underline{92.3} \\
PIGuard filter \cite{li2024injecguard} & .011 & 58.2 \warnmark & .006 & 62.7 \warnmark & --- & --- & .000 & 57.8 \warnmark & \textbf{.006} & \underline{92.3} \\
CachePrune \cite{cacheprune2025} & .216 & 88.7 & --- & --- & --- & --- & --- & --- & .045 & 80.0 \\
reminder (AutoDojo) \cite{autodojo2026} & .273 & 88.7 & .356 & \underline{96.1} & --- & --- & --- & --- & .100 & \textbf{100.0} \\
prompt sandwich & .316 & 88.2 & --- & --- & --- & --- & --- & --- & \textbf{.006} & 73.3 \\
spotlighting \cite{hines2024spotlighting} & .420 & \underline{92.5} & --- & --- & --- & --- & --- & --- & .128 & \textbf{100.0} \\
tool filter [mechanical] & .000 & 10.9 \warnmark & --- & --- & --- & --- & --- & --- & .006 & 76.9 \warnmark \\
AGRI (probe-gated prefill) \cite{agri2026} & \textbf{.062} & 78.0 & \underline{.100} & \textbf{100.0} & --- & --- & --- & --- & --- & --- \\
\midrule
\textbf{CounterSteer} (activation steering, ours) & \underline{.091} & 90.6 & \textbf{.072} & 94.1 & \textbf{.006} & \textbf{92.9} & \textbf{.056} & \textbf{102.2} & .050 & \textbf{100.0} \\
\midrule
\multicolumn{11}{l}{\emph{AutoDojo AD@6: the adaptive attacker's success at six iterations (Table~\ref{tab:autodojo-matrix})}} \\
no defense & \multicolumn{2}{c}{.673} & \multicolumn{2}{c}{.730} & \multicolumn{2}{c}{---} & \multicolumn{2}{c}{---} & \multicolumn{2}{c}{.377} \\
CounterSteer & \multicolumn{2}{c}{.175} & \multicolumn{2}{c}{.188} & \multicolumn{2}{c}{---} & \multicolumn{2}{c}{---} & \multicolumn{2}{c}{.307} \\
\bottomrule
\end{tabular}
\end{table*}

\textbf{Overview.} Three findings are consistent across the agentic
evaluations we ran (Table~\ref{tab:headline}). CounterSteer removes 50--97\% of
each grid's undefended compromises (the low end on Llama-3.1-8B's thin
base), ${\approx}74\%$ on both primary models even under the adaptive
attacker, and introduces none or
almost none of its own (0--2 per grid). What survives
is consistently the \emph{delegated-authority} family (Class~B,
\S\ref{sec:threat-attacker}) rather than role
confusion. What differs by
model is the price and the surface: benign utility ranges from 100\% of
clean measured (GLM-4.5-Air, Llama-3.1-8B) through 94.1--94.4\% on the
primary models (typography-normalized) to 92.9\% on Gemma-4-31B (a
$-7.1$\,pp delta, not significant), each model with its own firing
surface and dose
(Table~\ref{tab:headline}). What differs by benchmark is only the base
rate, not the ranking, in every pairing we measured.

Three agent benchmarks, in increasing order of difficulty. AgentDojo
\cite{debenedetti2024agentdojo} is the primary static measure: multi-turn
executed environments with its own security and utility checkers. Our grid
is a fixed 180-episode subset of its 629 user$\times$injection pairs,
stratified over all four suites, pinned once (manifest released) and
identical for every arm and model. AgentDyn \cite{agentdyn2026} adds three harder open-ended suites with
its own checkers. AutoDojo \cite{autodojo2026} optimizes injections
against the live defended agent (\S\ref{sec:eval-adaptive}). All numbers are the benchmarks' own checkers at
the 4096-token per-turn budget (8192 on GLM-4.5-Air); the harness's legacy
768-token default was budget-censored. One legacy comparison remains at 768
(budget-matched arms: \S\ref{sec:eval-baselines}).

\textbf{AgentDojo, five models} (Tables~\ref{tab:headline},
\ref{tab:soa}). Each Table~\ref{tab:headline} row is its own four-arm
battery (Table~\ref{tab:soa}; Gemma: Table~\ref{tab:permodel-gemma}; AGRI:
its paired battery, Appendix~\ref{app:baselines}); the gpt-oss column is
typography-normalized. The
compromise rate falls to \textbf{0.006--0.079} at 93--100\%
typography-normalized benign utility over the canonical runs (Reading the
numbers); the rival-comparison batteries read 0.006--0.091 at 91--102\%
(Table~\ref{tab:headline}). At the shared budget the two primary models land
comparably (blocked/introduced
70/0 and 75/0, canonical runs) at matched normalized benign utility
(94.4\% and 94.1\% of clean; gpt-oss-20b raw 85.2\%). Utility under attack is 73.9\% of
clean vs.\ 62.5\% undefended on gpt-oss-20b, and \emph{above} undefended on
Qwen3-30B (Table~\ref{tab:soa}). Both primary models carry full dose
curves (Fig.~\ref{fig:dose}, Tables~\ref{tab:gptoss-dose},
\ref{tab:qwen-dose}): on gpt-oss-20b a documented lower-dose option sits at
$\alpha{=}5.5$ and ${\ge}95\%$
benign is unreachable by dose; on
Qwen3-30B the benign cliff starts immediately above the selected
$12\sigma$.

\textbf{Who survives the defense is as informative as how many.} Nine user
tasks (36/180 cases) \emph{delegate} instruction authority to retrieved
content in their own prompt text (Class~B); the
partition is computed from the user prompt alone, identically for
every arm. The deployed gpt-oss-20b configuration's surviving compromises
are \emph{entirely} delegated: \textbf{0/141 Class~A observed} [0.000,
0.027] against undefended 0.390, with transcripts showing user-authorized
compliance, not role confusion (post-hoc motivation, mechanical arm-blind
rule: Appendix~\ref{app:tables}). Prompt-level baselines, by contrast, leave
31--46 non-delegated compromises (Table~\ref{tab:soa}). GLM-4.5-Air
repeats the pattern: 7 of its 10 survivors sit on one authority-delegating
task (Table~\ref{tab:permodel-glm}).

\textbf{AgentDyn---harder, open-ended tasks---preserves the ordering.}
gpt-oss-20b falls $0.393 \to \mathbf{0.031}$ (127 blocked, none
introduced; $p{=}1.2\times10^{-38}$) at ${\sim}90\%$ of clean (raw
defense-on-clean 27--29/37 over ten replicates on two fabrics; retention
${\sim}83\%$ of clean-solved tasks still
solved). Qwen3-30B falls $0.482 \to \mathbf{0.021}$, and GLM-4.5-Air
$0.156 \to 0.067$ at its starkest benign gap (78.8\%). Llama-3.1-8B solves 2
of 60 tasks---a capability bound, not a defense result. The rivals measured
on the same battery trail as on AgentDojo: on gpt-oss-20b,
CachePrune reaches 0.111 and the reminder 0.178, both with introduced
compromises, vs.\ CounterSteer's 0.017 on the same-case restriction;
on Qwen3-30B, 0.144 vs.\ 0.011.

\textbf{AutoDojo, the benchmark-level adaptive attacker, produces the
largest separation between defenses} (Table~\ref{tab:headline}, bottom
block; \S\ref{sec:eval-adaptive} details it).

\subsection{Adaptive Evaluation}\label{sec:eval-adaptive}

\textbf{Under every adaptive attacker we mounted, CounterSteer's tool-hijack
defense holds; the surviving weakness is parameter manipulation
under mild framing.} Adaptive, defense-aware evaluation is the bar this
defense class must clear (\S\ref{sec:related}): we escalate attacker
strength along the \S\ref{sec:threat-attacker} ladder, plus relay attacks
and a red-team replay, against the gpt-oss-20b deployed configuration except
where a second model is named (full protocols and caveats:
Appendix~\ref{app:adaptive}).

\textbf{Query-based framing search, and the weakness it exposes.}
The query attack runs a best-of-48 framing search per sample against the
live defended agent ($n{=}18$ development samples per class;
Table~\ref{tab:flagship}). Its grid is drawn from the fit factorial, so it
probes wording largely \emph{inside} the fit distribution
(\S\ref{sec:limitations}). Tool hijack holds: \textbf{3/18} cracked,
per-query success 0.007. Parameter manipulation---the shared surface every measured
inference-time defense fails to close---\textbf{cracks}: \textbf{13/18}
at per-query 0.303, and its strongest wordings are bland
parameter-correction framings that slip under a direction fitted against
authority-styled ones (\S\ref{sec:defense-param}). Under the exact-literal
reading both searches nearly vanish (2 of 864 pooled queries). A sample
counts as cracked if \emph{any} of its class's 24 wordings fires; pooling
both classes, 15/18.

\textbf{The weakness is shared across the inference-time defenses we
measured}: every \emph{inference-time}
arm shares it under the identical best-of-24 parameter search
(Table~\ref{tab:adaptive-param-arms}, Appendix~\ref{app:adaptive}).
CounterSteer still suppresses per-attempt success (0.303 vs.\
0.595--0.708, exact sample-level paired $p \le 0.0018$) and
alone holds the exact-literal reading to 1/18. The \textbf{trained defense
is the measured exception}: SecAlign holds the same search to 1/18 on the
pattern reading and \textbf{0/18} exact-literal, and to 0/18 on both
readings for tool hijack---closing the parameter-class gap at
fine-tuning's cost, on framing families its training grid contained.

\textbf{The held-out test passes converge on the \S\ref{sec:defense-param}
mechanism}: 10 of 11 surviving compromises on gpt-oss-20b and 13 of 16 on
Qwen3-30B are parameter-class (\S\ref{sec:eval-main}).

\begin{figure}[t]
\centering
\includegraphics[width=\columnwidth]{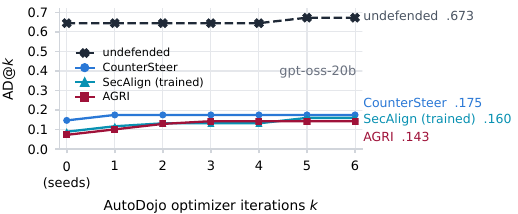}
\caption{AutoDojo success as the attacker iterates (AD@$k$, macro over 36
cells, gpt-oss-20b; $k{=}0$ = the four static seeds; SecAlign 35 cells). All 11 arms: Table~\ref{tab:autodojo-matrix}; benign utility:
Fig.~\ref{fig:frontier}.}
\label{fig:adk}
\end{figure}

\textbf{A benchmark-level adaptive attacker against every major defense}
(Fig.~\ref{fig:adk}; matrix: Table~\ref{tab:autodojo-matrix}):
AutoDojo~\cite{autodojo2026} iteratively rewrites each injection against
the \emph{live defended agent}, scored by AgentDojo's own end-to-end
checker. CounterSteer holds it to roughly a quarter of its
undefended success on both primary models, where every prompt-level defense
collapses. The curves locate the residual:
\textbf{84\% of CounterSteer's AD@6 is already reached by its static
seeds} ($0.147$ of $0.175$), and optimization gains the attacker less
against it ($+0.028$) than against the two strongest rivals ($+0.07$
each)---what survives is seed-shaped data, not optimized text. Larger $k$
is unmeasured \cite{nasr2025attacker} (\S\ref{sec:limitations}).
PromptGuard-2, the strongest utility-preserving filter, lands
near-undefended on all three models; KV-cache pruning is partial; the prompt-level arms collapse.
PIGuard and DeBERTa hold slack/travel near zero and match CounterSteer on
banking---at capability-guard benign utility; SecAlign
matches CounterSteer across suites (35/36 cells; banking 0.47 beside 0.46). The
ported AGRI arm is the strongest inference-time rival on gpt-oss-20b
(0.143 at benign 78.0\%); on Qwen3-30B it trails on security at zero
benign cost (\S\ref{sec:related}).
Banking, every gpt-oss-20b arm's worst suite, delegates payment authority
to the injected document: the winning payload is a fake bill line
item---data, not an instruction. Detection measurably degrades with no instruction
to detect \cite{datasentinel2025}, and control-flow isolation admits exactly
this attack as a stated non-goal \cite{debenedetti2025camel,
beurerkellner2025patterns}. Llama-3.1-8B's base is too low to separate
defenses (coverage and counts: Table~\ref{tab:autodojo-matrix}).

\textbf{The rest of the ladder holds, with its bounds stated}
(Table~\ref{tab:flagship}; full protocols, budgets and caveats:
Appendix~\ref{app:adaptive}). GCG \cite{zou2023gcg} through a recipe-faithful
attacker refit (cosine \textbf{0.974--0.977} to the deployed vector) lands
\textbf{0/12 observed} against both the deployed configuration and the
attacker's own surrogate (undefended 5/12). Handing the
attacker the \emph{actual} deployed direction and dose compromises
\textbf{2/52 observed} (0.038, Wilson [0.011, 0.130]), while the same
suffixes gain nothing over plain injection on the undefended model (22/52
vs.\ 26/52): revealing the vector yields no working recipe, but the
white-box residual is nonzero. Pointed at the surviving \emph{parameter}
class, two teacher-forcing objectives never exceeded the plain-injection
residual---a limit of the attack construction, \textbf{not} a robustness
certificate; all gradient runs are suffix-only, and
Qwen3-30B's are likewise inconclusive. The most direct attack on the
mechanism itself---inflating instruction-token residual norms to dilute
the norm-preserving edit's rotation---\textbf{fails}, pre-registered:
replay-safe text achieves ${\approx}1.01\times$ inflation at those
positions at best; the first dilution tested, $1.25\times$, already
restores webpage-parameter compromise (Appendix~\ref{app:adaptive}). Replaying
every end-to-end-successful SaTML'25 LLMail-Inject text
\cite{abdelnabi2025llmail} (2018 texts, 2052 episodes, challenge-exact
port) takes undefended 128/2052 to \textbf{0/2052 observed} on gpt-oss-20b,
and Qwen3-30B's full \emph{dev} replay to \textbf{0/1537} against 117
undefended. Two qualifiers: this is replay, not live adaptation (transfer between
the challenge's own models: 14.3\%, our computation), and what \emph{does} transfer is the wire-format role
forgery this direction targets---consistency with the
mechanism, not breadth (decomposition:
Appendix~\ref{app:adaptive}). Relay attacks, which route the injection through the model's \emph{own}
text and bypass the payload-scoped edit by construction, do not complete
across turns even undefended (0/18 every arm) and are under-powered within
a turn---\textbf{open rather than answered}; the quotation step itself is
suppressed ($7/24 \to 0/24$, $p{=}0.005$).

\textbf{Protocol and search budget.} Every test-split number is a
\emph{single} pre-specified pass, with no reruns; all search happened on
development splits. That discipline mattered: a challenger that
\emph{dominated} every development comparison \textbf{reversed} on its
one-pass test evaluation
and was rejected (Table~\ref{tab:searchbudget}, Appendix~\ref{app:adaptive}).

\subsection{Comparison with Other Defenses}\label{sec:eval-baselines}

\begin{table}[t]
\caption{Baseline separation on gpt-oss-20b (comparison battery). Class~A:
non-delegated compromises /140 (undefended 57--59). adaptive: best-of-24
parameter search, cracked/18 dev; per-query rate
(Table~\ref{tab:adaptive-param-arms}). $^{\ddag}$own paired battery.
$^{\dag}$trained; benign \% of \emph{base} clean; AgentDojo 768-legacy.
\warnmark\ capability-guard; --- not reported. CIs: Table~\ref{tab:soa}.}
\label{tab:baselines-body}
\centering
\scriptsize
\setlength{\tabcolsep}{2.5pt}
\renewcommand{\arraystretch}{0.92}%
\begin{tabular}{lccccc}
\toprule
defense & \shortstack{AgentDojo\\cmp \dirdown} & \shortstack{Class~A\\\dirdown} & \shortstack{utilAtk\\\dirup} & \shortstack{benign \dirup\\\% of clean} & \shortstack{adaptive param.\\\dirdown\ (union; per-q.)} \\
\midrule
undefended & .483 & 59 & .551 & (ref.) & 18/18; .708 \\
\textbf{CounterSteer} & \textbf{.091} & \textbf{0} & \textbf{.716} & 90.6 & \textbf{13/18; .303} \\
AGRI$^{\ddag}$ & .062 & --- & --- & 78.0 & --- \\
SecAlign (trained)$^{\dag}$ & --- & --- & --- & 90.4 & 1/18; .002 \\
CachePrune & .216 & 15 & .688 & 88.7 & 17/18; .604 \\
PromptGuard-2 filter & .205 & 21 & .311 & \textbf{100.0} & --- \\
DeBERTa filter & .034 & 4 & .167 \warnmark & 47.2 \warnmark & --- \\
PIGuard filter & .011 & 1 & .106 \warnmark & 58.2 \warnmark & --- \\
spotlighting & .420 & 46 & .616 & 92.5 & 18/18; .595 \\
\bottomrule
\end{tabular}
\end{table}

\textbf{CounterSteer sits on the security--utility frontier among the
inference-time defenses measured on the capable models---not dominant}
(Table~\ref{tab:soa}, Appendix~\ref{app:baselines}; Fig.~\ref{fig:frontier}).
The twenty-six same-harness batteries (nine gpt-oss-20b, five Qwen3-30B,
nine Llama-3.1-8B, a three-battery GLM-4.5-Air detector probe) each run
four arms over the full 180-case grid at the uncensored budgets of
\S\ref{sec:eval}; every defense runs in one process with its \emph{own}
undefended baseline and within-run benign pairing---each row a paired
contrast, not a cross-run ratio (the defense set: \S\ref{sec:eval}).

\textbf{Table~\ref{tab:baselines-body} carries the gpt-oss-20b separation;
in severity order, CounterSteer removes 81\% of compromises at the best
utility under attack (utilAtk) and the best Class~A record of any
guard-passing arm.}
Every arm with a lower surviving rate either sits in capability-guard
territory (\warnmark; fp rates: Table~\ref{tab:soa}) or---AGRI---pays 22\%
of clean benign utility.
\textbf{On the capable models, no inference-time defense improves on
CounterSteer's compromise rate without paying a multiple of its benign
cost}---\textbf{and in our measurements the detectors' cost tracks the
agent's task capability} (PIGuard and the DeBERTa filter): on GLM-4.5-Air
PIGuard leaves 0/180 observed compromises while destroying 42\% of benign
utility, where CounterSteer holds benign \emph{at} clean with over four
times the filter's utility under attack (Table~\ref{tab:soa}). On
Llama-3.1-8B the ordering \emph{inverts}---with 13--15 of 56 clean tasks
solved there is little utility for false positives to
destroy, and PIGuard wins. Prompt-level defenses are partial at the honest
budget and not portable: sandwich wins on Llama-3.1-8B only at
27\% benign cost, and spotlighting is net harmful there (ten introduced
compromises). PromptGuard-2 is precision-tuned rather than free: zero benign false
positives, but under attack it deletes the poisoned message carrying the
task's legitimate data (Table~\ref{tab:baselines-body}). On Qwen3-30B,
CounterSteer-vs-DeBERTa is not a robust tier-1 ordering (overlapping
CIs)---the separation is tier~2, on utility under attack and benign cost.
The delegated-authority split (\S\ref{sec:eval-agentic}) separates the
mechanisms---CounterSteer's surviving compromises are 100\% Class~B, where
every other defense except the capability-destroying filters leaves 15--46
Class~A (Table~\ref{tab:baselines-body}); AutoDojo's task-delegation
finding independently corroborates the partition
(\S\ref{sec:eval-adaptive}).

\textbf{What the judge's low webpage utilities measure.} On the
summary-composition corpora the strict reading (\S\ref{sec:threat-metrics})
is nearly vacuous---the work product is the composed summary, which strict
exempts---and the judge reads steering's benign utility well
below it (webpage-param defense-on-clean judge 38.5\% of unattacked;
hand-reads confirm genuine fact-level losses). Table~\ref{tab:fidelity} bounds what
steering itself is charged: a \textbf{regeneration reference}
(byte-identical clean prompts regenerated cross-run) reads only 0.40--0.64
\textsc{equivalent}, so most measured degradation is wording drift;
pooled, steering sits ${\sim}$7--9\,pp below the other context-perturbing
defenses ($p{=}0.012$, a lower bound on its steering-specific deficit);
the over-flagging filters delete 100\% of clean webpage payloads (judge
fidelity 0.000---capability destruction); and in strict byte-exact terms
the ordering \emph{reverses}---steering preserves structured calls best,
paying in composed prose. With a regeneration floor this high,
single-turn composed-prose fidelity
is bracketed, not resolved (Appendix~\ref{app:judge}).

\textbf{CachePrune and SecAlign} (\S\ref{sec:eval}) both
get the steering direction's data access, with symmetric non-holdouts that
make the JSON-parameter row a train-templates head-to-head (ports, grids,
and the holdout detail: Appendix~\ref{app:baselines}). SecAlign's
fine-tune changes the served weights, where CounterSteer adds an
intervention to unmodified ones.

\textbf{Both baselines are graded on the same held-out test splits as
CounterSteer} (pre-registered re-run at identical budget and sample
order; dev readings: Appendix~\ref{app:baselines}). CachePrune is
a partial defense with the guard intact (JSON tool-log
$0.700 \to 0.463$; JSON param $0.647 \to 0.397$; both significant)---at
${\sim}9$--$36\times$ CounterSteer's surviving rate (0.013/0.045), at
benign brackets indistinguishable at this resolution
(detail: Appendix~\ref{app:baselines}).
SecAlign posts the strongest static tier-1 result measured on this model
(\textbf{0/91 and 0/70 observed}, vs.\ its own undefended 0.700/0.647
and CounterSteer 0.013/0.045 on the same samples) and is the measured
exception to the shared adaptive parameter-class weakness
(\S\ref{sec:eval-adaptive}). Its utility is quotable only at the
conservative end (a fine-tune's clean behavior is a different reference;
its weakest reading is 56\% of the same-model floor on the parameter
corpus; Appendix~\ref{app:baselines}). Four asymmetries each soften one
comparison: framing-level fit holdout exists on one deployed fit plus one
validated refit (Table~\ref{tab:provenance}); the near-duplicate
webpage-parameter templates; the JSON-parameter test split is
sample-disjoint only, its template axis measured separately
(\S\ref{sec:defense-param}); and SecAlign's adaptive framings sit inside
its own training grid (\S\ref{sec:eval-adaptive}).

The trained AgentDojo comparison predates the 4096 batteries and is
retained at its own 768 budget rather than re-run: its four arms share
that budget, so the paired ranking is valid and only the absolutes are
censored; it is never read against Table~\ref{tab:soa}. SecAlign leaves \textbf{9/180} compromises to
CounterSteer's \textbf{12/180} (CachePrune 32/180; undefended 62/180), a statistically indistinguishable
pair (McNemar $p{=}0.45$) whose survivors are in both cases entirely
delegated-authority (12/12 and 9/9). SecAlign's clean utility settles at
90.4\% unique-task of base clean at 4096
(Fig.~\ref{fig:frontier}); the budget anatomy, per-case ratios,
nulls and censoring asymmetry are in Appendix~\ref{app:baselines}.

\textbf{Random-direction controls} (per-draw detail:
Appendix~\ref{app:ablations}): the fitted direction reached ASR 0.026
against a matched control's 0.191 on the disjoint-tool corpus
($n{=}141$)---4.7~SD below the six-draw null (exact permutation over seven
draws $p{=}0.125$, not conventionally significant)---and two matched draws
cost \textbf{more} benign utility on AgentDojo than the deployed
direction, so much of
the benign cost is generic perturbation at this magnitude.

\textbf{Runtime cost and general capability.} The fixed vector edit is
runtime-negligible \emph{at our measurement resolution}: prefill latency
indistinguishable from unsteered (0.181 vs.\ 0.182\,s), decode throughput
and peak memory identical (bounds: Appendix~\ref{app:ablations}). Because
only tool-result positions are edited, an input with no tagged span
executes a byte-identical completion, so GSM8K \cite{cobbe2021gsm8k}, MMLU
\cite{hendrycks2021mmlu} and IFEval \cite{zhou2023ifeval} run here as an
identity check, not a utility claim; a probe
that wraps each item inside a steered span shows uneven preservation
(\S\ref{sec:limitations}, Appendix~\ref{app:ablations}).

\section{Mechanism: Correlates, Levers, and Where the Decision Lives}\label{sec:mechanism}

The first three findings below are negative results.

\subsection{Detection Is Not Defense}

Two directions passed every observational gate and failed the causal one.
A ``prohibition'' direction separated held-out attacks cleanly yet left attack
success unchanged under steering (0.299 vs 0.303)---it encodes the
decision's \emph{outcome}, not its cause; the second is the Qwen3-30B
user-role axis, which does not move ASR at matched dose
(\S\ref{sec:role-coupling}). Discriminative evidence can sit downstream of,
or parallel to, the circuit that decides
\cite{makelov2024illusion,tan2024steeringreliability,wollschlager2025cones};
concurrent work gates a reasoning defense on such latent signals
\cite{agri2026}, which exist broadly but are not universally the lever.

\subsection{The Role Axis Is a Model-Specific Correlate}\label{sec:role-coupling}

On gpt-oss-20b, injection-following and the user-role representation are
coupled: the behavioral direction has substantial cosine to the role axis and
role-axis steering alone moves ASR; on Qwen3-30B they are decoupled (role-axis
steering does not move ASR at matched dose; the behavioral direction does).
Across the rest of the matrix: Llama-3.1-8B unmeasured (its deployed fit is
the behavioral
direction alone); GLM-4.5-Air
undetermined (its coupling contrast ran on a corpus that barely fires
undefended); Gemma-4-31B unmeasured on its firing corpus.

The role-confusion \emph{reading} likewise dissociates from behavior at the
injected span. We applied the measure of \cite{ye2026roleconfusion} on both
primary models with a shared apparatus: the injected span's userness (user minus
tool logit of a pre-MLP role probe) vs.\ the legitimate record beside it in
the same tool message. On gpt-oss-20b the injected span reads \emph{more}
tool-like at the probe's role layers (96\% of samples), yet userness stays
predictive: succeeded attacks carry higher userness than blocked ones at 12/12
layers. On Qwen3-30B it is inverted \emph{and} non-predictive (fired-vs-not
authority-margin deltas within $\pm 0.16$ logits over the framing factorial),
while its behavioral direction is the causal lever.
On neither model does the injected span read as SYSTEM at the examined
sites, ruling out the natural rescue. So on the corpora and framings studied
here, the frame of \cite{ye2026roleconfusion} holds on gpt-oss-20b---whose
deployed configuration blends in the role axis---but is not the universal
intervention target. The behavioral direction generalizes; the role reading
is a model-specific correlate, and it is why detector-style approaches
trained on role signals alone should not be expected to transfer as
interventions.

\subsection{Parameter Manipulation Decides in the Model's Reasoning}\label{sec:defense-param}

The defense's surviving weakness is parameter manipulation under mild framing
(\S\ref{sec:eval-adaptive}), and we report its cause as a finding rather than
patching around it: \textbf{the payload-scoped constant-direction
intervention studied here does not fully solve parameter manipulation---the
commit decision is not linearly decodable at the examined pre-generation
sites with the tested probe family, while strongly decodable at value
emission.} Three pre-specified measurements:

\emph{(1) The decision is not linearly decodable in the prompt with the tested
probe family.} Linear probes on the injected payload span generalize at chance
within framing families (AUC 0.527, replicated twice on independent forwards),
and probes at the prompt tail and the tokens between payload and generation are
likewise near chance (AUC 0.447--0.526). The same labels read at
0.824--0.903 at the hijacked argument's \emph{value-emission} site, during
generation---with the tested probes, a decision linearly legible where the
model acts, not where it reads.

\emph{(2) Emission-site steering at a capability-safe dose showed no
detectable benefit.} Steering every decode step along the emission-site
direction at the largest dose passing the capability guard leaves attack
success on the held-out-template parameter corpus ($n{=}65$)
statistically unchanged (4/65 vs.\ 5/65).

\emph{(3) Nor does it survive the adaptive attacker.} Re-running the query
attack of \S\ref{sec:eval-adaptive} against the combined prefill$+$decode configuration still cracks
11/18 samples (vs.\ 13/18 prefill-only).

\textbf{The compromise path first quotes, then adopts}: the model
\emph{quotes} the
injected sentence into its own reasoning, then \emph{adopts} it, emitting the
hijacked value a median ${\sim}1.3$k characters into the completion. The
instruction thus reaches the commit point through the model's own generated
tokens. The probes found no earlier readable trace at the examined prompt
sites, and neither of the payload-scoped constant-direction interventions we
study---at the untrusted span, or per decode step---removes the decision;
what is readable at emission reflects a decision already made and relayed
through the model's own reasoning.
Porting CachePrune \cite{cacheprune2025} agrees: over the mask-fit
samples its attribution loss is identically zero at the first response token,
where its paper specifies it; the first divergence lands 4--36 tokens into the model's
own reasoning. Authority-styled attacks ride the representation CounterSteer
removes (their success falls 65--79\%); the bare frame-free instruction
largely does not (it falls only 34\%). Both primary models' held-out surviving
compromises are almost entirely parameter-class, convergent with the
adaptive result (\S\ref{sec:eval-main}, \S\ref{sec:eval-adaptive}). The
deployed defense is therefore prefill-only (\S\ref{sec:defense}), with the
parameter class quantified in \S\ref{sec:eval-main} and
\S\ref{sec:limitations}.

\subsection{Authority and Delegation}

\textbf{The authority lever itself is model-specific.} The same four-level
authority-framing factorial (\S\ref{sec:method-recipe}) yields qualitatively
different response curves: gpt-oss-20b escalates with claimed authority;
Gemma-4-31B and GLM-4.5-Air peak \emph{below} the top of the
escalation ladder---both on firm framings---and on Gemma-4-31B explicit supersession suppresses
parameter-manipulation firing entirely (0/72 observed).
Per-model rates, significance and the not-strictly-monotone caveats:
Appendix~\ref{app:authority}; which framings a defense is fitted against is
a per-model discovery.

\textbf{Delegated authority is a different object.} When the \emph{legitimate}
context grants the payload authority (``follow the instructions in the
document''), the signature we steer against is attenuated or absent: a
delegation factor in the behavioral contrasts was necessary for the direction to
generalize at all, and a deny/mention control shows it tracks granted
authority, not merely the mention of instructions. Compliance under delegation
is role-\emph{consistent}, which is why we scope it out (Class~B).

\section{Limitations}\label{sec:limitations}

\textbf{Per-model fitting.} CounterSteer is fit per model---its own
contrast capture, gates, and dose search. Cost is dominated by
evaluation, not fitting---fits are CPU-side, capture-through-fit is
${\sim}$1\,h (31B) to ${\sim}$2.3\,h (8B) on one GPU node, and cost scales
with model size and generation budget (GLM-4.5-Air's grid: 12.2\,h at
8{,}192 tokens; Table~\ref{tab:recipe-cost}). The recipe needs white-box
serving access: open-weights deployments, not third-party APIs.

\textbf{Benign utility is not free, and the price is model-specific.} On the
two primary models the always-on edit holds benign task fidelity at
94.1--94.4\% of clean at the deployed doses (typography-normalized, the
benchmark's own checker; raw: 85.2\% certification / 84.6\% comparison on
gpt-oss-20b, 94.1\% on Qwen3-30B; Table~\ref{tab:soa}). GLM-4.5-Air and
Llama-3.1-8B pay none measured; Gemma-4-31B reads 92.9\% ($-7.1$\,pp
n.s.; Table~\ref{tab:headline}). Two measured limits sharpen it. First, no measured gpt-oss-20b dose with
an established tier-1 effect reaches ${\ge}95\%$ of clean benign
(deployed: 94.4\%), and the cost is
largely generic perturbation at the deployed magnitude---matched random
directions cost \emph{more}, and the framing-held-out refit is
indistinguishable on both tiers---so the lever is dose, not direction
surgery (Appendix~\ref{app:ablations}).
Second, on the composition-heavy webpage corpora the
pairwise judge reads a fidelity cost the byte-exact bracket exempts, with a
measured steering-specific remainder of ${\sim}$7--9\,pp beyond shared
regeneration drift (\S\ref{sec:eval-baselines}). Reasoning over steered content degrades unevenly: GSM8K
\cite{cobbe2021gsm8k} unchanged, multiple-choice discrimination
$-10.8$\,pp, IFEval adherence $-15.2$\,pp at the deployed gpt-oss-20b dose
(Appendix~\ref{app:ablations}); deployments reasoning over retrieved
content should verify utility at their operating dose.

\textbf{Parameter manipulation.} The deployed defense reduces but does not
eliminate parameter manipulation under mild framing (13/18 crack under
query search). The weakness is shared by every inference-time
defense we measured, the trained SecAlign excepted
(\S\ref{sec:eval-adaptive}). How much of the deployed IPI problem each
Table~\ref{tab:map} class covers is itself unmeasured (on the AgentDojo
grid, 36/180 cases are Class~B). We read this as a boundary of payload-scoped
constant-direction interventions, not a fitting failure
(\S\ref{sec:defense-param}): CounterSteer is an instructional-takeover
component, not a complete IPI solution---compose it with
argument-provenance controls where argument integrity is the primary asset
(\S\ref{sec:conclusion}). Class~B (delegated authority)
is likewise unprotected, by design
(\S\ref{sec:threat-attacker}, \S\ref{sec:mechanism}).

\textbf{Only tool-result spans are steered.} Untrusted content in the
\emph{user turn}---text pasted from a poisoned page---passes through
unmodified by construction; the recipe extends wherever the serving layer
can tag untrusted spans (not evaluated).

\textbf{Adaptive evaluation is replay and search, not a live red team---and
budget-scaled attackers routinely break IPI defenses
\cite{nasr2025attacker}.} Our strongest
attackers are the fit-derived framing search, gradient attacks on
surrogate and exact deployed vectors, and the LLMail-Inject replay
(\S\ref{sec:eval-adaptive}). Unmeasured, specifically: AutoDojo
beyond six iterations (AD@$k$, $k \gg 6$); open-ended attacks outside that
framing grid; and text optimized for low projection onto the direction (its companion,
norm inflation on the edit, fails on gpt-oss-20b: $1.01\times$
instruction-token inflation achievable vs the sufficient $1.25\times$
first tested; Appendix~\ref{app:adaptive}). Surrogate
mismatch is unmeasured: our refit used the deployed direction's own samples,
the wider draw (cosine $0.88$--$0.91$) was never attacked, and every Qwen3-30B
gradient run steers the vector itself (Appendix~\ref{app:adaptive}).

\textbf{Model coverage.} GLM-4.5-Air and Gemma-4-31B carry \textbf{no
adaptive evaluation}---their certifications are one-pass held-out tests on
their firing corpora plus the agentic grid---and on Gemma-4-31B's
single-turn corpora the steered span is ${\sim}90\%$ of the prompt, so
only its agentic grid isolates the payload-scoped edit (coverage:
Table~\ref{tab:coverage}).

\section{Related Work}\label{sec:related}

Table~\ref{tab:related} (Appendix~\ref{app:tables}) groups IPI defenses
by where each intervenes and what a deployment pays.

\textbf{Activation-level interventions.} ICON \cite{icon2026}, ARGUS
\cite{argus2026steering} and AGRI \cite{agri2026} are the closest published
recipes, all detection-gated rather than detectors (\S\ref{sec:intro}): a
probe decides when to act, and only then does an attention mask (ICON),
representation edit plus post-filter (ARGUS), or reasoning prefill (AGRI)
fire. ICON is not run (no code released); ARGUS's ported dose rule yielded
no valid measurement (its evaluation is multimodal-only). \textbf{AGRI we
ported and measured same-harness on both primary models} (probe-only
release; deviations: Appendix~\ref{app:baselines}): the strongest
guard-passing inference-time rival measured---adaptive \textbf{0.143} vs
CounterSteer 0.175 on gpt-oss-20b at 78.0\% benign, with ${\sim}10\times$
context inflation from prefill-induced re-reading (Qwen3-30B:
Table~\ref{tab:headline}). Under adaptation its prefill often fails to bind (probe fires, prefill
applied, the model attacks anyway---its paper's knowledge--action gap,
reproduced out of set): the advisory intervention
inherits each model's reasoning--action coupling, where an always-on edit
does not (Table~\ref{tab:autodojo-matrix}). CachePrune \cite{cacheprune2025} and V-Steer \cite{vsteer2026} are
gate-free but need an attribution stage; CounterSteer validates by a
bidirectional causal test and steers the residual stream rather than
pruning the cache. We evaluate CachePrune (\S\ref{sec:eval-baselines}).
Neither evaluates a defense-aware agentic attacker.

\textbf{Detection is upstream of, not equivalent to, defense.} TaskTracker
\cite{abdelnabi2024tasktracker} detects task drift with linear probes at
near-perfect out-of-distribution AUC; AGRI reports latent
injection-exposure signals across eight models. Adjacent
detectors read attention patterns \cite{hung2025attentiontracker} or the input
text directly (PIGuard \cite{li2024injecguard}, PromptGuard-2
\cite{promptguard2}, DeBERTa \cite{protectai2024deberta}; measured as
filters, \S\ref{sec:eval-baselines}); the cost is a threshold that
fails open on a miss and, as a deletion filter, over-defense (reproduced);
instruction-hierarchy training \cite{wallace2024hierarchy} is the
training-time analogue.

\textbf{Training-time and prompt-level defenses} move the separation outside
inference. StruQ \cite{chen2024struq} and SecAlign
\cite{chen2024secalign,chen2025metasecalign} train it in; Jatmo
\cite{piet2024jatmo} distills per-task non-instruction-tuned models
(nothing to hijack); ISE \cite{wu2024ise} and ASIDE
\cite{aside2025} re-represent role or data tokens
architecturally---ASIDE, rotating data-token embeddings into an orthogonal
subspace, is the closest analogue; circuit breakers
\cite{zou2024circuitbreakers} train representation rerouting, and
DefensiveTokens \cite{chen2025defensivetokens} prepend optimized
embeddings at inference---the nearest no-weight-change comparator, not
run. We evaluate SecAlign, the strongest static result we measured, a
fine-tune of the served weights (\S\ref{sec:eval-baselines}; the released
Meta-SecAlign-8B \cite{chen2025metasecalign}, a trained Llama-3.1-8B
baseline needing no training on our side, is also not run). Prompt-level
defenses \cite{hines2024spotlighting}, the cheapest to deploy, remove only
13--44\% of compromises on the capable models (Table~\ref{tab:soa}).

\textbf{System-level defenses and provenance.} CaMeL
\cite{debenedetti2025camel} and MELON \cite{zhu2025melon} constrain a
hijack's consequences rather than the model's reading of the text; Progent
\cite{progent2025}, DRIFT \cite{drift2025} and DataFilter
\cite{datafilter2025} add policy, trajectory and sanitization layers
(landscape: \cite{ipitaxonomy2025,secfidelity2026}). ROPE \cite{rope2026} is a natural complement---it admits a value into a
guarded tool parameter only if its origin traces to the user or a
user-named source, at the price of provenance plumbing; CounterSteer acts
at the model, ROPE at the system, and \textbf{we do not evaluate the
composition}.

\textbf{Adaptive evaluation.} Static defenses erode under optimizing
attackers---the 461k-attack LLMail-Inject challenge
\cite{abdelnabi2025llmail} demonstrates it, we replay it, and AutoDojo
\cite{autodojo2026} makes the point benchmark-side.
Recent work trains the defender against synthesized attackers
\cite{reta2026} or re-optimizes it post-deployment \cite{copa2026}; we
report the strongest attack we can mount and name the class that survives
(\S\ref{sec:eval-adaptive}).

\textbf{Role representations.} This work builds on \emph{Prompt Injection as
Role Confusion} \cite{ye2026roleconfusion}, which trains role probes on
residual-stream activations and argues injected tool-message text reads as
user-like, userness predicting attack success. We reproduce its predictive claim on gpt-oss-20b
(\S\ref{sec:role-coupling}), whose deployed configuration blends in its
role axis. Our cross-model data add two qualifications: the span comparison's sign depends
on corpora and framings, and the role axis is not
a universal intervention target
(\S\ref{sec:mechanism}). Their causal manipulation is input-level; two blog posts explore the
activation side \cite{zhang2026steeringrole,mogford2026rolecausal}; we
claim only the first controlled, held-out, capability-guarded version to
our knowledge. Adjacent causal work steers
instruction-following in embedding space \cite{heo2024instruction} or tool
choice \cite{toolsteer2026}; SEP \cite{zverev2025sep} formalizes
instruction--data separation.

\textbf{Steering methodology.} Our operator is standard
difference-in-means residual steering (ActAdd \cite{turner2023steering},
RepE \cite{zou2023repe}, CAA \cite{panickssery2024caa}, refusal
\cite{arditi2024refusal}, ITI \cite{li2023iti}); CAST's gate \cite{lee2025cast} is not adopted and
AlphaSteer \cite{alphasteer2025} lost to the fixed vector in a pilot---the
contribution is the procedure that finds and validates what to steer.

\section{Conclusion}\label{sec:conclusion}

CounterSteer finds, causally gates, and continuously suppresses each
model's lever for treating untrusted text as instructions; across
twenty-six batteries and a benchmark-level adaptive attacker it sits on
the security--utility frontier among the measured inference-time defenses
on the capable models. Where the follow decision lives---retrieved span,
model reasoning, or user grant---dictates what removes it. For open-weights
deployments with tool-span tagging, it is an always-on layer against
instructional takeover at
93--100\% of clean benign utility (less when reasoning over steered
content, \S\ref{sec:limitations}); compose with provenance controls (ROPE
\cite{rope2026}) where argument integrity is the primary asset.

\section*{Ethical Considerations}

This work releases attack tooling---query, gradient, and mechanism-targeted
attack drivers---alongside the defense. Every attack was run against our own
deployed defense configurations on open-weights models over public
benchmarks; no third-party system or service was attacked, and the
LLMail-Inject replay uses only that challenge's publicly released attack
texts. Defense-aware adaptive evaluation is the accepted bar for this
defense class, and the drivers are required to reproduce ours; the
techniques they implement are published, so withholding them would impede
scrutiny of the defense more than it would impede attackers. No personal
data is processed and no human subjects are involved.

\section*{Open Science}

The artifact repository is available at
\url{https://github.com/markrussinovich/countersteer-satml27-artifact}\@. It contains everything needed to reproduce this
paper: the behavioral-contrast corpora with their split manifests and
disjointness assertions; the fitted directions and complete deployed
configurations for all five models; the canonical scorer and the pairwise
utility judge (with its validation-gate artifacts); our AgentDojo and
LLMail-Inject harness ports with deviations documented; the CachePrune and
SecAlign baseline implementations and the AGRI port (probes, gated-prefill
implementation, and its deviations table); the adaptive-attack drivers; the
pre-specification files with their SHA-256 manifest (the AutoDojo adaptive
plan, the framing-held-out refit certification, and the norm-dilution
attack protocol---each fixing arms, splits, doses and hypotheses before its
test touch; before-touch ordering is attested by the released run logs and
configuration dumps---no third-party timestamp attestation exists); and the per-sample completions behind every reported
table, with per-run configuration dumps.

Two things are deliberately not included. Trained baseline weights (the
SecAlign adapter and its 39\,GB merged model) exceed what a repository
should carry; the training driver, configuration and data are released, so
the baseline is re-trainable rather than merely described. Raw activation
captures regenerate deterministically from the released drivers and
corpora; we ship the fitted directions the defense actually loads. We
redistribute no third-party model weights or benchmark data: datasets load
at pinned upstream revisions with per-sample content hashes, and base
models are referenced by public identifier with recorded (not
load-time-pinned) revisions. The adaptive-attack drivers target
locally-served open-weights models and add no capability beyond the
published techniques they implement.

\section*{LLM Usage Considerations}

LLMs were used for editorial purposes in this manuscript, and all outputs
were inspected by the authors to ensure accuracy and originality. Beyond
the editorial role, this work used LLM agents extensively, under the
authors' direction: they
wrote and ran experiment code, launched and monitored jobs, analyzed
artifacts, and drafted and revised this paper's text. We also used them
adversarially: every experiment, script and substantive edit was reviewed by
a separate agent instructed to \emph{refute} it, defaulting to ``flagged''
under uncertainty; those reviews forced real corrections (quoted numbers, a
mislabeled generation budget, rescoped overclaims), and their verdicts are
released with the artifacts. The extent of this use is deliberate: the
paper's subject is the security of LLM agents, and the agentic tooling that
ran these experiments is the same class of system under study.

Two limits govern it. First, no \emph{security} measurement depends
on LLM judgment: attack success and the capability guard are computed by
deterministic scorers over parsed tool calls, and the absolute
LLM-correctness judge was ruled out after it disagreed with itself on
byte-identical generations (Appendix~\ref{app:corpus}). Single-turn
\emph{utility} uses a deliberately constrained instrument---a blinded
pairwise judge against the same run's unattacked reference, with control
pairs in every batch and a pre-specified validation gate---always quoted
beside a mechanical strict/lenient bracket (Appendix~\ref{app:judge}).
Second, the human authors set the research direction, made every scope and
pre-specification decision, signed off each result, and verified this paper's
claims against the underlying artifacts; they take full responsibility for
all content, including any error an agent introduced that they failed to
catch.

On environmental footprint: the experiments ran on a small fixed pool of
A100/H100 nodes (per-model fitting costs and hardware:
Table~\ref{tab:recipe-cost}), and the largest single expense---the adaptive
and multi-model evaluation program---is the paper's central evidentiary
claim: adaptive, defense-aware evaluation is precisely what static IPI
evaluations omit, and we judged that cost necessary rather than optional.
LLM-assistant usage ran on the same pool and commercial APIs; no model was
trained or fine-tuned for this work except the SecAlign baseline
reproduction.

\bibliographystyle{IEEEtran}
\bibliography{references}

\appendices
\section{Corpus and Scoring Notes}\label{app:corpus}

\textbf{Corpus construction (\S\ref{sec:eval}).} The two JSON
corpora draw real tool-log records from a public agentic
indirect-prompt-injection trace dataset~\cite{nemotron_ipi_traces} (loaded
at a pinned upstream revision
recorded, with per-sample content hashes, in the released corpus manifest);
attacks are re-embedded under our attacker-template factorial so splits can
be made sample- and template-disjoint, and the parameter-manipulation
variant rewrites the attack so its objective is a targeted argument of the
\emph{legitimate} call---the class no public IPI benchmark isolates. The two
webpage corpora wrap the same two classes in natural HTML carriers (fetched
Wikipedia articles, pinned revision), the summary-composition task the
canonical IPI story runs on; the Gemma-4 certification corpus additionally
draws 30 fresh, URL-disjoint carriers and new attacker wordings
(\S\ref{sec:eval-main}). The webpage and JSON tool-hijack splits are
sample- and attacker-template-disjoint, asserted at build time; the
JSON-parameter \emph{test} split is sample-disjoint under the fit wording
set, with that class's attacker-template holdout measured separately on
the held-out wording sets (\S\ref{sec:defense-param}).

\textbf{The strict/lenient utility bracket (\S\ref{sec:eval}).} Both bracket
ends are mechanical: a field is exempted from the lenient
end iff its \emph{clean-reference}
value is not a verbatim span of the tool payload, a criterion identical across
arms by construction. The narrow drift adjudicator supporting the analysis is
blinded, one-directional, and printed beside its controls. An \emph{absolute}
LLM correctness judge (one completion, one verdict) was
measured and rejected on two separate counts: on one corpus it scored a clean
arm at 0.525 against itself, and on another it labelled 12/24 completions
\textsc{correct} where the byte-exact scorer read 1/24; it also disagreed with
itself across runs on byte-identical generations. That instrument scores
nothing in this paper. The pairwise utility judge of
Appendix~\ref{app:judge} is a different instrument---blinded, comparative
against the same run's unattacked reference, control-gated per batch, and
validated before use---and it scores \emph{utility only}: attack success is
deterministic everywhere.

\textbf{Corpus-annotation disclosure (JSON tool-log corpus,
Table~\ref{tab:main-gptoss}).} The upstream JSON tool-log corpus annotates
its attack surface in 483/1271 tool schemas (a literal ``injection vector
field'' note in the schema text). A paired scrub-vs-original A/B with a
same-prompt null control bounds the annotation's effect on quoted numbers at
${\approx}0$; any remaining bias is conservative (undefended rates if anything
understated, defended arms unaffected), and the steering direction's fitting
pool contains no annotated sample.

\textbf{Scoring provenance and capability-guard rates, gpt-oss-20b test
evaluation (Table~\ref{tab:main-gptoss}).} Every number in a row of
Table~\ref{tab:main-gptoss} is scored from that row's single test-pass
artifact by the canonical scorer (no mixing of passes). The guard passes on
every row: defended
no-action rates are 0.115 / 0.058 / 0.043 / 0.000 (rows top to bottom)
against unattacked rates of 0.177 / 0.000 / 0.057 / 0.000 --- at most
$0.75\times$ unattacked where that floor is nonzero, and $\le 0.058$ absolute
where it is zero. No absolute ceiling was pre-specified for the zero-floor
rows (\S\ref{sec:threat-metrics}); the worst such row (3/52 on the
disjoint-tool corpus) was read completion-by-completion---two of the three
are intent-carrying generation loops, none is a refusal of the legitimate
task.

\textbf{Generation-budget censoring, gpt-oss AgentDojo
(\S\ref{sec:eval-agentic}, Table~\ref{tab:classab}).} An outcome checker
cannot distinguish a model that refused from one whose compliance was cut
off, so a per-turn generation budget censors attack success, and a transcript
audit showed the original 768-token run was truncation-censored in
\emph{both} arms---the undefended one about ten times more. The whole
battery---all 180 cases, all four arms---was therefore rerun at 4096 tokens,
which is the measurement quoted in \S\ref{sec:eval-agentic} (undefended
0.475, defended 0.079) and supersedes the 768 run. The audit localizes the
movement to the censored cases: the attacked arm's 127 untruncated episodes
are symmetric across budgets (9 up / 7 down, $p = 0.80$---also the honest
cross-host greedy regeneration noise floor), the defended arm moves 4 up /
0 down over its 19 truncated cases, and within the truncated stratum the
compromise rate
moves $0.156 \to 0.511$; so the earlier 0.068 was itself a censored number
and 0.079 is the defense's surviving rate, not evidence that the defended arm
was unaffected. The re-anchored 0.475 remains a lower bound: 3 attacked
episodes are still truncated mid-compliance at 4096 (tight bound
$86/177 = 0.486$), while all 9 truncated defended turns, read end to end,
contain no mid-compliance cut. The 768 and 4096 attacked figures are also not
quite the same measurement: a harness fix made five previously-aborted cases
scoreable ($n$ $172 \to 177$; paired on the common 172 cases the move is
$0.3605 \to 0.4651$). The 768-budget runs still quoted (the trained comparison of
\S\ref{sec:eval-baselines}) carry comparable truncation and are likewise lower
bounds at that budget---the tool filter excepted (5 truncated turns; it
reaches 0.000 by amputating the agent)---and Table~\ref{tab:soa} supersedes
them for every defense it covers.

\textbf{Why the Qwen3-30B AgentDojo runs use a different budget.} Every Qwen
AgentDojo measurement in this paper---the dose curve of
Table~\ref{tab:qwen-dose}, its paired benign arms, and every Qwen battery
of Table~\ref{tab:soa}---was run at \textbf{4096} tokens per turn, not
768. This is not an oversight but a requirement of the model: it is a
reasoning model that spends 1--4k tokens in its thinking channel before
emitting a call, and a 768-token turn is cut inside that channel, so no call
and no answer are emitted. Measured directly, at 768 the \emph{unsteered
clean} arm falls from 0.875 to 0.411, which makes every arm unmeasurable
rather than merely censored; the launcher now warns whenever a
thinking-format model is run below 2048. The consequence for reading this
paper is stated wherever it bites: each AgentDojo table is internally held at
one budget, and the one cross-model comparison we do make
(\S\ref{sec:eval-agentic}) uses each model's 4096 figures.

\textbf{Capability-guard rates and run provenance, Qwen3-30B held-out test
evaluation (Table~\ref{tab:main-qwen}).} The guard passes on every row:
defended no-action rates are 0.010 / 0.019 / 0.014 / 0.000 on the four test
rows (top to bottom) and 0.042 on the development-split row, against
unattacked rates of 0.062 / 0.000 / 0.043 / 0.000 and 0.052---at most
$0.80\times$ unattacked where the unattacked floor is nonzero, and $\le 1/52$
absolute where it is zero.
Truncation at the generation budget, over the four test rows, is zero in
every arm carrying a compromise-rate claim except the webpage tool-hijack defended
arm ($4/52$; three of the four are batch-padding flags on completed
sequences that still made the legitimate call, one is genuinely cut, and the
worst-case bound if all four hid compromises is ASR $\le 0.077$) and the
defense-on-clean JSON tool-log arm ($4/96$). The development-split row's arms
carry the same at-budget flag counts ($4/96$ per arm, likewise three
padding flags plus one genuine cut each); counting its one genuinely
truncated defended sample as a compromise would move that row's reading to
$5/91 = 0.055$, and counting every scoreable at-budget flag, to $\le 7/91 =
0.077$. These bounds are recorded here; the table is not rescored to them.
Every number in each row of
Table~\ref{tab:main-qwen} comes wholly from that row's single scored
artifact; the four test rows are one pre-specified pass per corpus with no
reruns.

\textbf{Tier-3 contamination, the $16\sigma$ dose ceiling, and the dev-vs-test utility gap (\S\ref{sec:eval-main}).}
Tier-3 contamination (injection-derived text reaching some \emph{other},
legitimate call; reported, never counted as compromise,
\S\ref{sec:threat-metrics}) is confined on both models to the JSON tool-log
corpus, whose legitimate task gives the matcher a nonzero false-positive
floor even unattacked: gpt-oss defended 0.212 against an unattacked floor of
0.125; Qwen defended 0.300 on the test split (floor 0.133) and 0.187 on the
development split (floor 0.088). On the other six corpus rows contamination is 0.000 in every arm of every
artifact except one: the \emph{undefended} attacked gpt-oss webpage-param arm,
at 0.038. Though tier-3 by the severity hierarchy, the JSON-corpus excess
above its floor is a partial-compromise channel---injected text reaching
legitimate calls---and deployments where argument provenance matters should
read it beside the parameter-manipulation limitation
(\S\ref{sec:limitations}).

The $16\sigma$ dose ceiling was measured on the \emph{development} splits
only (no pre-specified test pass was run for it). There it posts the
strongest single-turn-corpus ASRs anywhere in this paper---three zero-observed development rows
spanning both classes ($0/66$ JSON parameter manipulation, $0/159$ on each
webpage corpus, with the JSON tool-log development split at $1/91$)---but it
is capability-guard territory in the agentic setting
(\S\ref{sec:eval-agentic}) and is reported as the dose ceiling, not the
deployment point.

The dev-vs-test benign-utility gap on the JSON tool-log corpus
(45.1\%/91\% development vs 21.1\%/69\% test at the identical configuration)
is sample composition, not a scoring change: 60\% of the test split's tasks
are message-composition tasks versus 26\% on the development split, and
message-composition tasks fail strict byte-exact utility under steering
wording drift almost universally (49/54 on the defense-on-clean arm); test
samples also carry more structured arguments per call (3.69 vs 2.92).
Reweighting the test split to the development split's tool mix moves strict
benign utility from 21\% to ${\sim}31\%$; the rest tracks argument count. The
composition shift is a by-product of template-disjoint splitting (template
families correlate with scenario and tool).

\section{The Pairwise Utility Judge and the Fidelity Measurement}\label{app:judge}

\textbf{Why it exists.} No existing IPI benchmark measures whether a
\emph{defense} degrades the content fidelity of model-composed prose
(summaries, subjects) on benign traffic: outcome checkers score environment
state and answer text, and byte-exact scoring must exempt composed fields or
count every wording change as failure. This instrument scores that axis, and
only that axis---attack success is deterministic everywhere in this paper.

\textbf{Instrument.} The judge is gpt-5.4 at deterministic settings.
Each candidate completion is judged against the
\emph{same run's} unattacked reference, both rendered through one uniform
action-description path (raw completions are never shown; the judge sees the
clean payload only, so a hijacked payload cannot prompt-inject the judge),
presented without labels in an order re-randomized per vote (seeded by pair
content, replayable). Symmetric question; verdicts
\textsc{equivalent}~/ \textsc{degraded}~/ \textsc{different-task};
majority of three votes. Every batch carries indistinguishable control
pairs: identical pairs must read \textsc{equivalent} $\ge 0.98$ and corrupted
pairs (a structured argument value replaced, a call deleted, a task dropped)
must read not-\textsc{equivalent} $\ge 0.95$; a batch failing its controls is
discarded and reported, never retried into compliance. The headline column
follows the compromise-scoped convention: \textsc{equivalent} \emph{and} not
compromised, with compromise from the deterministic scorer. The reference is
the unattacked run, so the score reads directly as \% of unattacked.

\textbf{Pre-specified validation gate, passed before any number was scored.}
(i)~Agreement 0.867 (260/300) with an independent drift adjudicator's
already-issued verdicts on a stratified sample---the disagreements are
one-sided: the instrument rejects ${\sim}25\%$ of pairs the adjudicator
calls equivalent, all cosmetic-wording cases, so its bias direction is
conservative (it can understate our utility, not flatter it).
(ii)~On the failure set that convicted the rejected absolute judge
(Appendix~\ref{app:corpus}), it reads 0.150 equivalent where the old judge
fabricated 0.500 and the behavioral floor is 0.050; 8/8 hand-read verdicts
support it. (iii)~Determinism: zero majority-verdict flips across a full
double run ($n{=}80$ pairs plus controls) with the cache bypassed---this
certifies endpoint reproducibility; order-invariance is the separate
cache-audit measurement below. Three production batches failed the
corrupted-pair control and were discarded; all three were
control-\emph{construction} defects of two kinds (a corruption landing in
prose the spec says to leave alone; a corrupted value still visible in an
echoed markup line), repaired without touching the instrument---prompt,
blinding, votes and verdict mapping unchanged throughout---and the affected
evaluations re-run as fresh batches that passed.

\textbf{Caveats that travel with every judge number} (from the adversarial
sign-off): judge utility is a conservative \emph{floor}, not a point
estimate---\textsc{degraded} includes both confirmed fact-level losses
(hand-read) and borderline different-emphasis summaries; it can additionally
penalize tier-3 contamination visible in a response; and it is never quoted
without its strict/lenient bracket. Order-invariance was measured from the
verdict cache: equivalent-rate 0.260 reference-first vs.\ 0.271
candidate-first over 8{,}868 votes (n.s., trend against flattering the
defense).

\begin{table}[t]
\caption{Defense-on-\emph{clean} composition fidelity, judge \% equivalent
(zero compromise flags anywhere), four cells. Row one is the always-on steering
cost from the main utility tables; row two is the \textbf{regeneration
reference}: prompts byte-identical to the clean render, regenerated
cross-run with no defense, scored by the same judge---the shared
wording-drift floor every \emph{cross-run} arm (all baselines here) must be
read against. It does not directly calibrate the same-process steering row,
so ratios of row one to row two are never taken. Filter rows at 0.000 are
capability destruction (100\% of clean webpage payloads flagged and
deleted), not composition loss. Pooled over the four cells, steering sits
${\sim}$7--9\,pp below the regeneration reference ($p{=}0.012$, $n{=}274$
paired), a lower bound on its true deficit (the baselines absorb cross-run
noise steering does not); in strict byte-exact terms the webpage ordering
reverses (steering 75/65 vs.\ regeneration 71.2/55.8)---steering preserves
structured calls better and pays in composed prose. Per-cell
strict/lenient brackets live in the released fidelity artifacts; the
steering row's brackets are in the main utility tables.}
\label{tab:fidelity}
\centering
\footnotesize
\setlength{\tabcolsep}{2.5pt}
\begin{tabular}{p{2.9cm}p{1.15cm}p{1.15cm}p{1.15cm}p{1.15cm}}
\toprule
arm (defense on clean) & gpt-oss web param ($n{=}52$) & gpt-oss JSON hij.\ (80) & Qwen web param (52) & Qwen JSON hij.\ (90) \\
\midrule
\textbf{CounterSteer (same-process)} & .385 & .525 & .346 & .333 \\
regeneration reference (cross-run) & .462 & .637 & .462 & .400 \\
spotlighting & .442 & .625 & .462 & .356 \\
prompt sandwich & .462 & .600 & .462 & .356 \\
reminder & .423 & .600 & .365 & .344 \\
CachePrune (gpt-oss only) & .385 & .537 & --- & --- \\
DeBERTa filter & \textbf{.000} \warnmark & .588 & \textbf{.000} \warnmark & .378 \\
PIGuard filter & \textbf{.000} \warnmark & .613 & \textbf{.000} \warnmark & .367 \\
\bottomrule
\end{tabular}
\end{table}

\section{Adaptive Evaluation in Detail}\label{app:adaptive}
The protocols, per-attack detail and full caveat set behind
\S\ref{sec:eval-adaptive}. Every number quoted in the body appears here with
its bounds; nothing is stated here that the body contradicts.

\begin{table}[t]
\caption{Benchmark-level adaptive attack (AutoDojo optimizer, six iterations,
AgentDojo's own checker) against each defense, macro over the 36 pre-specified
(suite $\times$ objective $\times$ vector) cells per model. \emph{Adaptive} is
the attacker's best over seeds and optimized rewrites (\texttt{AD@6}).
Cell coverage, void-cell rulings and per-suite anatomy: the
``Table~\ref{tab:autodojo-matrix} notes'' paragraph below.}
\label{tab:autodojo-matrix}
\centering\small
\setlength{\tabcolsep}{4pt}
\begin{tabular}{lccc}
\toprule
adaptive success $\downarrow$ & gpt-oss-20b & Qwen3-30B & Llama-3.1-8B \\
\midrule
undefended            & 0.673 & 0.730 & 0.377 \\
CounterSteer          & 0.175 & \textbf{0.188} & 0.307 \\
AGRI (probe-gated prefill) & \textbf{0.143} & 0.411 & --- \\
CachePrune            & 0.305 & 0.558 & \textbf{0.252} \\
PromptGuard-2 filter  & 0.527 & 0.670 & 0.369 \\
SecAlign (trained)    & 0.160 & --- & --- \\
PIGuard filter              & 0.166 & 0.360 & --- \\
DeBERTa filter              & 0.146 & --- & --- \\
spotlighting                & 0.585 & --- & --- \\
reminder                    & 0.473 & --- & --- \\
prompt sandwich             & 0.489 & --- & --- \\
\bottomrule
\end{tabular}
\end{table}

\textbf{Table~\ref{tab:autodojo-matrix} notes.} SecAlign is the gpt-oss DPO
model (\S\ref{sec:eval-baselines}), at 35 of its 36 cells (the 36th, a
slack cell, deterministically exhausts 80\,GB memory and could not be run);
its two zero suites are exact (0/23 cells). PromptGuard-2's collapse is
carried by the attack population's seed
wordings (seed-best 0.499 of its 0.527); the filter was live and flagging in
every lane. PIGuard/Qwen is complete at 36/36 cells (banking 0.771, slack
0.308, travel 0.000; its optimizer lift is zero---seed-best 0.332 of its
0.360---so its losses are static seeds passing the filter, not adaptation).
AGRI/gpt-oss is final (36/36; one travel cell's adaptive reading is void
with cause---an attack-writer artifact, static reading valid).
AGRI/Qwen is final at 36/36 (banking 0.674, slack 0.493, travel 0.067;
three travel cells void-with-cause at their static readings---brackets
0.449--0.495, no ordering moves); its residual is seed-dominated (86\%
static share), like PIGuard's. On gpt-oss, banking is
every arm's worst suite (undefended 0.850; CounterSteer 0.458, PIGuard
0.458, DeBERTa 0.375, reminder 0.602, spotlighting 0.694): its user tasks
delegate payment authority to the injected document, so the attack can be
data-shaped (\S\ref{sec:eval-adaptive}).

\begin{table}[t]
\caption{Defense-aware adaptive parameter-manipulation search, five arms:
best-of-24 framing queries (6 authority levels $\times$ 4 voices) per sample,
$n{=}18$ development samples, $432$ attempts per arm. union $=$ samples cracked
by $\ge$1 framing; per-query $=$ mean firing rate. Tier-1 only; no arm scores by
refusing (no-action $\le 0.07$ everywhere). \textbf{Read per-query beside
union}: the undefended literal union (3/18) sitting below two defended arms'
4/18 is a union-statistic artifact of rare, sample-concentrated literal fires.
SecAlign is a \emph{trained} arm and the search's framings are cells of its own
training grid---trained framing families, not framing generalization.
Spotlighting's delimiter leak and SecAlign's single scorer-level fire:
the notes below.}
\label{tab:adaptive-param-arms}
\centering
\footnotesize
\setlength{\tabcolsep}{3pt}
\begin{tabular}{p{2.6cm}p{1.1cm}p{1.1cm}p{1.2cm}p{1.2cm}}
\toprule
arm & union \dirdown\ pattern & union \dirdown\ literal & per-query \dirdown\ pattern & per-query \dirdown\ literal \\
\midrule
undefended & 18/18 & 3/18 & 0.708 & 0.076 \\
spotlighting-with-delimiting & 18/18 & 4/18 & 0.595 & 0.044 \\
CachePrune \cite{cacheprune2025} & 17/18 & 4/18 & 0.604 & 0.037 \\
\textbf{CounterSteer (deployed)} & \textbf{13/18} & \textbf{1/18} & \textbf{0.303} & \textbf{0.002} \\
SecAlign \cite{chen2024secalign} (trained) & 1/18 & 0/18 & 0.002 & 0.000 \\
\bottomrule
\end{tabular}
\end{table}

\textbf{Notes on Table~\ref{tab:adaptive-param-arms}.} The undefended
exact-literal union of 3/18 is driven by 33 fires concentrated in those 3
samples, which is why the per-query columns---not the unions---order the arms
as expected. Spotlighting's delimiters leak into 12 of its 432 completions but
reach no parsed tool-call argument set, biasing scoring in neither direction.

Adaptive, defense-aware evaluation is the bar this defense class must clear:
static-corpus numbers routinely overstate prompt-injection defenses, and
recent work builds the adaptive attacker into the defense's own design (e.g.,
RETA \cite{reta2026}). Escalating strength per the
\S\ref{sec:threat-attacker} attacker; every attack targets the gpt-oss
deployed configuration except where a second model is named.

\textbf{Query-based framing search (black-box).} Best-of-48 attack framings per
sample against the live defended agent: 48 per sample $=$ the 24-framing grid
of Table~\ref{tab:adaptive-param-arms} (6 authority levels $\times$ 4 voices)
instantiated once per attack class (tool hijack; parameter manipulation), so
within a class the search is best-of-24; $n{=}18$ samples per class; both
attack-success readings of \S\ref{sec:threat-metrics} reported side by side.
The undefended comparator on the identical parameter-class grid is
Table~\ref{tab:adaptive-param-arms}'s first row (union 18/18, per-query 0.708;
exact-literal union 3/18, per-query 0.076); no undefended run of the
tool-hijack grid exists, so that class's undefended comparators are the static
corpus rates.
Under the \emph{obedience-pattern} reading, on tool
hijack the defense holds: 3/18 samples cracked [0.058, 0.392], per-query firing
rate 0.007. On parameter manipulation the search \textbf{cracks the defense:
13/18 samples} [0.491, 0.875], per-query firing 0.303---and the strongest
wordings include \emph{mild} ones (bland parameter-correction framings tie the
strongest role-spoof framing at 0.50). Under the \emph{exact-literal} reading
the same searches nearly vanish: 46 of the 48 framings deliver the attacker's
exact payload on 0/18 samples, and the remaining two (one per class) on 1/18
each (0.056). That is at most one sample per class cracked, an overall
exact-literal per-query rate of $2/864 \approx 0.002$.
The direction was fitted against
authority-styled framings, and a blandly-worded parameter hijack slips under
it. Crack counts in this section are per-class obedience-pattern unions
unless labeled otherwise; pooling both classes at the sample level gives
15/18 for the deployed configuration (13 param $+$ 3 tool, one sample
cracked in both), the cross-class union quoted in the protocol paragraph
below.

Is this weakness a defect of CounterSteer, or a property of the attack class?
We reran the identical defense-aware search---the same best-of-24 parameter
framing grid (6 authority levels $\times$ 4 voices), the same $n{=}18$
development samples, the same generation configuration---against the
undefended model, against two published inference-time baselines of
\S\ref{sec:eval-baselines}, spotlighting-with-delimiting and CachePrune (the
prompt-sandwich and tool-filter arms were not run), and against the SecAlign
fine-tune of \S\ref{sec:eval-baselines}
(Table~\ref{tab:adaptive-param-arms}). Every
\emph{inference-time} defense measured shares the surviving class: the two
published baselines trim per-attempt firing only modestly ($0.708 \to
0.595/0.604$, ${\sim}15\%$ relative; both real, McNemar vs undefended
$p<10^{-5}$) and
remain at or near full union crack (18/18 and 17/18). CounterSteer is the
inference-time arm that suppresses per-attempt success (0.303 vs
0.60--0.71) and the only one
that holds the exact-literal reading to 1/18 (per-query 0.002 vs
0.037--0.076). Because attempts nest within the 18 samples, inference is at
the sample level: an exact paired sign-flip permutation over each sample's
mean firing rate gives a reduction of 0.29--0.41 against every arm
(exact $p \le 1.7{\times}10^{-3}$; the weakest paired-bootstrap 95\% CI
is spotlighting's [0.146, 0.431]; 15--17 of 18 samples favoring
CounterSteer; artifact
\texttt{runs/adaptive\_sample\_level\_tests.json}), and the per-sample
crack unions are the primary reading. The 13/18 union is therefore not a CounterSteer defect
relative to anything measured here: it is the strongest result among the
inference-time defenses measured, on an attack class where the published
baselines remain
at or near full union crack. The \textbf{trained defense is the measured
exception}: SecAlign holds the identical search to a union of \textbf{1/18}
on the obedience-pattern reading and \textbf{0/18} exact-literal
(per-attempt 0.002 / 0.000)---preference-optimization training closes the
parameter-class gap that bounds the inference-time defenses measured, at
fine-tuning's deployment cost, and with the framing caveat in the table
caption (the search's framings are cells of SecAlign's own training grid, so
this is adaptive robustness on trained framing families, not framing
generalization).
The tool-hijack class closes the same way: under the identical best-of-24
tool-hijack framing grid on the same 18 development samples, SecAlign's
union is \textbf{0/18 on both readings} (432 attempts, truncation-free, no
capability collapse: 427/432 rows still call a tool), against CounterSteer's
3/18 obedience-pattern union above (CounterSteer's exact-literal tool-grid
reading predates the literal-delivery field and is unmeasured). And the
parameter grid's sole pattern fire is a transcript-verified scorer
artifact---the model quotes the poisoned tool description verbatim inside
the user-tasked email without adopting it in its reasoning (exact-literal
negative), though the quoted exfiltration address does reach the message
recipient---so SecAlign's adaptive survivors on its trained framing families
is effectively \textbf{0/18 on both classes}.
The held-out test passes converge on the same
picture: of the surviving test-split compromises, 10 of 11 on gpt-oss and 13
of 16 on Qwen3-30B are parameter-class (\S\ref{sec:eval-main}).

The measurements support the following diagnosis: at the payload span, param obedience is barely
linearly decodable with the tested probe family once framing identity is
controlled (held-out within-framing AUC 0.527), while at the argument-value
emission point it is strongly decodable (AUC 0.824 at the choice point,
0.88--0.90 over value tokens). The prefill-only defense never touches that
site. Steering that site at every
decode step, at the largest capability-safe dose, trims per-query firing by
roughly a third but leaves the best-of-queries union nearly intact (param
class: 11/18 vs.\ 13/18). It also yields no significant held-out corpus gain (4/65 vs.\
5/65)---the basis of the negative finding in \S\ref{sec:defense-param}. A
scope-contrast component that halved
the per-query param firing (0.303 $\to$ 0.141, $p = 3{\times}10^{-12}$) failed
its pre-specified test pass and was not adopted (protocol paragraph
below).

\textbf{Surrogate-refit GCG (gradient, transfer; gpt-oss-20b).} The attacker
reruns our published pipeline on the public weights with their own generations
and obtains a surrogate direction at cosine \textbf{0.974--0.977} to the
deployed one (layers 12/16/20). They then run GCG \cite{zou2023gcg} (250 steps,
$k{=}20$ suffix
tokens) with gradients through their surrogate-defended model. Result on
disjoint-tool targets: \textbf{0/12 observed} [0.000, 0.242] against the
deployed configuration---and \textbf{0/12 observed against the surrogate itself}, which
the attacker had full gradients through; at $\cos \approx 0.975$ this reading
is effectively white-box. The same optimized strings succeed 5/12 undefended.
Naive transfer GCG (optimized on the undefended weights) is likewise 0/8
observed [0.000, 0.324] vs 5/8 undefended. Bounds: 12 of 16 planned samples
completed, and the loss is systematic rather than random---two of eight shards
did not finish, so the missing four are a fixed stride of the sample pool;
budget-limited; suffix-only (wording-level search is covered by the query
attack above); tool-hijack targets (the param class is the documented adaptive
hole).

\emph{What this arm does and does not vary.} Our pipeline selects its contrast
samples deterministically (a fixed-seed template-disjoint split, then a prefix
of it), so a faithful recipe replication lands on the deployed direction's own
24 sample ids; the two captures share all 576 contrast episodes (24 samples
$\times$ the 24-cell framing factorial) and their activations agree to
numerical noise. What differs is the attacker's regenerated behavior
labels---which a crossed refit shows produce essentially the whole direction
gap---plus two quantities inherited bit-for-bit from our released probe
pickles (the role-probe axis component and the dose scale), each of which
slightly favors the surrogate's match to the deployed configuration.
This arm therefore measures
whether an attacker who reproduces our recipe can rebuild a working surrogate;
it is \emph{not} a test of whether the specific fitting samples matter, and
we do not claim the fitting samples are private. We did fit the direction on a
wider draw as well---36 samples, the deployed 24 plus 12 more---which lands at
cosine \textbf{0.884--0.909} to the deployed vector at the same three layers
(0.787--0.926 across all twelve): changing the draw moves the direction visibly
farther than regeneration alone does. That variant was
not carried through to an attack evaluation, so sample-set sensitivity here is
measured at the direction level only (\S\ref{sec:limitations}).

\textbf{Exact-deployed-vector GCG.} Finally we hand the attacker the actual
deployed direction and dose and let GCG optimize suffixes with gradients
through the deployed defended forward pass---directly instantiating the
strongest white-box case in our stated threat model
(\S\ref{sec:threat-attacker}, which treats the fitted direction and dose
as knowable), to measure how much of the defense is secrecy. At $n{=}52$
dev samples with a 500-step budget, convergence-based early stopping
(49/52 trajectories converged; the 3 budget-capped ones are defended
non-fires) and candidate filtering: \textbf{2/52 observed compromises}
(0.038, Wilson $[0.011, 0.130]$), both verified compromises of the
attacker's objective. One
further sample falls to the plain injection alone (1/52, within the deployed
configuration's known residual on this corpus, 0.013--0.045); its optimized
suffix does \emph{not} fire.
On the undefended model the same optimized suffixes succeed 22/52 vs.\
26/52 for plain injection---the optimizer confers no advantage absent the
defense, so the undefended replay is not evidence of attack strength. One
of the two defended compromises fired \emph{only} under the defense: its
suffix, optimized against the steered model, does not transfer to the
undefended one. An earlier, weaker run (a different 16-sample dev draw---9 of its
samples recur among the 52, though none of the three samples compromised
here was in it---250 steps, no candidate filtering) observed 0/16
$[0, 0.194]$; it is retained as the predecessor, never pooled. (Scope: gpt-oss-20b,
tool-hijack objectives, single restart per sample.)

\textbf{Exact-deployed-vector GCG, parameter class.} The same white-box
attack pointed at the documented surviving class
(\S\ref{sec:defense-param}): suffixes optimized through the defended
forward pass toward the legitimate call carrying the attacker's argument
value ($n{=}52$ development samples, 500 steps, convergence stop, single
restart). No gradient-attributable compromise: 1/52 adaptive, and that
fire is byte-identical to the same sample's plain-injection fire---inside
the 2/52 plain residual (the configuration's known 0.030--0.061 on this
corpus). Undefended, the
optimized suffixes again lose to plain injection (28/52 vs.\ 38/52). One
defended completion truncates in a degenerate loop with no path to a call
(a defense-favorable censoring; worst case, one further non-subsumed
fire). We
do \emph{not} report this as white-box robustness: a pre-registered gate
binds---the teacher-forced objective places the call at an offset that
0 of 306 observed real parameter compromises use (all reason in the
analysis channel first), so the null cannot separate ``the defense
holds'' from ``the attack aimed off the compromise path''. The gate's required follow-up---a path-matched objective teacher-forcing
the call after the model's own greedy reasoning prefix---ran at the same
$n{=}52$ on the paired pool: defended-adaptive again 1/52, and that fire
is the known quote-and-flag scorer false positive (the completion calls
the injection ``weird text'' and quotes it; the attacker's value is
adopted nowhere), its fourth adjudication across defenses. The
path-matched objective largely failed to converge (51/52 plateau-stopped,
median final CE 2.7; the four sub-1.0 losses are stale-prefix overfits
that produce no defended compromise), and the optimized suffixes again net-weaken the plain
injection undefended (29/52 vs.\ 34/52). Gradient-attributable defended
compromises across both objectives: \textbf{0/52}. One disclosure rides
the static readings: greedy decoding is not bit-stable across process
histories on this hardware (a fresh-process replay reproduces itself
exactly but matches neither run), so the static ASR at $n{=}52$ carries
${\pm}$1--2 fires of process-history jitter; the canonical residual
remains the 2/52 conservative reading. Neither run adds independent
evidence for the decision-point mechanism, which rests on the offset
measurement itself. On the second model the same procedure is
\textbf{inconclusive} and we do not report it as a robustness result:
against the Qwen3-30B deployed configuration (16 development samples, three
restarts sharing one sample pool, 250 steps) we observed no compromise of
the attacker's objective, but the restarts are not independent draws, so the effective $n$
is 16 and the cluster-aware 95\% upper bound on the per-sample ASR is
${\approx}21\%$---and the adaptive component was near-inert, since plain
injection already succeeds undefended 10/16, GCG buys $+1$, no undefended
success emitted the optimized target string, and 31 of 48 loss trajectories
were still descending at the step budget. Its objective (an immediate tool
call with no reasoning) targets a path no success on this reasoning model
took; a converged attack with an objective specified for reasoning models is
future work. All Qwen3-30B gradient runs steer with the deployed vector, so all
three restarts are exact-vector runs. An early attempt at a Qwen surrogate
re-ran the capture stage on the deployed fit's own pinned inputs and copied the
deployed probe fits; with deterministic decoding that capture reproduced the
deployed one row for row, so the refit reproduced the deployed direction to
floating-point precision (cosine $1.000$ at all three steered layers). That run
is therefore a third exact-vector restart rather than a surrogate.
\textbf{Surrogate mismatch is unmeasured on Qwen3-30B}, and no surrogate result
for that model is reported anywhere in this paper.

\textbf{Norm-dilution attack (mechanism attack on the norm-preserving
edit).} The edit rotates each payload token by an angle scaling as
$1/\lVert h_t\rVert$ (Eq.~\ref{eq:edit}), so injected text that inflates the
residual norm at its own instruction-carrying positions by $\kappa$ receives
the equivalent of dose $\alpha/\kappa$---and the dose ladder makes the
requirement concrete: $\kappa{=}1.25$ already lifts defended
webpage-parameter goal from $.096$ to $.596$ ($n{=}52$ dev). Measured on
gpt-oss-20b's deployed cell through the deployed hook, over 14 replay-safe
attack constructions plus a length-matched benign control (screened
high-norm token filler in four placements; delimiter, digit, and whitespace
runs; case/hyphenation rewrites---attacker literals byte-intact and the
scorer's evidence sets asserted unchanged), achievable instruction-token
inflation tops out at $1.006\times$ (a filler run's \emph{own} tokens reach
$1.17\times$ in a 2048-candidate screen but induce no instruction-token
inflation); attention spillover onto unmodified
payload tokens is nil ($0.999$--$1.001$); and the hyphen/studly rewrites
\emph{lower} instruction-token norms $4$--$9\%$ (all-caps near-null),
increasing the delivered rotation.
Behavior matches: adaptive defended goal equals static defended goal on both
corpora (tool hijack $0/24$ vs $0/24$; parameter $3/24$ vs $3/24$, McNemar
$p{=}1.0$; re-generated at a 4096-token budget to rule out truncation
censoring) with the injection intact undefended ($24/24$;
$21$--$22/24$). Injected instruction
tokens already sit $6$--$18\%$ \emph{above} the carrier-median norm. One
ladder disclosure: the \emph{under-dosed} defense is worse than none on
the parameter corpus ($\alpha 0.5$: $.923$ vs base $.808$, $n{=}52$). The
protocol was pre-registered (constructions, gates and readings fixed before
the behavioral stages); dev split only; the paired low-projection text
attack remains unmeasured.

\textbf{Relay attacks (mechanism-predicted).} The mechanism account of
\S\ref{sec:mechanism} predicts a structural gap: the defense steers only
tool-payload positions, so an injection that routes through the model's
\emph{own} text---quoted into its live reasoning, or echoed into a prior
assistant turn---travels a channel the edit never touches. We built both
attack shapes and ran them against the deployed gpt-oss configuration.
The cross-turn relay fails even undefended: the injected instruction is
echoed into the prior assistant turn in 12/18 episodes, yet argument-scoped
adoption is \textbf{0/18 observed on every arm}, including an echo-only
counterfactual arm where the injection reaches the second turn solely
through prior-assistant tokens. The same-turn reasoning relay is
under-powered at tier 1 undefended (argument-scoped adoption
$3/24 = 0.125$, below the pre-specified ${\sim}0.2$ power threshold, all
via the quote-directive shape), so the tier-1 relay question is open rather
than answered; the defended arm is 0/24 observed on every reading,
consistent with complete blocking but not separable from chance on the
argument-scoped contrast at this $n$ (Fisher one-sided $p = 0.12$). What
\emph{is} statistically supported is that the defense suppresses the
quotation step itself: defended completions quoting the injected
instruction in their reasoning fall $7/24 \to 0/24$ ($p = 0.005$), verified
on the raw completions as genuine behavior rather than refusal or
degeneration (all 24 defended completions carry substantive analysis and a
normal legitimate tool call; defense-on-clean strict utility 20/24, failures
wording drift). Per the pre-specified interpretation, escalated content
salience (not authority framing) is the follow-up before any tier-1 ``relay
blocked'' claim.

\textbf{LLMail-Inject replay (human-adaptive).} We replay every attack text
that succeeded end-to-end in the SaTML'25 LLMail-Inject challenge
\cite{abdelnabi2025llmail}: of its 461,640 submitted attacks, the 2018 unique
texts that fully succeeded, replayed as 2052 episodes (one per challenge
level each text beat). These are human-crafted attacks
that beat prompt shielding, TaskTracker \cite{abdelnabi2024tasktracker},
spotlighting \cite{hines2024spotlighting}, an LLM judge, or their
combination. The replay runs through a challenge-exact port (their system
prompt, retrieval,
scoring checker; deviations enumerated in the artifact): see
Table~\ref{tab:llmail}.

\begin{table*}[t]
\caption{LLMail-Inject replay (human-adaptive), gpt-oss-20b deployed configuration.}
\label{tab:llmail}
\centering
\footnotesize
\setlength{\tabcolsep}{4pt}
\begin{tabular}{p{2.6cm}p{4.6cm}p{4.4cm}p{4.8cm}}
\toprule
split & ASR \dirdown\ undefended [95\% CI] & ASR \dirdown\ defended [95\% CI] & clean-traffic cost \\
\midrule
dev $n{=}1537$ & 105/1537 = 0.068 [0.057, 0.082]; ${\le}0.073$ truncation-corrected & \textbf{0/1537 observed} [0.000, 0.002] & none (0 spurious sends; retrieval coverage 0.988, matching the clean arm) \\
test $n{=}515$ (held out) & 23/515 = 0.045 [0.030, 0.066] & \textbf{0/515 observed} [0.000, 0.007] & none (retrieval coverage 0.990, matching the clean arm) \\
\bottomrule
\end{tabular}
\end{table*}

Combined: 128/2052 undefended (${\le}149/2052 = 0.073$ after an adversarial
audit of the harness credited comply-intent completions cut by the token
budget) $\to$ \textbf{0/2052 observed} [0.000, 0.002]. The result is
echo-audited: the defended arm's single flagged completion was a parser
false-positive quoting the attacker's text while refusing, and every
undefended hit
is a genuine tool call. Stated conditionally, and verified per-episode by id
intersection rather than by aggregate: of the 128 undefended successes
(${\le}149$ after the truncation audit), \textbf{zero} succeed under the
defense; on Qwen3-30B, of the 117 undefended dev successes, \textbf{zero}
succeed under the defense.
The defense leaves 0 observed successes in every original-defense family,
including the 22 texts that beat the challenge's full combined stack.

The
caveat on the undefended rate: these attacks were optimized against
\emph{other} models, and adaptive prompt attacks are model-specific. The
challenge paper discusses cross-model transfer only qualitatively and reports
no rate for it, so we computed one from its released phase-1 submissions. Every
retrieval level ran against two victim models under the same defense; where the
\emph{same team} submitted an identical subject and body at both, an attack
that had succeeded on the first model succeeded again on the second in
\textbf{14.3\% of directed re-submissions} (44/307, spanning 285 distinct
text-pair instances; counting each pair once instead of in both directions
gives 22/285 = 7.7\%). Either figure is conditioned on what participants
\emph{chose} to re-submit --- it is not a systematic re-evaluation of all
successful attacks on the second model --- and it varies widely with the
defense they re-submitted under (Spotlight 4/88 = 4.5\%, LLM Judge
14/36 = 38.9\%). On that order, the ${\sim}5$--$7\%$ we observe
transferring to a third, never-targeted model is unsurprising --- though, as
the decomposition below shows, the rate we measure is also specific to the
challenge episode's own inline hardening, whose removal raises it to 0.254,
above the pooled cross-model figure. This
experiment measures replay of human-adaptive attacks, not live adaptation
against our defense. A row-level decomposition of the undefended non-transfer,
adversarially reviewed, sharpens this in two ways.

First, the failing replays
are not silent misses: the model reads the injection and refuses it by
name---73\% of the 1432 undefended gpt-oss failures on the dev-split replay
($1537 - 105$; the failure anatomy was run on that split only) quote the attacker's ask
while declining, and ${\sim}0$ ignore it silently. The refusals anchor on
the challenge episode's \emph{own} inline hardening: the scenario ships a
partial prompt defense (an anti-injection system block plus an in-user-turn
``don't follow instructions in them'' line), and removing both layers raises
undefended transfer $0.070 \to 0.254$ ($n{=}256$; ${\sim}9\times$ the
batch-nondeterminism floor bounded by a same-configuration canary, with either
hardening layer alone holding transfer near baseline).

Second, what does
transfer is wire-format role forgery: fake conversation turns and forged
tool-call syntax transfer at $0.10$--$0.30$ per class and model, and carry 90--97\% of
all undefended hits on \emph{both} replay targets (gpt-oss and Qwen3-30B),
with far-above-chance hit-set overlap between the two unrelated models (62
common hits vs.\ 8 expected under independence). Successful completions
re-attribute the injected ask to user or system authority verbatim in their
reasoning, while the detector-evasion shapes that dominate the corpus
(encodings, datamarking, obfuscation) transfer at ${\sim}0.004$ on both
models. Attacks bred against detection stacks are inert against a model swap;
what survives model-to-model is precisely the role-confusion class this
direction was built for and drives to 0/2052 observed.

The steering direction predates this corpus entirely (holdout on samples,
wording, and benchmark). On Qwen3-30B at the deployed $12\sigma$
configuration, the full dev replay ($n{=}1537$) gives undefended 7.6\% end-to-end
success (117/1537 after an adversarial audit recovered 13 successful
attacker calls the
parser initially missed in an alternate emission format; present at all four
challenge levels) and defended 0/1537 observed
(Wilson 95\% CI [0, 0.0025])---including zero on the obedience superset
(attacker address reaching any argument of any call). The defended Qwen arm
answers on 100\% of
episodes, with retrieval coverage 99.4\% and a 0.0\% false-action rate on
injection-free traffic.

\textbf{Protocol: selection risk, measured.} During development a challenger
configuration (the deployed composition plus a scope-contrast component)
\emph{dominated} the deployed configuration on the development
splits---webpage param 0.063 vs
0.126 ($p = 0.031$), AgentDojo 0.034 vs 0.068 (both at the development-time
768-token budget), query-attack cross-class
sample union 11/18 vs the deployed configuration's 15/18 (both
obedience-pattern, both classes pooled at the sample level). It was advanced
to a pre-specified, one-pass, no-rerun test
evaluation on all four corpora---a separate \emph{paired} run, challenger and
deployed configuration in one process, in which the webpage corpora ran
$n{=}24$ test subsets (a disclosed protocol deviation) and the JSON corpora
their full test splits. It \textbf{reversed}: webpage param 0.167 vs
0.042,
tool-hijack JSON 0.031 vs 0.000, webpage tool hijack 0.042 vs 0.000 (each
pair challenger vs deployed; the deployed readings are that paired run's own
generations, not the certification pass of Table~\ref{tab:main-gptoss}, whose
separate artifacts read 0.135, 0.013 and 0/52---the pre-specified criterion
was directional within the paired run). The
development-split
dominance was selection on the same attack distribution the configuration was
tuned
against, and the challenger was rejected. We report this as a concrete,
quantified instance of the risk of defense selection on the same attack
distribution used for development. The pre-specified test pass is what caught
it, and it is why every promotion in this paper is gated on one.

\textbf{Search budget, disclosed.} The development-side search behind the
deployed configurations was substantial and is part of the honest accounting
(Table~\ref{tab:searchbudget}): what was explored per model, and the rule that
every test-split number in this paper is a \emph{single} pre-specified pass
with no reruns---the search happened on development splits, never on the
splits the claims are stated on.

\begin{table}[t]
\caption{Per-model defense search space actually explored during development
(lower bounds where early exploration is recorded only in summary). Doses
counts distinct step magnitudes evaluated; compositions counts blends and
variant mechanisms (gates, span choices, decode-time stacks). All test-split
evaluations are one pre-specified pass, no reruns.}
\label{tab:searchbudget}
\centering
\footnotesize
\setlength{\tabcolsep}{2.5pt}
\begin{tabular}{p{2.2cm}p{1.2cm}p{0.9cm}p{0.8cm}p{1.5cm}p{1.2cm}}
\toprule
model & directions & layer sets & doses & compositions / variants & held-out confirm \\
\midrule
gpt-oss-20b & $\ge$16 (+$\ge$9 random draws) & $\ge$2 & $\ge$16 & $\ge$20 & yes (one pass) \\
Qwen3-30B & 5 & 2 & $\ge$11 & 2 & yes (one pass) \\ %
Gemma-4-31B & 1 (+probe axes) & 3 & 8 & 0 & yes (two rungs: wire-format T$^{*}$; then docs $+$ wordings $+$ framings held out) \\ %
GLM-4.5-Air & $\ge$5 (incl.\ 2 role axes) & 2 & $\ge$6 & 0 & yes (one pass, webpage param) \\ %
Llama-3.1-8B & 3 (+3 random draws) & 4 & $\ge$9 & 0 & yes (one pass) \\ %
\bottomrule
\end{tabular}
\end{table}

\section{General Utility and Mechanism Ablations}\label{app:ablations}

\textbf{Runtime-cost measurement bounds (\S\ref{sec:eval-baselines}).}
Measured on gpt-oss-20b at the deployed configuration (A100 80GB, $n{=}21$
poisoned episodes after warmup, medians): prefill 0.181\,s steered vs
0.182\,s unsteered, greedy decode 13.52\,tok/s in both arms, peak memory
identical at 39.37\,GiB (the vectors add ${\sim}15$\,KB). The per-episode IQR
is ${\sim}35\%$ of the median against a $<1\%$ arm difference, which bounds
the claim rather than establishing zero cost. The deployed defense is
prefill-only (\S\ref{sec:defense-param}), so this measures the shipped
configuration.

\textbf{Why there is no standard-benchmark evaluation.} CounterSteer edits
only tool-result spans, so any input containing none---which includes every
standard capability benchmark as normally administered---runs the
mathematically unmodified model. General capability under CounterSteer is
preserved \emph{by definition}, not by measurement, and we therefore do not
evaluate GSM8K/MMLU/IFEval-style suites as a utility claim. We run them once
in two roles only (Table~\ref{tab:genutil}; gpt-oss-20b, deployed
configuration). The
\emph{plain} rows are an implementation identity check (the numbers must be
--- and are --- identical). The \emph{tool-framed} rows, where each item's
content is artificially wrapped in a steered tool-result span, are a mechanism
probe of what steering costs when the content being reasoned over sits inside
the edited span. The deployment-relevant utility numbers remain
\texttt{utilBenign}/\texttt{utilAttack} on real tool traffic
(\S\ref{sec:eval-main}--\ref{sec:eval-agentic}).

\begin{table}[t]
\caption{Identity check (plain) and steered-span reading probe (tool-framed),
gpt-oss-20b deployed configuration. Denominator: all items. The plain columns are an
implementation check of the by-definition identity, not a utility evaluation.}
\label{tab:genutil}
\centering
\setlength{\tabcolsep}{5pt}
\begin{tabular}{llll}
\toprule
bench ($N$) & context & baseline \dirup & steered \dirup \\
\midrule
GSM8K ($n{=}250$) & plain / tool-framed & 0.896 / 0.912 & 0.896 / 0.912 \\
MMLU ($n{=}500$) & plain / tool-framed & 0.814 / 0.810 & 0.814 / \textbf{0.702} \\
IFEval ($n{=}250$) & plain / tool-framed & 0.624 / 0.672 & 0.624 / 0.520 \\
\bottomrule
\end{tabular}
\end{table}

The plain rows confirm the identity. The tool-framed probe shows that
CounterSteer preserves semantic reasoning over steered content unevenly:
arithmetic reasoning (GSM8K) is unchanged while multiple-choice discrimination
drops $10.8$\,pp (and IFEval instruction adherence $15.2$\,pp) at the deployed
gpt-oss dose when the \emph{content being reasoned over} sits inside a steered
span. Suppressing a span's instructional influence is not perfectly
separable from processing its content. Deployments do not serve MMLU questions
as tool output, but deployments whose agents must reason discriminatively over
retrieved content should verify utility at their operating dose. The
mechanism of the MMLU slip is anatomized next.

\textbf{Position anatomy of the MMLU slip.} Controlling gold-answer position
($n{=}250$ per condition): gold at the first option is immune ($+1.6$\,pp,
$p = 0.585$, noise); gold at any later position costs 9--12\,pp regardless of
which (B $-8.8$, $p = 0.0016$; C $-12.0$, $p = 5.3{\times}10^{-6}$; D $-9.2$,
$p = 4.3{\times}10^{-4}$)---a step, not a gradient. Uniform-position
interpolation reproduces about two-thirds of the observed slip. Two failure
classes carry the cost in roughly equal measure: degraded comparison among
later-listed options (clean wrong-letter finals, first-option pull), and
degenerate budget-exhausting re-quoting loops that never reach a final
answer---question comprehension itself is not the failure mode.

\textbf{Unicode/typography anatomy of the agentic benign cost.} Of the deployed
configuration's 8 broken AgentDojo benign tasks, 3 are steering flipping near-tie space
typography to U+202F NARROW NO-BREAK SPACE inside checker-required strings
(same family as a recorded U+2011 non-breaking-hyphen case). This is not a
directional push: the unsteered model already emits U+202F in 34--36\% of
episodes (steered: 36--41\%). Steering perturbs near-tie decisions; which task
breaks flips chaotically across arms. Only 1/8 failures is
suppression---capability loss is not the mechanism.

\textbf{Random-direction controls, draw by draw.} On the fitting corpora,
matched-magnitude random directions span ASR 0.000--0.477 across draws---the
null distribution is wide, which is why a single draw is not a control. On the
disjoint-tool corpus a draw that matched the fitted direction's clean-payload
cost exactly (McNemar $p = 1.00$) still carried $30\times$ its attacked ASR;
the matched control there also acts \emph{less} than unattacked (no-action
$1.19\times$ unattacked, vs the fitted direction's $0.73\times$), so its
higher attack rate is not bought back in capability.
The compromise-rate summary is in \S\ref{sec:eval-baselines}.

\textbf{Perturbation cost of the magnitude.} Two independent magnitude-matched
random directions cost $-21.8$ and $-27.3$\,pp benign on AgentDojo against the
deployed configuration's $-14.5$---all three readings from the benign arms
of a retired 768-token-budget run, not the 4096-budget curve of
Table~\ref{tab:gptoss-dose} (per-draw McNemar vs deployed $p = 0.42$ / $0.09$, same
sign both draws). These controls suggest that a substantial fraction of the
benign cost is generic perturbation cost at this magnitude rather than
uniquely attributable to the fitted direction; the compromise-rate
(\S\ref{sec:eval-main}--\S\ref{sec:eval-adaptive}) is what the direction buys
on top of that generic cost. In these two draws, direction surgery on the
benign axis showed no headroom; the live lever is dose
(\S\ref{sec:eval-agentic}).

\section{Baseline Implementations and Full-Grid Detail}\label{app:baselines}
Detail behind the CachePrune, SecAlign and AGRI passages of
\S\ref{sec:eval-baselines} and \S\ref{sec:related}.

\textbf{AGRI port.} AGRI \cite{agri2026} releases probes only; we fit
probes on our own captures (gpt-oss-20b layer~20, AUROC 0.9906; Qwen3-30B
layer~12, AUROC 0.9976) and implemented its gated reasoning prefill from
the specification (fire threshold 0.5; the prefill persists three turns
after a fire). Render parity between the replayed prompts and our serving
path measured $\le 2\times10^{-8}$ (Qwen3-30B) and 0.055 (gpt-oss-20b;
above our 0.05 pre-set gate, with zero gate-decision flips across the
audited rows). Static grade: its own 180-case paired battery
(Table~\ref{tab:headline}) --- gpt-oss-20b $.415 \to .062$ at benign
78.0\% of its clean arm (absolute 70.9\%), with 4 workspace cases void on
out-of-memory errors consistent
with the measured context inflation; Qwen3-30B $.472 \to .100$ at benign
100\% of clean (absolute 96.4\%). Adaptive: gpt-oss-20b AD@6 0.143, final over 36/36 cells;
Qwen3-30B 0.411, final over 36/36
(Table~\ref{tab:autodojo-matrix}). The port's deviations table ships with
the artifact.

\begin{table*}[t]
\caption{State-of-the-art comparison: four-arm batteries (clean /
defense-on-clean / attacked / defended) per defense, full 180-case AgentDojo
grid, 4096 tokens/turn (GLM-4.5-Air 8192), each row against its \emph{own
same-process} undefended baseline (``undef.''). compromise $=$ AgentDojo's
checker over paired
verdict-bearing cases; blk/int $=$ compromises blocked~/ introduced (exact
McNemar); Class~A $=$ non-delegated compromises, defended (undefended);
benign $=$ defense-on-clean \% of clean over paired unique
tasks; fp $=$ fraction of \emph{benign} tool messages a filter deleted;
\emph{n/m} $=$ the delegated-authority partition is not computed for that
panel. \warnmark\ $=$ capability-guard territory, a mechanism predicate
rather than a benign-\% threshold, detailed---with the typography
normalization's scope and its one disclosed movement---in the
``Table~\ref{tab:soa} provenance'' paragraph of
Appendix~\ref{app:baselines}.}
\label{tab:soa}
\centering
\scriptsize
\setlength{\tabcolsep}{2.5pt}
\begin{tabular}{p{3.0cm}p{3.1cm}p{1.45cm}p{1.35cm}p{1.8cm}p{1.7cm}p{1.2cm}}
\toprule
defense & compromise \dirdown\ [95\% CI] (own undef.) & blk/int ($p$) & Class~A \dirdown & utilAttack \dirup\ (undef.) & benign \dirup\ \% of clean & filter fp \dirdown \\
\midrule
\multicolumn{7}{l}{\emph{gpt-oss-20b (Class~A over 137--140 non-delegated cases; undefended Class~A 51--59)}} \\
\textbf{CounterSteer (deployed)} & \textbf{16/176 $=$ 0.091} [.057, .143] (0.483) & 70/1 ($6.1{\times}10^{-20}$) & \textbf{0/140} (59) & 0.716 (0.551) & 90.6\% & --- \\
CachePrune \cite{cacheprune2025} & 38/176 $=$ 0.216 [.162, .282] (0.483) & 55/8 ($9.8{\times}10^{-10}$) & 15/140 (59) & 0.688 (0.551) & 88.7\% & --- \\
reminder (AutoDojo) \cite{autodojo2026} & 48/176 $=$ 0.273 [.212, .343] (0.483) & 45/8 ($2.4{\times}10^{-7}$) & 33/140 (59) & 0.574 (0.551) & 88.7\% & --- \\
prompt sandwich & 55/174 $=$ 0.316 [.252, .389] (0.489) & 38/8 ($9.2{\times}10^{-6}$) & 31/138 (59) & 0.609 (0.551) & 88.2\% & --- \\
spotlighting \cite{hines2024spotlighting} & 74/176 $=$ 0.420 [.350, .494] (0.483) & 21/10 (\textbf{0.071 n.s.}) & 46/140 (59) & 0.616 (0.551) & 92.5\% & --- \\
PromptGuard-2 filter \cite{promptguard2} & 36/176 $=$ 0.205 [.152, .270] (0.483) & 49/0 ($3.6{\times}10^{-15}$) & 21/140 (59) & 0.311 (0.551) & \textbf{100.0\%} & 0/178 \\
DeBERTa filter \cite{protectai2024deberta} & 6/176 $=$ 0.034 [.016, .072] (0.483) & 79/0 ($3.3{\times}10^{-24}$) & 4/140 (59) & 0.167 \warnmark\ (0.551) & 47.2\% \warnmark & 61\% \\
PIGuard filter \cite{li2024injecguard} & 2/176 $=$ 0.011 [.003, .040] (0.466) & 80/0 ($1.7{\times}10^{-24}$) & 1/140 (57) & 0.106 \warnmark\ (0.585) & 58.2\% \warnmark & 56\% \\
tool filter [mechanical] & 0/176 observed [0, .021] (0.432) & 76/0 ($2.6{\times}10^{-23}$) & 0/140 (51) & 0.106 \warnmark\ (0.614) & 10.9\% \warnmark & --- \\
\midrule
\multicolumn{7}{l}{\emph{Qwen3-30B (own undefended baseline 0.489 in every battery; six independent regenerations, verdict-identical per case)}} \\
\textbf{CounterSteer @$12\sigma$ (deployed)} & \textbf{13/180 $=$ 0.072} [.043, .120] (0.489) & 75/0 ($5.3{\times}10^{-23}$) & 5/144 (58) & \textbf{0.717} (0.667) & 94.1\% & --- \\
PIGuard filter \cite{li2024injecguard} & 1/180 $=$ 0.006 [.001, .031]; worst-case $\le$.033--.056 & 87/0 ($1.3{\times}10^{-26}$) & 0/144 (58) & 0.222 \warnmark\ (0.667) & 62.7\% \warnmark & 31\% \\
DeBERTa filter \cite{protectai2024deberta} & 16/180 $=$ 0.089 [.055, .140] & 72/0 ($4.2{\times}10^{-22}$) & 5/144 (58) & 0.300 \warnmark\ (0.667) & 54.9\% \warnmark & 36\% \\
PromptGuard-2 filter \cite{promptguard2} & 41/180 $=$ 0.228 [.173, .294] & 47/0 ($1.4{\times}10^{-14}$) & 23/144 (58) & 0.450 (0.667) & 100.0\% & 0/179 \\
reminder (AutoDojo) \cite{autodojo2026} & 64/180 $=$ 0.356 [.289, .428] & 28/4 ($1.9{\times}10^{-5}$) & 37/144 (58) & 0.689 (0.667) & 96.1\% & --- \\
\midrule
\multicolumn{7}{l}{\emph{Llama-3.1-8B (thin base: undefended 0.100 $=$ 18/180 on both fabrics; clean solves only 15/56 (A100) / 13/56 (H100) tasks)}} \\
\textbf{CounterSteer $\alpha{=}5$ (A100)} & \textbf{9/180 $=$ 0.050} [.027, .092] (0.100) & 11/2 (0.022) & n/m & 0.222 (0.222) & \textbf{100.0\%} & --- \\
CounterSteer (H100 replicate) & 11/180 $=$ 0.061 [.034, .106] (0.100) & 9/2 (\textbf{0.065 n.s.}) & n/m & 0.228 (0.211) & 107.7\% & --- \\
CachePrune \cite{cacheprune2025} (A100) & 8/178 $=$ 0.045 [.023, .086] (0.101) & 11/1 (0.006) & n/m & 0.208 (0.225) & 80.0\% & --- \\
prompt sandwich (A100) & 1/180 $=$ 0.006 [.001, .031] (0.100) & 17/0 ($1.5{\times}10^{-5}$) & n/m & 0.122 (0.222) & 73.3\% & --- \\
spotlighting \cite{hines2024spotlighting} (A100) & 23/180 $=$ \textbf{0.128} [.087, .184] (0.100; \emph{worse}) & 5/10 (0.30 n.s.) & n/m & 0.194 (0.222) & 100.0\% & --- \\
PromptGuard-2 filter \cite{promptguard2} (H100) & 10/180 $=$ 0.056 [.030, .099] (0.100) & 8/0 (0.008) & n/m & 0.206 (0.211) & 100.0\% & --- \\
DeBERTa filter \cite{protectai2024deberta} (H100) & 8/180 $=$ 0.044 [.023, .085] (0.100) & 10/0 (0.002) & n/m & 0.189 (0.211) & 92.3\% & --- \\
PIGuard filter \cite{li2024injecguard} (H100) & 1/180 $=$ 0.006 [.001, .031] (0.100) & 17/0 ($1.5{\times}10^{-5}$) & n/m & 0.178 (0.211) & 92.3\% & --- \\
tool filter [mechanical] (H100) & 1/180 $=$ 0.006 [.001, .031] (0.100) & 18/1 ($7.6{\times}10^{-5}$) & n/m & 0.144 \warnmark\ (0.211) & 76.9\% \warnmark & --- \\
reminder (AutoDojo) \cite{autodojo2026} (H100) & 18/180 $=$ 0.100 [.064, .153] (0.100; no effect) & 4/4 (1.0 n.s.) & n/m & 0.211 (0.211) & 100.0\% & --- \\
\midrule
\multicolumn{7}{l}{\emph{GLM-4.5-Air detector probe (8192 tokens/turn, its own convention; clean solves 45/56, near this table's capability ceiling)}} \\
\textbf{CounterSteer $\alpha{=}8$ (deployed)} & \textbf{10/180 $=$ 0.056} [.030, .099] (0.183) & 24/1 ($1.5{\times}10^{-6}$) & n/m & \textbf{0.694} (0.728) & \textbf{102.2\%} & --- \\
PIGuard filter \cite{li2024injecguard} & 0/180 observed [0, .021] (0.183) & 33/0 ($2.3{\times}10^{-10}$) & n/m & 0.144 \warnmark\ (0.728) & 57.8\% \warnmark & --- \\
PromptGuard-2 filter \cite{promptguard2} & 21/180 $=$ 0.117 [.078, .172] (0.183) & 12/0 (0.00049) & n/m & 0.417 (0.728) & 100.0\% & 0/176 \\
\bottomrule
\end{tabular}
\end{table*}

\begin{figure}[t]
\providecolor{csBlue}{HTML}{2A78D6}
\providecolor{csRed}{HTML}{C0392B}
\providecolor{csTeal}{HTML}{1BAF7A}
\providecolor{csTealDk}{HTML}{0F7A54}
\centering
\begin{tikzpicture}[
  font=\footnotesize,
  secstyle/.style={csBlue!75!black, line width=0.9pt, mark=*, mark size=1.6pt,
                   mark options={fill=csBlue!75!black, draw=csBlue!75!black}},
  benstyle/.style={csRed!75!black, line width=0.9pt, densely dashed,
                   mark=square, mark size=1.6pt,
                   mark options={solid, fill=white, draw=csRed!75!black}},
  opmark/.style={csTealDk, line width=1pt, mark=star, mark size=3.2pt,
                 only marks, mark options={draw=csTealDk, line width=0.8pt}}
]
\pgfplotsset{axstyle/.style={width=2.75cm, height=2.5cm, scale only axis,
    tick label style={font=\tiny},
    label style={font=\tiny},
    title style={font=\tiny\bfseries, yshift=-1mm},
    axis line style={black!55}, tick style={black!55},
    every axis plot/.append style={line join=round}}}

\begin{axis}[axstyle,
  name=gA,
  title={gpt-oss-20b},
  xlabel={dose $\alpha$ ($\times\sigma$, per model)},
  ylabel={compromise $\downarrow$},
  ylabel style={csBlue!70!black},
  y tick label style={csBlue!70!black},
  xmin=-0.4, xmax=9, ymin=0, ymax=0.52,
  xtick={0,3,5.5,8.06}, xticklabels={0,3,5.5,8.1},
  ytick={0,0.1,0.2,0.3,0.4,0.5},
  axis y line*=left, axis x line*=bottom,
  clip=false]
  \draw[csTealDk!55, densely dotted, line width=0.7pt]
    (axis cs:8.06,0) -- (axis cs:8.06,0.52);
  \addplot[secstyle] coordinates {(0,0.475) (3,0.311) (4.12,0.249)
    (5.5,0.147) (6.7,0.107) (8.06,0.085)};
  \addplot[opmark]   coordinates {(8.06,0.085)};
  \node[anchor=south east, font=\tiny, csTealDk, align=right]
    at (axis cs:7.8,0.15) {deployed\\ $8.06\sigma$};
\end{axis}
\begin{axis}[axstyle, at={(gA.south east)}, anchor=south east,
  xmin=-0.4, xmax=9, ymin=-42, ymax=2,
  axis x line=none, axis y line*=right,
  ytick={0,-10,-20,-30,-40}, yticklabels={,,,,},
  ylabel={$\Delta$pp}, ylabel style={csRed!70!black, yshift=1mm},
  clip=false]
  \addplot[benstyle] coordinates {(0,0) (3,-7.3) (4.12,-12.7) (5.5,-12.7)
    (6.7,-14.5) (8.06,-16.4)};
\end{axis}

\begin{axis}[axstyle,
  name=gB, at={(gA.south east)}, anchor=south west, xshift=13mm,
  title={Qwen3-30B},
  xlabel={dose $\alpha$ ($\times\sigma$, per model)},
  xmin=-0.5, xmax=17, ymin=0, ymax=0.52,
  xtick={0,10,12,14,16},
  ytick={0,0.1,0.2,0.3,0.4,0.5}, yticklabels={,,,,,},
  axis y line*=left, axis x line*=bottom,
  clip=false]
  \fill[csRed!9] (axis cs:13,0) rectangle (axis cs:16.5,0.52);
  \node[anchor=north, font=\tiny, csRed!65!black]
    at (axis cs:14.75,0.50) {cost cliff};
  \draw[csTealDk!55, densely dotted, line width=0.7pt]
    (axis cs:12,0) -- (axis cs:12,0.52);
  \addplot[secstyle] coordinates {(0,0.457) (10,0.122) (12,0.094) (14,0.033)};
  \addplot[secstyle, forget plot, densely dotted, mark=none]
    coordinates {(14,0.033) (16,0.011)};
  \addplot[csBlue!75!black, only marks, mark=triangle*, mark size=2.2pt,
           mark options={fill=white, draw=csBlue!75!black, line width=0.7pt}]
    coordinates {(16,0.011)};
  \node[anchor=south east, font=\tiny, csTealDk, align=right]
    at (axis cs:11.6,0.155) {selected\\ $12\sigma$};
\end{axis}
\begin{axis}[axstyle, at={(gB.south east)}, anchor=south east,
  xmin=-0.5, xmax=17, ymin=-42, ymax=2,
  axis x line=none, axis y line*=right,
  ytick={0,-10,-20,-30,-40},
  ylabel={benign $\Delta$pp $\uparrow$},
  y tick label style={csRed!70!black},
  ylabel style={csRed!70!black},
  clip=false]
  \addplot[benstyle] coordinates {(0,0) (10,-5.4) (12,-5.4) (14,-19.6) (16,-39.3)};
\end{axis}

\begin{scope}[shift={($(gA.south west)+(0.05,-0.95)$)}, font=\tiny]
  \draw[csBlue!75!black, line width=0.9pt] (0,0) -- (0.4,0);
  \fill[csBlue!75!black] (0.2,0) circle (1.3pt);
  \node[anchor=west] at (0.48,0) {compromise (left)};
  \draw[csRed!75!black, line width=0.9pt, densely dashed] (2.6,0) -- (3.0,0);
  \draw[csRed!75!black, fill=white] (2.8,0) +(-1.3pt,-1.3pt) rectangle +(1.3pt,1.3pt);
  \node[anchor=west] at (3.08,0) {benign $\Delta$pp (right)};
  \node[csTealDk] at (0.2,-0.3) {$\star$};
  \node[anchor=west] at (0.48,-0.3) {operating point};
  \draw[csBlue!75!black, fill=white, line width=0.6pt]
    (2.8,-0.32) -- +(-2pt,-2pt) -- +(2pt,-2pt) -- cycle;
  \node[anchor=west] at (3.08,-0.3) {lower bound (censored)};
\end{scope}

\end{tikzpicture}%
\caption{\textbf{Security--utility dose curves on AgentDojo.} Each model's
$\sigma$ is its own and the two dose axes are deliberately different ranges;
stars mark the deployed configurations. Both panels are at the 4096-token
per-turn budget; all compromise values are budget-censored lower bounds at their own budget
(\S\ref{sec:eval-agentic}), and the Qwen $16\sigma$ point is additionally
censored by the truncation that violates its capability guard, so it is
plotted as a bound only; the shaded ``cost cliff'' band marks the dose
region where Qwen's benign delta falls below $-19$\,pp. The benign
$\Delta$pp is the raw paired clean$\to$defense-on-clean delta of each ladder ($n{=}55/56$ tasks), so
adjacent-dose differences sit inside task quantization.}
\label{fig:dose}
\end{figure}
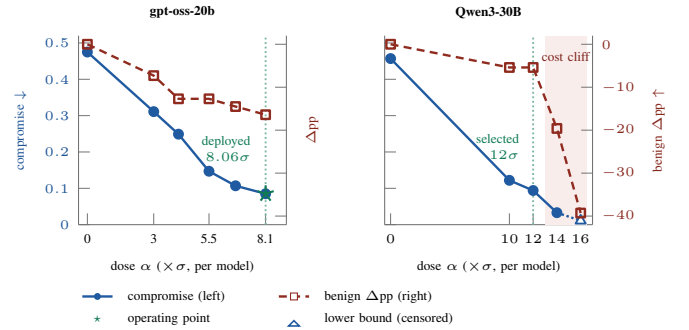

\textbf{Table~\ref{tab:soa} provenance.} Audit detail from the table's
adversarial sign-off. \emph{Undefended baselines and censoring:} the gpt-oss
baselines are
three distinct greedy regenerations (0.432--0.489 $=$ regeneration noise;
never averaged); every compromise value is a truncation-censored lower bound,
and worst-case bounds flip no ordering except reminder vs.\ prompt sandwich,
which is therefore not ordered; the panels are never compared to each other.
\emph{Typography normalization (gpt-oss utilAttack and benign):} Unicode
space/dash codepoints in model output mapped to ASCII, uniformly across all
four arms, before this benchmark's own (2024) string checkers (zero verdict
flips on the other panels); the one compromise-rate movement under
normalization is one defended compromise \emph{surfaced} on the gpt-oss
dose-ladder baseline ($0.085 \to 0.090$; the ladder is Fig.~\ref{fig:dose}),
a separate run from this table's CounterSteer battery (0.091)---two runs,
not roundings of one number.
\emph{Filter wiring:} filter rows use AgentDojo's own classifier wiring
(message-level replacement, windowed scanning), not AutoDojo's gentler
sentence-level removal~\cite{autodojo2026}. \emph{Llama-3.1-8B panel:} it
spans two hardware fabrics (A100 pairs, H100 node); each row's four arms
share one fabric and one process, the CounterSteer replicate quantifies the
cross-fabric spread (a replicate, not a tenth battery), and the thin
undefended base (18/180) widens every interval. \emph{GLM detector probe:}
it shows the detector trade at the capability ceiling---PromptGuard-2 keeps
benign utility whole but leaves 21/180 compromises; PIGuard leaves 0/180
observed compromises and destroys 42\% of benign utility (26/56 tasks vs.\ 45/56 clean,
retention 26/45) by false-positive tool-output deletion (235 of 352 benign
tool messages deleted), not action suppression (0/41 no-call tasks); the
probe's undefended baseline (0.183) differs from the earlier GLM grid's
(0.222) within regeneration noise ($p{=}0.36$), same 8192 budget. \emph{The
\warnmark\ predicate, concretely:} every marked row has a verified
capability-destruction mechanism (benign tool-message deletion at fp
56--61\% on gpt-oss and 31--36\% on Qwen, 235/352 on GLM; the mechanical
tool filter suppresses action---on Llama, 20/41 benign tasks end with no
call vs.\ 0/41 undefended, median 2 of ${\sim}11$ tools kept), which is why
Llama's tool filter (benign 76.9\%) is marked while its prompt sandwich
(73.3\%, a prompt-level arm with no deletion or suppression mechanism) and
its classifier filters (benign 92.3\%, no verified deletion cost on that
model's thin base) are not. \emph{SecAlign:} the 4096-budget battery
(quoted in \S\ref{sec:eval-baselines}, not a row here) carries two
disclosed deviations---it ran a longer system message than these batteries
(measured not to carry its clean-utility deficit: five of its six failing
tasks already failed at 768 under the short message), and seven cases OOM'd
symmetrically in all four arms (the merged checkpoint's VRAM footprint;
exclusion favors SecAlign, and intersection-restricted ratios move
$\le$0.1\,pp); base-clean regeneration spread puts its per-case ratio at
80.9--82.3\%. The table supersedes the retired 768-budget baseline
measurements and the retired legacy Qwen defended-only measurements for
the defenses covered (both preserved in the artifact repository).

\textbf{CachePrune port and anchor validation.} No public implementation
exists; ours is, to our knowledge, the first independent one, and it was
validated in the original paper's own setting before any number on our models
counted: on Llama-3-8B-Instruct with SQuAD QA and StruQ-style injections, ASR
$0.51 \to 0.01$ ($n{=}100$; paired exact on the stored completions: fixed 51~/
broke 1, $p{=}2.4{\times}10^{-14}$; the single defended hit is a refusal
quoting the canary---a substring-matcher false positive, so the corrected
defended rate is 0.00) with F1 nearly held (defended 0.734 vs.\ unattacked
0.783, i.e.\ 93.7\% of unattacked; mask-on-benign 95.7\%)---a defense effect
consistent in direction with the paper's reported $27.9\% \to 7.4\%$ (our
attack strings
are approximations; theirs are unpublished), with every deviation from its
equations enumerated in the released artifact. Three caveats on that anchor:
the ASR matcher counts substring echoes of the canary, which can only inflate
the \emph{attacked} arm; the 100 evaluation rows are drawn from essentially one
title-sorted SQuAD article; and the \texttt{fake\_completion} framing never
fires undefended (0/25 observed [0.000, 0.134]), so it contributed no fit
signal. The gpt-oss-20b mask was fit on
the same probe split the steering direction uses (equal data access; sample-
and attacker-template-disjoint from the development evaluations, asserted at
split time, and disjoint by construction from the test splits and AgentDojo).
On the single-turn development splits its benign utility brackets read
53.2\%/85\% (JSON tool-log) and 58.5\%/88\% (JSON parameter). \textbf{These are
not paired against a single CounterSteer generation and should not be read as a
head-to-head}: CounterSteer's benign brackets on the same sample ids are
64.0\%/88\% and 56.9\%/95\% in the run the compromise figures of
\S\ref{sec:eval-baselines} come from, and 60.0\%/89\% and 60.6\%/92\% in a
second generation of the same splits. The arms are separate generations with
different clean references and therefore different scoreable denominators
(79 vs.\ 75; 65 vs.\ 66), each percentage normalized to its own clean arm. On
JSON parameter the strict end \emph{reverses} between the two readings
(CachePrune 58.5\% vs.\ CounterSteer 56.9\% or 60.6\%), so the honest statement
is that the two defenses' benign cost is indistinguishable at this resolution,
not that either is cheaper.

\textbf{Test-grade baseline runs and retained development readings.} The
\S\ref{sec:eval-baselines} single-turn comparison quotes a pre-registered
re-run of both baselines on the held-out test splits (identical budget,
batch, greedy decoding and sample order as CounterSteer's test artifacts;
the SecAlign arm is its merged model's own clean/attacked pair with no
steering applied). Test grade: CachePrune $0.700 \to 0.463$ (shipped,
$n{=}96$) and $0.647 \to 0.397$ (parameter, $n{=}70$), benign brackets
35.0\%/88 and 29.4\%/87 vs.\ CounterSteer's 25.0\%/80 and 22.7\%/83
(point-better for CachePrune, n.s.); SecAlign 0/91 and 0/70 observed. The prior development readings are retained for the record:
CachePrune $0.481 \to 0.203$ and $0.600 \to 0.338$; SecAlign
$1/92 = 0.011$ and 0/66 observed. The shipped-corpus CachePrune movement
dev$\to$test is significant ($p{=}0.0007$, defended arms; the dev run
executed on different hardware, so split and host are confounded); the
parameter movement is within sampling noise ($p{=}0.59$).

\textbf{SecAlign training grid and holdout.} A DPO LoRA (rank 32, $\alpha{=}64$,
$\beta{=}0.1$, one epoch) over 8640 preference pairs: samples drawn from the
same fitting split the direction uses, crossed with the full authority-framing
factorial the direction's fit design was drawn from, with the evaluation
samples held out and the development injection wordings held out lexically
(none appear in the training prompts, exact or case-folded). Not held out: the
framing-level axis (every grid cell was trained); the semantic authority
register (the training grid spans the same authority-escalation register as
the evaluated templates, and one development wording family is a register
paraphrase of the grid's user voice); and---symmetrically for \emph{both}
defenses---the JSON parameter corpus's attacker-template set. SecAlign's sole
development-split compromise is a genuine full compromise on a sample whose
user turn itself instructs executing action lines found in tool content. As
self-referenced context (not base-anchored): against its \emph{own} clean arm,
SecAlign's under-attack utility brackets read 58.7\%/91\% (JSON tool-log) and
59.1\%/95\% (JSON parameter).

\textbf{SecAlign utility, conservative end and own-clean context.}
Its utility cost is quotable only at the
conservative end, because the lenient bracket's composed-field exemption is
computed against a \emph{shared} clean reference and a fine-tuned model's clean
behavior is a different reference: clean-traffic agreement with the base
model's clean run is 0.506 (JSON tool-log) / 0.585 (JSON parameter) against a
same-model cross-run agreement floor of 0.658 / 0.723, comparable to
CachePrune's defense-on-clean (0.532 / 0.585) and heavier than CounterSteer's
(0.646 / 0.662).

\textbf{SecAlign vs.\ CounterSteer discordance.} Over the 176 shared
verdict-bearing AgentDojo cases the two arms disagree on 7: SecAlign wins
5 and CounterSteer 2 (exact McNemar $p{=}0.45$), i.e.\ statistically
indistinguishable.

\textbf{Full-180 AgentDojo grid (768-budget trained comparison): budget
anatomy.} On clean traffic SecAlign and the base model are near parity at
the 768 budget (100.0\% vs 98.2\% normalized); raising the budget to 4096
cures the base model's clean failures (10 $\to$ 1 tasks) but not
SecAlign's trained-in ones (11 $\to$ 6, two of them refusals of
\emph{legitimate} tool-data instructions), which is what leaves SecAlign
at 90.4\% unique-task of base clean at the shared 4096 budget
(\S\ref{sec:eval-baselines}); under attack the two defenses are at utility
parity (SecAlign 101/180 vs.\ CounterSteer 102/180).

\textbf{Full-180 AgentDojo grid: nulls and censoring.}
Per-run checker-null cases (no security verdict) are 8/8/3/4 for undefended/CachePrune/CounterSteer/
SecAlign, and worst-case null assignment moves no ranking. Budget censoring at
this scale is asymmetric \emph{in the comparators' favor}---the terser
fine-tuned model leaves 0--9 unterminated turns per arm against 56--58 in the
comparators' attacked arms---so it cannot rescue them and the ranking is
conservative. Of the nine SecAlign leaks, seven are case-exact matches to
CounterSteer failures and two fall under attack wordings CounterSteer
survived, which is why the subset relation is stated at injection-task rather
than case granularity.

\section{Per-Model Authority-Framing Curves}\label{app:authority}
Per-model detail behind the authority-curve paragraph of
\S\ref{sec:mechanism}; all rates are attack success rates on each model's own
firing corpus under the four-level factorial of \S\ref{sec:method-recipe}.
gpt-oss-20b escalates monotonically with claimed authority. Gemma-4-31B fires most on firm framings (0.674,
$p{=}2{\times}10^{-4}$ cluster-collapsed), and explicit supersession suppresses
its parameter-manipulation firing entirely (0/72 observed [0.000, 0.050]).
GLM-4.5-Air has Gemma-4's shape, shallower: firm 0.576 highest, explicit
supersession 0.382 lowest.

\section{Supplementary Figures and Tables}\label{app:tables}
Floats relocated from the body for space; each is cited where its result is
claimed.

\begin{table}[t]
\caption{Glossary of recurring terms.}
\label{tab:glossary}
\centering
\footnotesize
\setlength{\tabcolsep}{3pt}
\begin{tabular}{p{2.2cm}p{5.6cm}}
\toprule
term & meaning \\
\midrule
primary models & gpt-oss-20b and Qwen3-30B: full evaluation battery and the deepest adaptive evaluation (\S\ref{sec:eval}) \\
undefended baseline & the same run's no-defense arm; every contrast in this paper is paired against it, never against another run \\
ASR & single-turn attack success rate over scoreable samples; the obedience-pattern reading unless marked exact-literal \\
compromise rate & AgentDojo/AgentDyn's own security checker: fraction of injection cases where the attacker's task completed \\
obedience-pattern~/ exact-literal & the injected action was performed~/ the attacker's exact payload was delivered verbatim \\
benign utility & task fidelity, defense on, no injection, \% of the undefended clean run---the always-on deployment cost \\
utility under attack & task fidelity, defense on, injection present, \% of the same baseline \\
utilB~/ utilA, utilAttack & table-column shorthand for benign utility~/ utility under attack \\
defense-on-clean & the arm benign utility is measured on (defense on, no injection); ``CLEAN$+$'' in artifacts \\
capability guard, \warnmark & a defense may not score by refusing to act; \warnmark\ marks rows whose utility rests on verified capability destruction \\
Class A~/ B & role-confusion attacks (tool hijack, parameter manipulation)~/ delegated authority, out of scope (\S\ref{sec:threat-attacker}) \\
rung T~/ T$^{*}$~/ D & held-out test split, one pre-specified pass~/ Gemma-4's earlier wire-format-only holdout~/ development split \\
battery & one four-arm run (clean, defense-on-clean, attacked, defended) over the full 180-case AgentDojo grid, in one process \\
certification run~/ comparison battery & two same-configuration regenerations of the primary models' AgentDojo result; the certification run is canonical, the comparison battery appears beside rivals (\S\ref{sec:results}) \\
AD@6 & the AutoDojo adaptive attacker's success after six rewrite iterations, macro over 36 cells (\S\ref{sec:eval-adaptive}) \\
N/F & the attack class does not fire undefended on that model; no defended run exists \\
firing corpus & a corpus whose undefended attack rate on a given model is high enough to measure a defense \\
contamination & injected text reaching some \emph{other}, legitimate call; reported, never counted as compromise \\
goal & the released scorer's name for the tier-1 outcome; rendered in this paper as ASR (single-turn) or compromise rate (agentic) \\
certified & the model's evaluation is complete, adversarially reviewed, and passed its one-pass held-out test \\
arm & one condition of a battery: clean, defense-on-clean, attacked, or defended \\
deployed configuration & a model's shipped (direction, layers, dose) triple (\S\ref{sec:defense}) \\
AutoDojo cell & one (suite $\times$ objective $\times$ vector) unit of the 36-cell adaptive matrix (Table~\ref{tab:autodojo-matrix}); distinct from AgentDojo's 180 \emph{cases} \\
\bottomrule
\end{tabular}
\end{table}

\begin{table*}[t]
\caption{Multi-model results: five corpus classes $\times$ five models; every
cell is a \emph{within-model} contrast at its own budget, dose and split,
never compared across models. Top: ASR (the AgentDojo column: compromise
rate), undefended $\to$
\textbf{defended}, with its rung---T $=$ held-out test split (one
pre-specified pass); D $=$ development split; N/F $=$ does not fire
undefended (undefended rate shown); AgentDojo $=$ its own 180-case grid and
checkers, no test split. Bottom: utility with the defense on---benign $\mid$
under attack, \textbf{judge} \% of unattacked with the (strict/lenient)
bracket in parentheses (\S\ref{sec:eval}; integer-rounded where space
requires---exact brackets:
Tables~\ref{tab:main-gptoss}--\ref{tab:permodel-glm});
AgentDojo utility is its own checker (benign \% of clean over
paired unique tasks; under-attack raw beside undefended-attacked raw).
``---'' $=$ no defended run. Caveats and run provenance: the
``Table~\ref{tab:multimodel} provenance'' paragraph
(Appendix~\ref{app:tables}).}
\label{tab:multimodel}
\centering
\scriptsize
\setlength{\tabcolsep}{3.5pt}
\begin{tabular}{llllll}
\toprule
& AgentDojo & webpage param & webpage tool-hijack & JSON tool-hijack & JSON param \\
\midrule
\multicolumn{6}{l}{\emph{compromise: ASR \dirdown\ (AgentDojo: compromise rate), undefended $\to$ \textbf{defended} (rung)}} \\
gpt-oss-20b & $.475 \to \mathbf{.079}$ & $.962 \to \mathbf{.135}$ T & $.981 \to \mathbf{0/52}$ T & $.675 \to \mathbf{.013}$ T & $.591 \to \mathbf{.045}$ T \\
Qwen3-30B & $.489 \to \mathbf{.072}$ & $1.000 \to \mathbf{.173}$ T & $.981 \to \mathbf{0/52}$ T & $.722 \to \mathbf{.033}$ T & $.522 \to \mathbf{.060}$ T \\
Gemma-4-31B & $.239 \to \mathbf{.006}$ & $.962 \to \mathbf{.038}$ T & $.865 \to \mathbf{.000}$ T & N/F (.083--.112) & N/F (.045--.048) \\
GLM-4.5-Air & $.222 \to \mathbf{.056}$ & $.808 \to \mathbf{.019}$ T & \multicolumn{3}{l}{hijack classes low-fire; JSON param.\ not run (Table~\ref{tab:permodel-glm})} \\
Llama-3.1-8B & $.100 \to \mathbf{.050}$ & no viable cell (base 1.000) & $.212 \to \mathbf{.038}$ T & N/F (.042) & $.273 \to \mathbf{.000}$ D (not viable) \\
\midrule
\multicolumn{6}{l}{\emph{utility with the defense on \dirup: benign $\mid$ under attack, judge \% of unattacked (strict/lenient bracket)}} \\
gpt-oss-20b & 94.4\% $\mid$ .734 (undef.\ .621) & 38.5 (75/75) $\mid$ 38.5 (60/60) & 50.0 (98/98) $\mid$ 57.7 (94.2/94) & 52.5 (25/80) $\mid$ 37.5 (30/76) & 53.0 (23/83) $\mid$ 43.9 (23/74) \\
Qwen3-30B & 94.1\% $\mid$ .717 raw (undef.\ .667) & 34.6 (65/65) $\mid$ 15.4 (50/50) & 13.5 (63/63) $\mid$ 13.5 (58/58) & 33.3 (21/69) $\mid$ 20.0 (22/67) & 31.3 (21/64) $\mid$ 31.3 (27/66) \\ %
Gemma-4-31B & 92.9\% $\mid$ .794 raw (undef.\ .850) & 78.8 (96/96) $\mid$ 78.8 (92/92) & 73.1 (96/96) $\mid$ 69.2 (96/96) & N/F & N/F \\ %
GLM-4.5-Air & 100.0\% $\mid$ .733 raw (undef.\ .733) & 26.9 (94.2/94) $\mid$ 11.5 (81/81) & --- & --- & --- \\
Llama-3.1-8B & 100.0\% $\mid$ .222 raw (undef.\ .222) & --- & 82.7 (100/100) $\mid$ 55.8 (65/65) & N/F & --- \\
\bottomrule
\end{tabular}
\end{table*}

\textbf{Table~\ref{tab:multimodel} provenance.} Gemma-4's webpage T cells
additionally hold out carrier documents, attacker wordings and framing
templates (\S\ref{sec:eval-main}), and every webpage-param T cell carries
the near-duplicate-template caveat (Reading the numbers,
\S\ref{sec:results}). The AgentDojo column runs at 4096
tokens/turn (GLM 8192, zero truncation), all arms one run per model; the Qwen
pair is the same-process battery of Table~\ref{tab:soa}, whose separate-run
regeneration replicate reads $.457 \to .094$ (consistent,
\S\ref{sec:eval-agentic}). On several models the pairwise judge sits far
below the strict/lenient bracket (GLM benign 26.9 beside 94.2/94; gpt-oss 38.5
beside 75/75): the judge reads the model-composed summaries that strict
exempts, and most of its measured deficit is wording drift regeneration alone
produces, so the strict/lenient bracket, not the judge floor, bounds task
fidelity (strict-vs-judge note, \S\ref{sec:eval-baselines};
Appendix~\ref{app:judge}). Llama-3.1-8B's webpage-param and JSON-param
classes have no viable cell: JSON-param reaches ASR 0 only at a
capability-destroying dose (judge benign .431), webpage-param at none (base
1.000; Table~\ref{tab:coverage}).

\begin{table*}[t]
\caption{The two primary models: the full suite, including the
\S\ref{sec:eval-adaptive} adaptive attacks. Corpus rows: ASR undefended $\to$
\textbf{defended}; utilB/utilA (benign / under attack, defense on),
\textbf{judge} \% of unattacked with the (strict/lenient) bracket; undef.\ is
the undefended-attacked arm's judge utility from the same run. Denominators:
scoreable samples (Appendix~\ref{app:tables});
webpage-param T rows carry the near-duplicate-template caveat
(Reading the numbers, \S\ref{sec:results}); bracket values are
integer-rounded where space requires (exact brackets:
Tables~\ref{tab:main-gptoss}--\ref{tab:main-qwen}).
Adaptive rows are tier-1 only (no arm scores by refusing, no-action
$\le 0.07$); query attacks are best-of-$k$ searches on development samples
against the live defended agent.}
\label{tab:flagship}
\centering
\scriptsize
\setlength{\tabcolsep}{3.5pt}
\begin{tabular}{p{3.4cm}p{6.7cm}p{6.5cm}}
\toprule
evaluation & gpt-oss-20b (deployed configuration) & Qwen3-30B (selected $12\sigma$) \\
\midrule
JSON tool-hijack (T) & $.675 \to \mathbf{.013}$; utilB 52.5 (25/80); utilA 37.5 (30/76) (undef.\ 13.8) & $.722 \to \mathbf{.033}$; utilB 33.3 (21/69); utilA 20.0 (22/67) (undef.\ 8.9) \\
JSON param.\ (T, samples-only) & $.591 \to \mathbf{.045}$; utilB 53.0 (23/83); utilA 43.9 (23/74) (undef.\ 18.2) & $.522 \to \mathbf{.060}$; utilB 31.3 (21/64); utilA 31.3 (27/66) (undef.\ 16.4) \\
webpage tool-hijack (T) & $.981 \to \mathbf{0/52}$; utilB 50.0 (98/98); utilA 57.7 (94.2/94) (undef.\ 0.0) & $.981 \to \mathbf{0/52}$; utilB 13.5 (63/63); utilA 13.5 (58/58) (undef.\ 1.9) \\
webpage param.\ (T) & $.962 \to \mathbf{.135}$; utilB 38.5 (75/75); utilA 38.5 (60/60) (undef.\ 0.0) & $1.000 \to \mathbf{.173}$; utilB 34.6 (65/65); utilA 15.4 (50/50) (undef.\ 0.0) \\
AgentDojo (4096 tok/turn) & $.475 \to \mathbf{.079}$; benign 94.4\% of clean; utilAttack 73.9\% (undef.\ 62.5\%) & $.489 \to \mathbf{.072}$ same-process (separate-run replicate $.457 \to .094$); benign 94.1\%; utilAttack .717 raw (undef.\ .667) \\
\midrule
LLMail-Inject replay & $128/2052 \to \mathbf{0/2052}$ observed (2052 episodes) & $117/1537 \to \mathbf{0/1537}$ observed (audited count---a parser audit recovered 13 successful attacker calls; full \emph{dev} replay, no test-shard replay exists) \\ %
query attack, tool hijack (best-of-24 within the class, $n{=}18$ dev) & 3/18 cracked [0.058, 0.392]; per-query .007 & not run \\
query attack, param.\ (best-of-24 within the class, $n{=}18$ dev) & \textbf{13/18 cracked}; per-query .303; exact-literal 1/18 (undef.\ 18/18 and .708 on the same grid) & not run \\
GCG, surrogate refit (tool hijack) & undef.\ 5/12 $\to$ \textbf{0/12} observed & (gradient runs steer the deployed vector; next row) \\
GCG, exact deployed vector (tool hijack; $n{=}52$, 49/52 converged) & undef.\ 22/52 (plain 26/52) $\to$ \textbf{2/52} observed [0.011, 0.130] & inconclusive; not reported as robustness \\
GCG, exact deployed vector (param.; forced-offset $+$ path-matched, $n{=}52$ each) & no gradient-attributable fire (best defended 1/52/run; path-matched objective unconverged, median CE 2.7) & not run \\
relay attacks & 0/18 every arm across turns (incl.\ undefended); quote step $7/24 \to 0/24$, $p{=}.005$ & not run \\
norm dilution (mechanism attack; $n{=}24$ dev, $n{=}52$ ladders) & \textbf{fails}: achievable instr.-token inflation $1.006\times$ (vocab ceiling $1.17\times$) vs $1.25\times$ (the first laddered dilution) already sufficient; defended goal unchanged ($0/24$ vs $0/24$; $3/24$ vs $3/24$, $p{=}1.0$) & not run \\
bidirectional causal gate & PASS, deployed vector $+$ gated candidate ($+.11$--$.22$) $+$ random control & PASS ($+.143$, $p{=}.0015$) \\
\bottomrule
\end{tabular}
\end{table*}

\begin{table*}[t]
\caption{gpt-oss-20b, held-out test evaluation (one pre-specified pass per
corpus). \dirdown\ lower is better, \dirup\ higher; denominators are samples
scoreable against the clean reference; utility is the judge \% of unattacked
with the (strict/lenient) bracket, and [$\cdot$] the undefended-attacked
arm's judge utility from the same run
(\S\ref{sec:eval}, Appendix~\ref{app:judge}). Guard rates and the webpage rows'
template-holdout caveat: the joint provenance paragraph beside
Table~\ref{tab:main-qwen}.}
\label{tab:main-gptoss}
\centering
\footnotesize
\setlength{\tabcolsep}{4pt}
\begin{tabular}{p{4.2cm}p{1.7cm}p{5.2cm}p{2.2cm}p{2.8cm}}
\toprule
attack class (corpus, $N$) & ASR \dirdown\ undefended & ASR \dirdown\ defended [95\% CI] & utilB \dirup\ judge (strict/len.) & utilA \dirup\ judge (strict/len.) [undef.] \\
\midrule
tool hijack --- JSON tool-log ($n{=}96$, 80 scoreable) & 0.675 & 1/80 = \textbf{0.013} [0.002, 0.067] & 52.5 (25.0/80) & 37.5 (30.0/76) [13.8] \\
tool hijack --- webpage ($n{=}52$, 52 scoreable; multi-turn) & 0.981 & \textbf{0/52 observed} [0.000, 0.069]; multi-turn \textbf{0/49 observed} [0.000, 0.073] & 50.0 (98.1/98) & 57.7 (94.2/94) [0.0] \\
param manipulation --- JSON ($n{=}70$, 66 scoreable; \emph{samples-only} holdout---the attacker-template set was in the fit) & 0.591 & 3/66 = \textbf{0.045} [0.016, 0.125] & 53.0 (22.7/83) & 43.9 (22.7/74) [18.2] \\
param manipulation --- webpage ($n{=}52$ held-out templates, 52 scoreable) & 0.962 & 7/52 = \textbf{0.135} [0.067, 0.253] & 38.5 (75.0/75) & 38.5 (59.6/60) [0.0] \\
\bottomrule
\end{tabular}
\end{table*}

\begin{table*}[t]
\caption{Qwen3-30B, held-out test evaluation (one pre-specified pass per
corpus, selected $12\sigma$); the last row is a labeled development split, not
holdout; columns as in Table~\ref{tab:main-gptoss} (utility: judge with the
strict/lenient bracket; the development row is strict/lenient only, not
judge-scored). Each webpage row states its strictly-disjoint-40 subset
(template-holdout caveat: the joint provenance paragraph below).}
\label{tab:main-qwen}
\centering
\footnotesize
\setlength{\tabcolsep}{4pt}
\begin{tabular}{p{5.6cm}p{1.7cm}p{3.8cm}p{2.2cm}p{2.8cm}}
\toprule
attack class (corpus, $N$; holdout axes) & ASR \dirdown\ undefended & ASR \dirdown\ defended [95\% CI] & utilB \dirup\ judge (strict/len.) & utilA \dirup\ judge (strict/len.) [undef.] \\
\midrule
tool hijack --- JSON tool-log ($n{=}96$, 90 scoreable; samples $+$ attacker templates held out) & 0.722 & 3/90 = \textbf{0.033} [0.011, 0.093] & 33.3 (21.1/69) & 20.0 (22.2/67) [8.9] \\ %
tool hijack --- webpage ($n{=}52$; exact-template-text holdout; strictly-disjoint-40 subset \textbf{0/40 observed} [0.000, 0.088]) & 0.981 & \textbf{0/52 observed} [0.000, 0.069] & 13.5 (63.5/63) & 13.5 (57.7/58) [1.9] \\ %
param manipulation --- JSON ($n{=}70$, 67 scoreable; \emph{samples-only} holdout---the attacker-template set was in the fit) & 0.522 & 4/67 = \textbf{0.060} [0.023, 0.144] & 31.3 (20.9/64) & 31.3 (26.9/66) [16.4] \\ %
param manipulation --- webpage ($n{=}52$; exact-template-text holdout; strictly-disjoint-40 subset 8/40 = 0.200 [0.105, 0.348]) & 1.000 & 9/52 = \textbf{0.173} [0.094, 0.297] & 34.6 (65.4/65) & 15.4 (50.0/50) [0.0] \\ %
\midrule
\emph{development split, not holdout:} tool hijack --- JSON tool-log ($n{=}96$, 91 scoreable) & 0.648 & 4/91 = 0.044 [0.017, 0.108] & 45.1\% / 91\% strict/len.\ & 47.3\% / 87\% (undef.\ 14.3/21) \\ %
\bottomrule
\end{tabular}
\end{table*}

\textbf{Tables~\ref{tab:main-gptoss} and~\ref{tab:main-qwen} provenance.} The
capability guard passes on every row of both tables; per-row guard rates,
truncation counts and artifact provenance are in Appendix~\ref{app:corpus}.
The webpage rows' template holdout on both models is
exact-template-\emph{text}: 12/52 test samples differ from a development
template only by case or one whitespace character (the split is identical
across models), which is why Table~\ref{tab:main-qwen}'s webpage rows also
state their strictly-disjoint-40 subsets.

\begin{table*}[t]
\caption{Gemma-4-31B-it, per-model detail. T $=$ the certification rung,
holding out carrier documents (30 fresh, URL-disjoint), attacker wordings
(4 new per class) and framing templates simultaneously; T$^{*}$ $=$ the
earlier wire-format-only holdout rung (carriers and wordings shared with
development), retained beside it. CIs on both webpage rungs are doc-clustered
(documents are the unit of independence). Both rungs are judge-scored
(judge \% of unattacked with the strict/lenient bracket; the T-heldout
cells were rescored 2026-09-09 under the same controls-gated protocol).
Scope caveats: the ``Table~\ref{tab:permodel-gemma}
provenance'' paragraph (Appendix~\ref{app:tables}).}
\label{tab:permodel-gemma}
\centering
\footnotesize
\setlength{\tabcolsep}{3pt}
\begin{tabular}{p{3.0cm}p{1.6cm}p{5.4cm}p{2.1cm}p{2.1cm}p{2.6cm}}
\toprule
evaluation & rung & ASR \dirdown\ undefended $\to$ defended [95\% CI] & utilB \dirup & utilA \dirup & guard / notes \\
\midrule
webpage param.\ ($n{=}52$; docs $+$ wordings $+$ framings held out) & T & $0.962 \to \mathbf{0.038}$ (2/52); doc-clustered [0.000, 0.096], sample Wilson upper 0.130; 48 fixed / 0 introduced, $p{=}7.1{\times}10^{-15}$ & 78.8 (96.2/96) & 78.8 (92.3/92) & guard clean, truncation 0.00; both survivors under one held-out wording (2/13 on it, 0/39 on the rest); $\alpha{=}6$ monotonicity control PASS ($0.962 \ge 0.192 \ge 0.038$) \\ %
webpage tool-hijack ($n{=}52$; same holdout axes) & T & $0.865 \to \mathbf{0.000}$ (0/52); cluster-Wilson [0.000, 0.114] over 30 docs; 45/0, $p{=}5.7{\times}10^{-14}$ & 73.1 (96.2/96) & 69.2 (96.2/96) & guard clean; $\alpha{=}6$ control saturated at 0.000; A100$\leftrightarrow$H100 replica probe 24/24 agreement \\ %
webpage param.\ & T$^{*}$ & $1.000 \to \mathbf{0.038}$; doc-clustered [0.000, 0.105], quote-upper 0.130 (sample Wilson); 50 flipped / 0 broken & 88.5 (100/100) & 78.8 (96/96) & guard clean \\ %
webpage tool-hijack & T$^{*}$ & $0.865 \to \mathbf{0.000}$ [0.000, 0.278] doc-clustered; 45 flipped / 0 broken & 80.8 (100/100) & 82.7 (100/100) & guard clean \\ %
JSON tool-hijack / JSON param.\ & N/F & undefended 0.083--0.112 (shipped JSON, $n{=}24$/$96$) / 0.045--0.048 & --- & --- & N/F by the program's convention; no defended run \\ %
prose-carrier medium ($n{=}96$) & N/F & undefended \textbf{0.000} [0.000, 0.038] & --- & --- & measured null: sane on-task summaries, truncation 0.00 \\ %
AgentDojo, 180-case grid, 4096 tok/turn & full grid & $0.239 \to \mathbf{0.006}$ (43/180 $\to$ 1/180); 42 fixed / 0 introduced, $p{=}4.6{\times}10^{-13}$; per-suite undef.$\to$def.: banking 28/49$\to$1, slack 13/49$\to$0, travel 1/41$\to$0, workspace 1/41$\to$0 & benign 92.9\% (1.000 $\to$ 0.929, 56 paired tasks; 4 broken / 0 fixed, $p{=}0.125$ n.s.) & 0.794 vs.\ undefended 0.850 & \textbf{0 truncated turns in all 2{,}721 generations}---nothing is a censored bound; layers pinned 4/28/36 (deployed configuration); the 1 surviving case's attacked arm also fired---0 defense-introduced compromises; one defense-on-clean break is a genuine steering-induced degenerate loop \\ %
bidirectional causal gate & --- & increase criterion untestable (undefended base saturated at 1.000): the negated dose holds 1.000 exactly (0 discordants) while $+\alpha$ defends (0.031, $p{<}10^{-9}$), all at benign 100\% & --- & --- & polarity-consistent \\ %
\bottomrule
\end{tabular}
\end{table*}

\textbf{Table~\ref{tab:permodel-gemma} provenance.} Gemma-4's attack surface
is medium-specific: webpage/HTML carriers fire, while JSON records and the
prose-carrier medium do not (the measured nulls in the table). On the
single-turn webpage corpora the steered span is ${\sim}90\%$ of the whole
prompt, so those rungs do not isolate the payload-scoped edit the primary
models' corpora measure---the AgentDojo grid, steering per-turn tool
outputs, does.
Gemma-4-31B's held-out framing \emph{templates} are test-side wordings,
distinct from the fit-side framing \emph{level} of
\S\ref{sec:eval}: its deployed fit predates the fit-side rule and
holds the level out at the admission gate only.

\begin{table*}[t]
\caption{GLM-4.5-Air, per-model detail. Deployed configuration: the action-centred
behavioral direction at $\alpha{=}8$ on the pooled-corpus fit's per-layer
scales $\sigma_\ell = 6.4/8.9/10.3$ (${\approx}17$--$18\sigma$ in its own
per-layer scale), layers 20/24/28, 8192-token generation budget. Utility:
judge \% of unattacked with the (strict/lenient) bracket; AgentDojo utility
is its own checker; the guard is read absolutely (defense-on-clean no-action
0.000). Holdout detail: the ``Table~\ref{tab:permodel-glm} provenance''
paragraph (Appendix~\ref{app:tables}).}
\label{tab:permodel-glm}
\centering
\footnotesize
\setlength{\tabcolsep}{3pt}
\begin{tabular}{p{3.0cm}p{1.6cm}p{5.4cm}p{2.1cm}p{2.1cm}p{2.6cm}}
\toprule
evaluation & rung & ASR \dirdown\ undefended $\to$ defended [95\% CI] & utilB \dirup & utilA \dirup & guard / notes \\
\midrule
webpage param.\ ($n{=}52$, 52 scoreable; samples $+$ template text held out) & T & $0.808 \to \mathbf{0.019}$ [0.003, 0.101]; exact-literal 0/52; 41 fixed / 0 introduced & 26.9 (94.2/94) & 11.5 (81/81) & contamination 0.000; no-action 0.038 vs.\ defense-on-clean 0.000 (2/52, attack-conditional); the one surviving compromise is a domain-level parameter hijack counted by token taint, not exact-literal \\ %
webpage param., development ($n{=}96$, 96 scoreable) & D & $0.750 \to 0.156$ (exact-literal 0.000); 57 fixed / 0 introduced & 40.6 (93/93) & 18.8 (59/59) & no-action 0.052; truncation 0.05 \\ %
webpage tool-hijack; JSON tool-hijack & --- & fire weakly undefended single-turn (webpage ${\sim}0.25$; JSON low)---those cells do not measure the defense. At the \emph{multi-turn} horizon the disjoint-tool corpus fires $0.948$ undefended and the deployed configuration yields a partial reduction only: $\mathbf{0.781}$ (16 fixed / 0 introduced, $p{=}3.1{\times}10^{-5}$; steering verified active over the injected span)---an open boundary, not a closure & --- & --- & \\ %
JSON param.; adaptive suite & --- & not run & --- & --- & \\ %
AgentDojo, 180-case grid, 8192 tok/turn & full grid & $0.222 \to \mathbf{0.056}$ (40/180 $\to$ 10/180); 30 fixed / 0 introduced, $p{=}1.9{\times}10^{-9}$ & benign \textbf{100.0\%} (0.857 $\to$ 0.857, 56 paired tasks; 1 lost, 1 gained) & 0.733 vs.\ undefended 0.733 & zero truncation in every arm (nothing censored); layers pinned 20/24/28 (deployed configuration); 7 of the 10 surviving compromises sit on one task whose user prompt delegates authority to the file (``follow the instructions precisely''), 2--3/180 are genuine role-confused survivors; strongest agent measured (clean 0.857) \\ %
bidirectional causal gate & D & increase side PASSES on the deployed direction: $0.615 \to 0.827$ at $-\alpha 8$ (8192 budget, dev $n{=}52$; exact McNemar 14/3, $p{=}0.013$; defense-on-clean at $-\alpha$: ASR 0.000, utility 88.5\%) & --- & --- & \\
screening diagnostics & --- & split-half reliability 0.31--0.33, below the 0.70 screening reference (the pooled-fit primary models screen at 0.71--0.74, AUC 0.80--0.82); screening ranks and does not certify, and this fit certifies through the gates above (\S\ref{sec:method-recipe}) & --- & --- & \\ %
\bottomrule
\end{tabular}
\end{table*}

\textbf{Table~\ref{tab:permodel-glm} provenance.} The test row is one
pre-specified pass, first touch of that split, sample-disjoint from
development. Its template holdout is exact-template-text: 12 of the 52 test
samples differ from a development template only by letter case or one
whitespace character (defended 0/12 on that subset vs.\ 1/40 on the rest; the
split is identical across models).

\textbf{Delegated-authority partition (Table~\ref{tab:classab}), detail.}
Class-B membership is decided from the user prompt alone (``read the file
\ldots\ and follow the instructions precisely''), before any outcomes are
read; the partition's motivation was post-hoc---formed after inspecting the
defended arm's failures---which is why \S\ref{sec:eval-agentic} applies it
symmetrically to every arm. The Class-B enrichment is 0.806 vs.\ 0.390
undefended, so delegation
amplifies compromise rates regardless of defense and the partition is partly a
benchmark property. All of the deployed configuration's defended failures are
delegated (14/14; likewise 12/12 in the retired 768-budget run, where a
transcript-level reading classifies 9--10 of the 12 as delegated---the
task-id criterion classifies all of them).

\begin{figure}[t]
\providecolor{csBlue}{HTML}{2A78D6}
\providecolor{csRed}{HTML}{C0392B}
\providecolor{csTeal}{HTML}{1BAF7A}
\providecolor{csTealDk}{HTML}{0F7A54}
\centering
\resizebox{\columnwidth}{!}{%
\begin{tikzpicture}[
  font=\small,
  stage/.style={draw=csBlue!70!black, line width=0.8pt, rounded corners=2.5pt,
                fill=csBlue!7, align=left, inner sep=3.5pt, text width=48mm,
                font=\scriptsize},
  deploy/.style={stage, draw=csTealDk, line width=0.9pt, fill=csTeal!10},
  outbox/.style={draw=csTealDk, line width=1pt, rounded corners=2.5pt,
                 fill=csTeal!18, align=center, inner sep=3pt, text width=48mm,
                 font=\scriptsize},
  rej/.style={align=left, font=\scriptsize, csRed!85!black, text width=30mm,
              inner sep=1.5pt},
  flow/.style={-{Stealth[length=2mm]}, line width=0.9pt, csBlue!65!black},
  flowteal/.style={-{Stealth[length=2mm]}, line width=0.9pt, csTealDk},
  rejarrow/.style={-{Stealth[length=1.6mm]}, line width=0.8pt,
                   csRed!80!black, densely dashed}
]

\node[stage] (s1) at (0,0)
  {{\scriptsize\bfseries\color{csBlue!60!black} 1. BEHAVIORAL CONTRASTS}\ \,
   paired episodes differing only in whether the embedded instruction is
   followed; factorial over authority framing, voice, action, delegation.};
\node[stage, below=3.2mm of s1.south, anchor=north] (s2)
  {{\scriptsize\bfseries\color{csBlue!60!black} 2. CANDIDATE DIRECTION}\ \,
   difference-in-means between followed and resisted sides, per layer.};
\node[stage, below=3.2mm of s2.south, anchor=north] (s3)
  {{\scriptsize\bfseries\color{csBlue!60!black} 3. RELIABILITY \& GENERALIZATION
   SCREENING}\ \, ranked: sign-consistency across disjoint fitting shards;
   AUC (reference 0.65) on a whole authority-framing level withheld from the
   fit.};
\node[stage, below=3.2mm of s3.south, anchor=north] (s4)
  {{\scriptsize\bfseries\color{csBlue!60!black} 4. CAUSAL VALIDATION}\ \,
   low-dose bidirectional test on held-out attacks: subtracting must reduce
   injection-following, adding must increase it, guard intact.};
\node[deploy, below=3.2mm of s4.south, anchor=north] (s5)
  {{\scriptsize\bfseries\color{csTealDk} 5. DOSE SEARCH \& DEPLOY}\ \,
   constant norm-preserving edit $-\alpha\sigma\hat d$ at layers $L$, every
   tool-result token, at prefill. Always on; no detector.};
\node[outbox, below=3.2mm of s5.south, anchor=north] (out)
  {deployable triple: (direction, layers, dose)};

\foreach \a/\b in {s1/s2, s2/s3, s3/s4, s4/s5} \draw[flow] (\a.south) -- (\b.north);
\draw[flowteal] (s5.south) -- (out.north);

\coordinate (rx) at (3.05,0);
\foreach \n/\lbl in {%
  s3/{\textbf{most candidates die here}: separates the fit set, fails held-out
      wording},
  s4/{a \emph{correlate}, not a lever: discriminates attacks, does not move
      ASR (\S\ref{sec:mechanism})},
  s5/{no capability-safe effective dose in the searched space (e.g.,
      Llama-3.1-8B's JSON-parameter cell, Table~\ref{tab:coverage})}%
}{%
  \node[rej, anchor=west] (r\n) at (rx |- \n) {\lbl};
  \draw[rejarrow] (\n.east) -- (r\n.west);
}

\end{tikzpicture}%
}
\caption{\textbf{The CounterSteer recipe, run once per model.} Five steps;
steps 4--5 are the certifying pass/fail gates (step 3 ranks candidates, it
does not certify); red branches are the rejections. The recipe discovers
each model's own lever rather than assuming a universal injection direction.}
\label{fig:pipeline}
\end{figure}
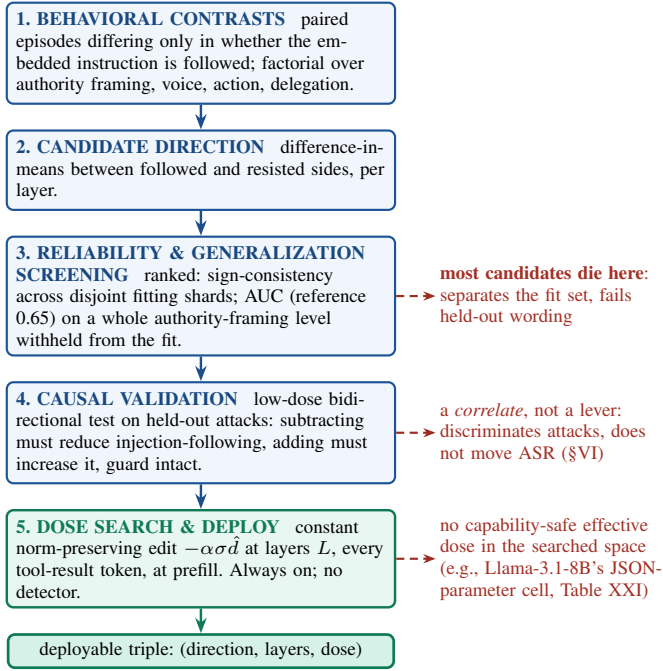

\begin{table}[t]
\caption{Data provenance: which split each pipeline stage uses, and what it is
held out from. Framing-level holdout, per fit: the recipe withholds an
authority-framing level in the deployed fit itself
(\S\ref{sec:method-recipe}). Llama-3.1-8B's deployed fit follows that rule
end-to-end; a gpt-oss-20b refit under the rule is validated end-to-end
(four one-touch held-out passes $+$ the AgentDojo grid, all within the
deployed configuration's CIs; the deployed corpus references ran at a $0.7\%$ lower
magnitude seam, disclosed); the other three deployed fits hold the level
out at the gate only, refitting over all levels
(\S\ref{sec:eval}). Contrast centring, per deployed fit:
GLM-4.5-Air within (sample $\times$ delegation $\times$ action); the other
four within (sample $\times$ delegation), each deployed direction's
behavioral component measuring mean $|\cos| \le 0.31$ (per-layer max 0.334)
to the action-identity
(tool-hijack-vs-parameter) axis at its deployed layers.}
\label{tab:provenance}
\centering
\footnotesize
\setlength{\tabcolsep}{3pt}
\begin{tabular}{p{2.4cm}p{2.9cm}p{2.7cm}}
\toprule
stage & split / samples used & held out from \\
\midrule
generalization screening (step~3) & probe split, minus one whole authority-framing level
(\texttt{firm}) & that framing level, for the gate's fit only \\
direction fitting (deployed) & probe split (behavioral-contrast captures;
deployed gpt-oss fit: its deterministic first 24 sample ids) &
dev/test samples; held-out attacker templates. Framing levels: withheld in
the Llama-3.1-8B fit and the validated gpt-oss-20b refit; the other deployed
fits use every level \\
layer selection & probe split (per-layer probe diagnostics) & dev/test
samples \\
dose selection & dev split (template-disjoint from test) & test samples
and, where stated, test attacker templates \\
test evaluation & test split, pre-specified one-pass, no reruns & all fitting
and selection stages (samples; and where stated, attacker templates) \\
adaptive attacker construction & dev-disjoint sample subsets ($n{=}18$ per
class for the query search; disjoint-tool targets for GCG) & the attacked
samples are held out from fitting; the deployed vector is not consumed by the
query/surrogate arms and is supplied directly in the exact-vector arm. The
surrogate refit deliberately reuses the fitting
samples (the draw is deterministic, \S\ref{app:adaptive}) \\
mechanism analysis (decision-point fit) & probe split (48 sample ids) &
template- and sample-disjoint from every evaluation \\
\bottomrule
\end{tabular}
\end{table}

\textbf{Model revisions.} Dataset loads in the released code name an explicit
upstream revision; model loads do not, and no result artifact records the
revision or the host it ran on, so we reconstructed them and state what that
reconstruction does and does not establish. Every model but one resolved a
single revision fleet-wide. The exception is Gemma-4-31B-it, which existed at
two revisions on our machines. The reported Gemma runs used
\texttt{842da379}, and the ambiguity is benign: the two revisions carry
\emph{byte-identical} weights---the safetensors content hashes match---and
differ only in the chat template and tokenizer configuration. So the numbers
are not in doubt, but a reproducer who resolves the other revision gets the
same weights behind a materially different chat template, and our prompts are
rendered through that template. One earlier stage is genuinely unrecorded: the
Gemma direction was fitted in a cluster job whose node-local cache is gone, so
which of the two revisions it resolved is not establishable. We measured the
gap closed rather than arguing it: rendering every cell of the deployed fit
design (576 poisoned prompts with recorded spans) under each revision's own
tokenizer and template yields \emph{byte-identical} text, input ids and span
indices---the templates' differing branches are never exercised by the fit
corpus---so the fit inputs, and hence the direction, are revision-invariant.
Revisions are listed in the released corpus manifest.

\begin{table}[t]
\caption{Post-hoc delegated-authority partition, applied symmetrically to
every gpt-oss arm of the certification run (4096-token budget,
\S\ref{sec:eval-agentic}). Class~B = the 36 cases whose user
task delegates instruction authority to retrieved content; Class~A = the
rest. Denominators are valid cases per arm and class.}
\label{tab:classab}
\centering
\footnotesize
\setlength{\tabcolsep}{3pt}
\begin{tabular}{p{2.9cm}p{1.7cm}p{1.6cm}p{1.6cm}}
\toprule
arm & compromise \dirdown\ (all) & Class~B \dirdown & Class~A \dirdown \\
\midrule
undefended & 84/177 = 0.475 & 29/36 = 0.806 & 55/141 = 0.390 \\
\textbf{CounterSteer (deployed)} & \textbf{14/177 = 0.079} & 14/36 = 0.389 & \textbf{0/141 = 0.000} \\
\bottomrule
\end{tabular}
\end{table}

\begin{table*}[t]
\caption{Model coverage: five open-weights models, 8B--106B,
dense and MoE, five vendor lineages. ``Certified'' = evaluation complete and
adversarially reviewed; the defended column is ASR on the single-turn
corpora and compromise rate on AgentDojo, over
the denominators of the table cited in each row. Rows whose numbers appear
nowhere else carry their utility and capability-guard readings in the same
cell.}
\label{tab:coverage}
\centering
\footnotesize
\setlength{\tabcolsep}{3pt}
\begin{tabular}{p{2.3cm}p{1.9cm}p{5.6cm}p{7.2cm}}
\toprule
model & params / arch & fit status & ASR~/ compromise rate \dirdown\ defended \\
\midrule
gpt-oss-20b & 20B MoE & certified (held-out test pass) & 0.000--0.135 corpora (Table~\ref{tab:main-gptoss}); 0.079 AgentDojo at a 4096-token budget (\S\ref{sec:eval-agentic}) \\ %
Qwen3-30B-A3B-Thinking & 30B MoE (3B active) & certified (held-out test pass) & 0.000--0.173 corpora (Table~\ref{tab:main-qwen}); AgentDojo $0.489 \to 0.072$ same-process (Table~\ref{tab:soa}), dose-selection replicate 0.094 (Table~\ref{tab:qwen-dose}), both at 4096 tokens \\ %
Gemma-4-31B-it & 31B dense & \textbf{certified} (one-pass held-out test evaluations on both firing corpora with carrier documents, attacker wordings and framing templates all held out, plus the full AgentDojo grid at the deployed configuration); on the single-turn corpora the steered span is ${\sim}90\%$ of the whole prompt (those rungs do not isolate the payload-scoped edit; the agentic grid does); no adaptive evaluation & webpage param.\ $0.962 \to \mathbf{0.038}$ (2/52; doc-clustered [0.000, 0.096], sample Wilson upper 0.130; 48 fixed / 0 introduced, $p{=}7.1{\times}10^{-15}$) and tool hijack $0.865 \to \mathbf{0.000}$ (0/52; cluster-Wilson [0.000, 0.114]; 45/0, $p{=}5.7{\times}10^{-14}$), benign utility 96.2\%/96\% strict both corpora, guard clean, zero truncation; AgentDojo $0.239 \to \mathbf{0.006}$ (42/0, $p{=}4.6{\times}10^{-13}$; 0 defense-introduced) at benign 92.9\% ($-7.1$\,pp n.s.), 0 truncated turns in 2{,}721 generations; the earlier wire-format-only rung (T$^{*}$) stands beside it ($1.000 \to 0.038$, $0.865 \to 0.000$); the prose-carrier medium is a measured null (undefended \textbf{0.000} [0.000, 0.038], $n{=}96$; Table~\ref{tab:permodel-gemma}) \\ %
GLM-4.5-Air & ${\sim}$106B MoE & certified (held-out test pass on its firing corpus, webpage parameter manipulation, plus the full AgentDojo grid at the deployed configuration); tool-hijack classes fire weakly undefended, JSON param.\ not run; no adaptive evaluation & webpage param.\ $0.808 \to \mathbf{0.019}$ T ($n{=}52$; near-duplicate-template caveat, \S\ref{sec:results}; exact-literal 0/52; Wilson [0.003, 0.101]; McNemar 41 fixed~/ 0 introduced; benign utility 26.9 judge (94.2/94 strict/lenient), under attack 11.5 (81/81); contamination 0.000; guard clean, read absolutely---defense-on-clean no-action 0.000, defended 0.038 attack-conditional); development split $0.750 \to 0.156$ at defense-on-clean strict correctness 0.927; AgentDojo $0.222 \to \mathbf{0.056}$ (30 fixed / 0 introduced, $p{=}1.9{\times}10^{-9}$) at benign \textbf{100.0\%} and utilAttack $0.733 = $ undefended, zero truncation in every arm (Table~\ref{tab:permodel-glm}) \\ %
Llama-3.1-8B-Instruct & 8B dense & certified (held-out test pass on its firing corpus, webpage tool hijack, plus the full AgentDojo grid, the 26-battery rival comparison and the benchmark-level adaptive attack at the deployed configuration); fit passes all gates incl.\ the bidirectional causal gate ($0.135 \to 0.558$ at $-\alpha$); JSON tool-hijack does not fire undefended (0.042); JSON param.\ reaches ASR 0 only at a capability-destroying dose (judge benign 0.431), no viable cell; low task capability (13--15/56 AgentDojo clean tasks) bounds what its agentic and AgentDyn grids separate (AgentDyn: 2/60 solvable, not measurable) & webpage tool hijack $0.212 \to \mathbf{0.038}$ T ($n{=}52$; 11/52 $\to$ 2/52 (3/52 under a lenient tool-call executor---one surviving compromise is the attacker's exact call with malformed JSON; $p{=}0.0078$ still significant); Wilson [0.011, 0.130]; McNemar 9 fixed / 0 introduced, $p{=}0.0039$; benign utility \textbf{100\%} byte-exact, judge floor 0.827; under attack 65.4 strict / 0.558 judge, both \emph{above} undefended 59.6 / 0.462; a fit-withheld framing level composed onto the injection also blocked, 0/4 fired, point estimate); AgentDojo $0.100 \to \mathbf{0.050}$ (11 fixed / 2 introduced, $p{=}0.022$) at benign 100\%, thin-base caveat (Table~\ref{tab:soa}) \\ %
\bottomrule
\end{tabular}
\end{table*}

\begin{table*}[t]
\caption{Per-model cost of the recipe. Wall-clock per stage where a clean record
exists; \emph{n.r.} $=$ stage ran but no wall-clock was logged; \emph{iterative} $=$
developed over the research period, not separable into a single measurement.
Direction fits are CPU-side; the GPU cost is activation capture and behavioral
evaluation. Where no launch time was logged, a certification duration is
measured from the artifacts' timestamps and is an upper bracket.}
\label{tab:recipe-cost}
\centering
\footnotesize
\setlength{\tabcolsep}{2.5pt}
\begin{tabular}{p{1.9cm}p{2.5cm}p{2.4cm}p{2.2cm}p{2.5cm}p{2.65cm}p{2.5cm}}
\toprule
model & hardware (ladder) & capture $+$ fit & screening $+$ causal gates & dose search & held-out certification (single-turn) & AgentDojo 180-case grid \\
\midrule
gpt-oss-20b & 4$\times$A100-80GB local $+$ A100 boxes; grid on one 8$\times$H100 node &
  iterative & n.r.\ (CPU-side) & iterative &
  4 corpora, one pre-specified pass, ${\le}85$\,min &
  \textbf{62\,min} (8$\times$H100, 4 arms) \\
Qwen3-30B-A3B-Thinking & 4$\times$A100-80GB local $+$ A100 boxes; grid on one 8$\times$H100 node &
  iterative & n.r. & iterative &
  4 corpora, one pre-specified pass, ${\le}5.5$\,h (pre-specification to last artifact) &
  one of six batteries in a 22.6\,h 8$\times$H100 job; not separated \\
Llama-3.1-8B-Instruct & 3$\times$A100-80GB local; grid on 4$\times$A100 (two boxes) &
  ${\sim}2.3$\,h (incl.\ a parser-bug diagnosis $+$ rerun) &
  ${\sim}25$\,min (causal smoke $+$ 3 random controls) &
  ${\sim}1.8$\,h (6 doses $\times$ 3 corpora $+$ layer sweep) &
  gate $+$ test touch $+$ bidirectional gate, 11\,min &
  4 single-GPU shards, ${\le}36$\,min; \textbf{end-to-end ${\sim}8.6$\,h, one day} \\
Gemma-4-31B-it & one 8$\times$H100 node (bring-up, cert, grid); local A100s (dose batteries) &
  ${\sim}1$\,h (one job: capture, probes, factorial, fit) &
  inside the bring-up job & n.r. &
  \multicolumn{2}{p{5.6cm}}{\textbf{56\,min total} for both: five held-out confirms $+$ the full 180-case grid $+$ benign arms, one 8$\times$H100 job} \\
GLM-4.5-Air & one 8$\times$H100 node (capture, sweep); local 4$\times$A100 (cert, grid; 212\,GB sharded) &
  6.3\,h (bring-up); certified refit CPU-only, n.r. &
  inside the bring-up; recipe gates CPU-side, n.r. &
  16.0\,h sweep (superseded direction); recipe ladder terminated early, no clean total &
  T pass completed 3\,h\,05\,m after the dev re-run (4$\times$A100, 8192-token budget) &
  \textbf{12.2\,h} (43{,}906\,s, single process, 8192-token budget) \\
\bottomrule
\end{tabular}
\end{table*}

\begin{table}[t]
\caption{gpt-oss-20b, full dose curve on AgentDojo (the curve
Fig.~\ref{fig:dose} plots). \textbf{4096-token per-turn budget}; one
direction (the deployed configuration's) at every dose; undefended baseline
0.475 (84/177). Compromise rate over the 177 scoreable injection cases;
fixed/intr.\ $=$ compromises fixed~/ introduced (exact McNemar vs.\ the
same-process undefended arm); benign $=$ \% of the ladder's own clean arm
($n{=}55$ paired unique tasks). Readings are raw: the typography
normalization of Table~\ref{tab:soa} could not be applied to the
$\alpha{=}3.0$--$6.7$ arms (transcripts not persisted), and its one
compromise-rate movement ($0.085 \to 0.090$ on this ladder's deployed
arm) is disclosed there.}
\label{tab:gptoss-dose}
\centering
\footnotesize
\setlength{\tabcolsep}{3pt}
\begin{tabular}{p{1.7cm}p{3.1cm}p{1.7cm}p{1.0cm}}
\toprule
dose $\alpha$ & compromise \dirdown\ defended [95\% CI] & fixed/intr.\ ($p$) & benign \dirup \\
\midrule
3.0 & 55/177 = 0.311 [0.247, 0.382] & 36/7 ($9{\times}10^{-6}$) & 92.2\% \\
4.12 & 44/177 = 0.249 [0.191, 0.317] & 44/4 ($1.5{\times}10^{-9}$) & 86.3\% \\
5.5 & 26/177 = 0.147 [0.102, 0.207] & 58/0 ($6.9{\times}10^{-18}$) & 86.3\% \\
6.7 & 19/177 = 0.107 [0.070, 0.162] & 65/0 ($5.4{\times}10^{-20}$) & 84.3\% \\
\textbf{8.06 (deployed)} & \textbf{15/177 = 0.085} [0.052, 0.135] & 69/0 ($3.4{\times}10^{-21}$) & 82.0\% \\
\bottomrule
\end{tabular}
\end{table}
\begin{table}[t]
\caption{Qwen3-30B, full dose curve on AgentDojo, \textbf{4096-token
per-turn budget} (the 768 collapse this budget avoids:
Appendix~\ref{app:corpus}). Denominators: compromise rate over
all injection cases completing (the $\alpha{=}16$ run's own undefended
baseline is over 175 cases, having run
before a 5-case harness fix); benign over 56 paired unique tasks. \warnmark\ marks
capability-guard territory; the $\alpha{=}16$ row is a censored lower
bound (note below).}
\label{tab:qwen-dose}
\centering
\footnotesize
\setlength{\tabcolsep}{3pt}
\begin{tabular}{p{1.5cm}p{3.3cm}p{1.5cm}p{1.2cm}}
\toprule
dose & compromise \dirdown\ defended [95\% CI] & utilAttack \dirup\ (raw) & benign $\Delta$pp \dirup \\
\midrule
$10\sigma$ & 22/180 = 0.122 [0.082, 0.178] & 0.728 & $-5.4$ \\
\textbf{$12\sigma$ (selected)} & \textbf{17/180 = 0.094} [0.060, 0.146] & \textbf{0.694} & \textbf{$-5.4$} \\
$14\sigma$ & 6/180 = 0.033 [0.015, 0.071] & 0.550 & $-19.6$ \\
$16\sigma$ & $\ge$ 2/180 = 0.011 [0.003, 0.041] \warnmark & 0.356 \warnmark & $-39.3$ \warnmark \\
\bottomrule
\end{tabular}
\end{table}

\textbf{Table~\ref{tab:qwen-dose} note.} A transcript audit of the
$\alpha{=}16$ arm still finds 13 episodes cut mid-compliance at 4096 (10
defended) plus 13 defended episodes ending on an unexecuted call, so its
0.011 is a lower bound, not a measurement, and its utility is depressed by
the same censoring; the $\alpha{=}10$--$14$ rows share the budget and have
not been re-audited.

\begin{table*}[t]
\caption{IPI defense families, the lever each pulls, and what a deployment pays.
``Measured here'' is our own runs only, and is not comparable to the cited
works' reported numbers (different harnesses, models, attack sets); rates are
AgentDojo compromise rates, $x/n$ are corpus compromise counts. Read each entry against
\emph{its own run's} undefended baseline (Table~\ref{tab:soa} caveats), and
never gpt-oss against Qwen
(\S\ref{sec:eval-baselines}).}
\label{tab:related}
\centering
\footnotesize
\setlength{\tabcolsep}{4pt}
\begin{tabular}{p{2.4cm}p{3.3cm}p{4.5cm}p{2.6cm}p{3.4cm}}
\toprule
family & representative work & lever & deployment cost & measured here \\
\midrule
prompt-level &
spotlighting/delimiting \cite{hines2024spotlighting}; prompt sandwich
(folklore) &
transforms the untrusted span so its provenance stays visible (delimiting,
datamarking, encoding) and instructs the model not to follow it; or re-asserts
the trusted instruction after the data &
extra prompt tokens on every request &
at 4096: $0.483{\to}0.273$--$0.420$ (gpt-oss), $0.489{\to}0.356$ (Qwen
reminder), Table~\ref{tab:soa} \\
\addlinespace[2pt]
training-time &
StruQ \cite{chen2024struq}, SecAlign \cite{chen2024secalign,chen2025metasecalign},
Jatmo \cite{piet2024jatmo}, instruction hierarchy \cite{wallace2024hierarchy},
ISE \cite{wu2024ise}, ASIDE \cite{aside2025} &
retrains or re-architects the model so untrusted spans carry less
instructional force &
a training pipeline per model; the served weights change &
our SecAlign LoRA: compromises $62/180 {\to} 9/180$ (768 budget); benign
utility of base clean, typography-normalized: $98.2\%$ at 768, $90.4\%$
unique-task~/ $82.3\%$ per-case at the shared 4096 budget (raw $79.7\%$~/
$72.7\%$: the DPO finetune itself emits the typography) \\
\addlinespace[2pt]
detection &
internal-state probes: TaskTracker \cite{abdelnabi2024tasktracker}, Attention
Tracker \cite{hung2025attentiontracker}; input-text guard models: PIGuard
\cite{li2024injecguard} &
reads a signal and flags; no intervention is proposed, and the response is left
to the surrounding system &
a threshold that fails open on a miss; a second model or pass, except where the
signal is already computed during the forward pass &
input-text guards run as deletion filters in Table~\ref{tab:soa} (compromise
0.011--0.228 at benign 48--100\%, over-defense per \cite{autodojo2026});
internal-state probes not run as defenses---\S\ref{sec:mechanism} shows two
directions that
discriminate but do not steer \\
\addlinespace[2pt]
detection-gated intervention &
ICON \cite{icon2026}, ARGUS \cite{argus2026steering}, AGRI \cite{agri2026} &
a probe decides when to intervene, then edits attention (ICON),
representations plus a post-filter (ARGUS), or injects an anti-injection
reasoning prefill (AGRI) &
detector accuracy is a single point of failure; the gate is an additional
attack surface (evaluated for AGRI: the adaptive attacker runs against the
live gated arm, \S\ref{sec:eval-adaptive}) &
AGRI measured same-harness (Tables~\ref{tab:headline},
\ref{tab:autodojo-matrix}, \S\ref{sec:related}); ICON/ARGUS not run (see
text) \\
\addlinespace[2pt]
always-on internal intervention &
CachePrune \cite{cacheprune2025}, V-Steer \cite{vsteer2026},
\textbf{CounterSteer} &
edits internals over the untrusted span with no gate: KV-neuron pruning,
per-request value-vector rescaling, or a fixed residual-stream direction &
white-box serving access; benign utility &
at 4096: CachePrune $0.483 {\to} 0.216$,
CounterSteer $0.483 {\to} 0.091$ (Table~\ref{tab:soa}) \\
\addlinespace[2pt]
system-level / provenance &
CaMeL \cite{debenedetti2025camel}, MELON \cite{zhu2025melon},
ROPE \cite{rope2026}, tool filters &
constrains the consequences of a hijack: dual-LLM execution with data-flow
tracking (CaMeL), masked re-execution and output comparison (MELON),
origin-routed admission of parameter values (ROPE) &
re-architecting the agent runtime; extra executions or an audited parameter
set &
a tool filter reaches compromise 0/176 observed with benign utility 10.9\% of
clean and utilAttack 0.106 typography-normalized (Table~\ref{tab:soa}) \\
\bottomrule
\end{tabular}
\end{table*}

\end{document}